\documentclass[12pt]{article}

\usepackage{graphicx, verbatim}
\usepackage{amsmath}
\usepackage{epsfig}
\usepackage{amsfonts}
\usepackage{amssymb}
\usepackage{subfigure,color}

\usepackage{float}
\restylefloat{table}

\newcommand{\be}{\begin{equation}}
\newcommand{\ee}{\end{equation}}
\newcommand{\bea}{\begin{eqnarray}}
\newcommand{\eea}{\end{eqnarray}}

\newcommand{\nn}{\nonumber}

\newcommand{\lp}{\left(}
\newcommand{\rp}{\right)}
\begin{document}

\begin{flushright}
DESY 14-127\\
NSF-KITP-14-089\\
\end{flushright}

\vskip 8pt

\begin{center} {\bf \LARGE {From Boltzmann equations  \\  to steady
      wall velocities}}
\end{center}

\vskip 12pt

\begin{center}
 {\bf Thomas Konstandin$^a$, Germano Nardini$^{a,b}$, Ingo Rues$^a$ }\\[4mm]
{\em $^a$DESY, Notkestr.~85, 22607 Hamburg, Germany } \\
{\em $^b$KITP, University Road 552, CA-93106 Santa Barbara, USA} \\
\end{center}

\vskip 20pt

\begin{abstract}
\vskip 3pt
\noindent

By means of a relativistic microscopic approach we calculate the
expansion velocity of bubbles generated during a first-order electroweak phase
transition. In particular, we use the gradient expansion of the
Kadanoff-Baym equations to set up the fluid system.  This turns out to
be equivalent to the one found in the semi-classical approach in the 
non-relativistic limit. Finally, by including hydrodynamic
deflagration effects and solving the Higgs equations of motion in the
fluid, we determine velocity and thickness of the bubble walls.  Our
findings are compared with phenomenological models of wall velocities.
As illustrative examples, we apply these results to three theories
providing first-order phase transitions with a particle content in the thermal 
plasma that resembles the Standard Model.
\end{abstract}

\newpage

\section{Introduction\label{sec:intro}}
Strong first-order cosmological phase transitions predict a variety of
interesting phenomena: gravitational waves \cite{Witten:1984rs,
Kosowsky:1991ua, Kosowsky:1992rz, Kosowsky:1992vn,
Kamionkowski:1993fg, Huber:2008hg}, baryogenesis \cite{Kuzmin:1985mm},
magnetic fields \cite{Vachaspati:1991nm} and many more. The
thermodynamic features of a phase transition, as e.g.~its critical
temperature, latent heat and order parameter, can be easily determined
using standard techniques \cite{Coleman:1977py, Callan:1977pt,
Linde:1980tt}. Its (out-of-equilibrium) dynamic properties are instead
more difficult to predict. Among these properties, for a first-order
phase transition the speed of the expanding bubbles and their wall
thickness are probably the most relevant.

In particular, little is known about the wall velocity in most
models. Determining it hinges on quantifying the friction that is
exerted by the fluid on the bubble wall. This requires a framework
that captures out-of-equilibrium features of the plasma. On a
technical level this can be achieved by solving Boltzmann
equations. This route has been followed for the Standard Model
(SM)~\cite{Dine:1992wr,Liu:1992tn, Moore:1995ua, MoAndPro} and for the Minimal Supersymmetric SM
(MSSM)~\cite{John:2000zq}.

A second way to quantify the friction is to use a phenomenological
approach~\cite{Ignatius:1993qn, Megevand:2009ut, Megevand:2009gh,
Sopena:2010zz, Huber:2011aa, Megevand:2012rt, Huber:2013kj, Megevand:2013hwa}. 
In this case, the friction is modeled by an additional
dissipative term in the Higgs equation of motion. This involves a free
friction coefficient that is inferred by matching to the full
Boltzmann treatment. Even though this approach gives reasonable
results in the small wall velocity limit, it has its limitations. For
example, it is not clear whether these results can be extrapolated to
supersonic wall speeds. Extensive numerical simulations of the phase
transitions have followed this approach~\cite{Ignatius:1993qn,
Hindmarsh:2013xza, Giblin:2014qia}.

In the present work we attempt to fill the gap between the full
Boltzmann treatment and the phenomenological approach. Rather than
adding an {\em ad hoc} term to the Higgs equation of motion, we solve
numerically the Boltzmann equations. Subsequently, we fit the obtained
friction in terms of the parameters characterizing a first-order phase
transition. The interpolations we provide can be easily applied
to models with first-order phase transitions.

From a technical point of view, the parametrization we provide assumes
a given set of particle species contributing to the friction. We
indeed consider a SM-like framework, in which the friction is
dominated by the electroweak gauge bosons and top
quarks~\cite{MoAndPro}. In extensions of the SM, however, any particle that is not too heavy 
and is strongly coupled to the Higgs contributes. For instance, in the parameter region of the MSSM
suitable for electroweak
baryogenesis~\cite{Carena:2008vj,Laine:2012jy} (but in tension with
LHC date~\cite{Cohen:2012zza,Curtin:2012aa,Carena:2012np} and possible
magnetogenesis~\cite{DeSimone:2011ek}), also stops participate in the
friction~\cite{John:2000zq}. Our parametrization then underestimates the friction in the
MSSM. On the other hand, it well applies to models beyond the SM with not too
many new degrees of freedom coupled to the Higgs. The gauge-singlet
extension belongs to this class of theories. It is weakly constrained
by present collider
measurements~\cite{Barger:2007im, Ashoorioon:2009nf, No:2013wsa, Ram:LHCsingl} and can
provide very strong two-stage phase transitions if the singlet acquires a vacuum
expectation value (VEV) before the electroweak symmetry breaking~\cite{Espinosa:2011ax}. 

The paper is organized as follows. From section \ref{sec:eom} to
section \ref{sec:transport} we rederive the fundamental Boltzmann and
Higgs equations in the Schwinger-Keldysh formalism. Compared to the
semi-classical derivation in Moore and Prokopec's paper~\cite{MoAndPro}
(called M\&P in the following), the resulting Kadanoff-Baym equations
have the advantage to allow for a systematic inclusion of quantum 
corrections. 
This should be relevant to develop a unifying framework
to determine the wall velocity and baryogensis~\cite{Kainulainen:2001cn, Prokopec:2003pj, 
Prokopec:2004ic, Konstandin:2013caa}. It also permits to
derive the transport equations for relativistic wall velocities. In
particular, we assume that the system is close enough to equilibrium
such that the flow ansatz we consider can be linearized in terms of
deviations from equilibrium. This approximation does not automatically
imply a small wall velocity at all. For example, if all particles are
weakly coupled to the Higgs, the wall speed is large although the
fluid remains close to equilibrium.  In section \ref{sec:beyond}
we compare our results with phenomenological
approaches to the wall velocity. In section \ref{sec:resu} we apply our
Boltzmann approach to several models. We start with the SM with a
(experimentally excluded) small Higgs mass in order to facilitate the
comparison between our results and M\&P. Then, we study the SM with a
low cutoff (including additional $\phi^6$ operators) and a singlet
extension of the SM. Conclusions are given in section \ref{sec:dis}.

\section{Equations of motion\label{sec:eom}}
The dynamics of the particles in the plasma is described by the
Kadanoff-Baym equations whose gradient expansion reduces to the usual
Boltzmann equations. The advantage of Kadanoff-Baym equations over
Boltzmann equations is two-fold. First, the Kadanoff-Baym equations
naturally provide the forces that act on the particles due to the
Higgs background. Second, the Kadanoff-Baym equations allow for
accurate treatment of spin in the case of fermionic particles.

In the context of the Kadanoff-Baym formalism, certain two-point
functions encode the dynamics of the system. 
In particular, the Wightman function $G^<$ encodes the particle distribution
functions.  At leading order, the Kadanoff-Baym equations for a scalar
degree of freedom in the gradient expansion read
\bea
\label{eq:KB}
(p^2 - m^2 ) \, G^<(p,x) &=& 0 \label{eq:KB_con}~, \\
\lp p_\mu \partial^\mu + \frac12 \partial_\mu m^2 \partial_{p_\mu} \rp 
G^<(p,x) &=& \, {\rm coll} \label{eq:KB_kin}~,
\eea
where the term ``coll'' summarizes the collision contribution.

The first equation is the so-called constraint equation and encodes
  the fact that the Wightman function $G^<$ can be expressed in terms
  of the particle distribution function $f( \vec p, x)$:
\be
G^<(p,x) = 2\pi \, f( \vec p, x) \, \delta (p^2 -m^2 )~.
\ee
Using this ansatz and the identity
\be
\lp p_\mu \partial^\mu + \frac12 \partial_\mu m^2 \partial_{p_\mu} \rp 
\lp p^2 -m ^2 \rp = 0~,
\ee
eq.~\eqref{eq:KB_kin} leads to the relation
\be
 \lp p_\mu \partial^\mu + \frac12 \partial_\mu m^2 \partial_{p_\mu}
\rp f( \vec p, x) = \, {\rm coll}~. 
\ee
The first term corresponds to free floating of the particles. In
the non-relativistic limit it reduces to the usual kinetic term of the
Boltzmann equation:
\be
p_\mu \partial^\mu \to m \lp \partial_t + \vec v\, \nabla \rp~.
\ee
The second term describes the force acting on the particles. To
understand its effect, we can imagine a Higgs background that is
constant in space and only depends on time. In this case, we expect
the three-momentum of the particles to be conserved. So the energy $p_0$ has
to change in order to ensure the $t$-dependent on-shell condition $p^2
= m^2(t)$. This behavior is well reflected by the force term as it admits solutions 
of the form $f(\vec p, x)
= g(E)$ with $E=\sqrt{\vec p^2+ m^2(t)}$.

In the analysis of the wall velocity, it is the force contribution
that drives the plasma out of equilibrium. The complexity of the
problem however lies in the collision terms. They depend on the
interactions between the particle species and will be discussed in
more detail in section~\ref{sec:transport}.

\section{Symmetries and conservation laws\label{sec:sym}}
Before studying the effect of the collision terms, it is useful to discuss
some symmetries of the problem. To this aim, we consider the
energy-momentum tensor and the particle current. By the conservation
of the former we deduce the equation of motion of the background.

\subsection{Four momentum} \label{sec:momentum}

The spatial variation of the classical background can be seen as a
bubble wall separating the inner (electroweak broken) and external
(electroweak unbroken/symmetric) phases. Ultimately, we are interested in the
velocity of this wall once the growing bubble reaches a steady
expansion regime. In order to quantify this speed, one needs the
equation of motion of the Higgs in the plasma~\footnote{Hereafter we
  assume only the Higgs field to acquire a VEV and to act as a classical
  background. See section~\ref{sec:singlet} for a more general
  discussion.}.  A simple way of achieving it is to use the
energy-momentum conservation of all particles in the plasma and the
Higgs background (the expansion of the Universe can be neglected
during the phase transition):
\be
\partial^{\mu} \, T_{\mu\nu}^{\rm total} =\partial^{\mu} \,(\sum_n T_{n,\mu\nu}^{\rm plasma}+T_{\mu\nu}^{\phi})=0~,
\label{eq:Tcons}
\ee
with $n$ running over each species in the plasma.

The energy momentum tensor of the plasma of each species $n$ can be
expressed as~\footnote{In the following we focus on bosonic degrees of
  freedom in the plasma. However, the conclusion does not change for
  fermions.} 
\bea
T^{{\rm plasma}}_{n, \mu\nu} &=& \int \, \frac{d^4p}{(2\pi)^4} \, p^\mu p^\nu \, G_n^<(p,x) \nn \\
&= & \int \left. \frac{d^3p}{(2\pi)^3} \, p^\mu p^\nu \frac{1}{E_n} \, f_n(\vec p, x) \right|_{p_0 = E} \ .   
\eea
Its divergence yields [cf.~eq.~\eqref{eq:KB_kin}]
\bea
\partial^{\mu} \, T_{n, \mu\nu}^{\rm plasma} + \, {\rm coll}_n&=& 
- \frac12 \partial_\mu m_n^2 \int \, \frac{d^4p}{(2\pi)^4} \, p_\nu \partial_{p_\mu} G_n^<(p,x) \nn\\
&=& \frac12 \partial_\nu m_n^2 \int \, \frac{d^4p}{(2\pi)^4} \,  G_n^<(p,x) \nn \\
&= & \frac12 \partial_\nu m_n^2 \int \,  \frac{d^3p}{(2\pi)^3} \frac{1}{E_n} \, f_n(\vec p, x)~. 
\label{eq:div1}
\eea

On the other hand, the energy momentum tensor of the classical field
background is
\be
T_{\mu\nu}^\phi = \partial_\mu \phi \partial_\nu \phi 
- g_{\mu\nu} \lp \frac12 \partial_\rho \phi \partial^\rho \phi - V(\phi)\rp.
\ee
Hence, the divergence of this energy-momentum tensor reads
\be
\partial^\mu T_{\mu\nu}^\phi = \partial_\nu \phi 
\lp \square \phi + \frac{dV}{d\phi} \rp.
\label{eq:div2}
\ee
Finally, by plugging eqs.~\eqref{eq:div1} and \eqref{eq:div2} into
\eqref{eq:Tcons}, one obtains the Higgs equation of motion
\be
\label{eq:higgs}
\square \phi + \frac{dV}{d\phi} + \sum_n \frac{d m^2_n}{d\phi} 
\int \frac{d^3 p}{ (2 \pi)^3} \frac{1}{2 E} f_n(\vec p, x)=0~.
\ee
Note the collision terms are absent. They indeed cancel out when the sum
over all species is performed.  This is a consequence of the
energy-momentum conservation in the decay and scattering amplitudes
and can be checked explicitly once the model is specified.

In thermal equilibrium, the last term in eq.~\eqref{eq:higgs} can be
easily related to the thermal contribution to the Higgs
finite-temperature effective potential $V(\phi,T)$.  This can be
verified by using the relations
\be
\Delta V^{\rm plasma} = -{\rm pressure}= 
-\frac13 \int \frac{d^3p}{(2\pi)^3} \frac{p^2}{E} f(E)
\ee
and
\bea
\frac{\Delta V^{\rm plasma}}{d\phi} = \frac{d m^2}{d \phi} \frac{d \Delta V^{\rm plasma}}{dm^2} &=&
-\frac 13 \frac{d m^2}{d\phi} \int \frac{d^3 p}{ (2 \pi)^3}  
\frac{p^2}{E} \frac{d}{dE} \frac{f(E)}{2 E} \nn \\
&=& - \frac 13  \frac{d m^2}{d\phi} \int \frac{d^3 p}{ (2 \pi)^3}  
\frac{p}{2} \frac{d}{dp} \frac{f(E)}{E} \nn \\
&=& \frac{d m^2}{d\phi} \int \frac{d^3 p}{ (2 \pi)^3} \frac{f(E)}{2 E}~,
\eea
where ``pressure'' stands for the pressure of a bosonic
gas in the plasma and $f(E)$ does for the
Boltzmann distribution (i.e. $f(\vec p, x)\to f(E)$ in the equilibrium
limit). Therefore, splitting the distribution functions into an equilibrium part 
plus some deviations $\delta f_n$ yields
\be
\label{eq:higgs2}
\square \phi + \frac{dV(\phi,T)}{d\phi} + \sum_n \frac{d m^2_n}{d\phi} 
\int \frac{d^3 p}{ (2 \pi)^3} \frac{1}{2 E} \delta f_n(\vec p, x)=0~.
\ee
The last contribution is the so-called {\it friction} term.

We remark that this result, which is based on the Kadanoff-Baym
equations, reproduces the finding in M\&P where the Higgs equations was
obtained in the WKB approximation.

\subsection{Charges}

The system conserves electric charge and also iso-spin in the symmetric
phase. This should be reflected in the equations. In particular, the
particle current
\bea
J^{\mu}_n &=& \int \, \frac{d^4 p}{ (2 \pi)^4} \, p^\mu \, G_n^<(p,x) \nn \\
& = & \int \frac{d^3 p}{ (2 \pi)^3} \, p^\mu \frac{1}{E_n} \, f(\vec p,
x)  
\eea
should be conserved in the sum over species (weighted by the
corresponding charges). Notice that in the conservation equation of
the charge, the force term does not enter. Indeed, after partial
integration, it turns out that
\be
\partial_{\mu} \, J^{\mu}_n + \, {\rm coll}_n= 0~. 
\ee
This reflects the physical picture that the force modifies the
trajectory of the quasi-particles but does not change their charge.
The collision terms on the other hand contain decay and annihilation processes
that change the individual particle numbers but also conserve the total charge.

\section{Transport equations of the plasma components\label{sec:transport}}
\subsection{The fluid approximation}\label{sec:fluid}

In order to solve the equation of motion~\eqref{eq:higgs2} one needs
to determine the friction contribution. This requires to identify the
correct particle distribution function. As in M\&P, we consider the
{\em flow ansatz} 
\be
f(\vec p,x) = \frac{1}{\exp[X] \pm 1}
 = \frac{1}{\exp[\beta (x) ( u^\mu (x) p_\mu + \mu(x) )] \pm 1}~,\label{eq:ansatz}
\ee
where the four-velocity $u^\mu(x)$, the chemical
potential $\mu(x)$ and the inverse temperature $\beta(x)$ are space
dependent. In the limit of negligible space dependence, it reproduces
the usual Boltzmann distribution in the frame boosted by the
four-velocity $u^\mu$.

Contrarily to ref.~\cite{MoAndPro}, in which this ansatz was first
used, here we do not require small fluid velocity $u^\mu$,
although we still assume small spatial dependence (the consistency
of this assumption will be checked {\em a posteriori}).  We can hence
use
\be
X \simeq ( u^\mu + \delta u^\mu(x) + \delta \tau u^\mu ) \beta p_\mu + \delta \mu(x)
\ee
for the individual particle species and linearize in the following in the 
fluctuations
$\delta \tau$, $\delta u$ and $\delta \mu$ when necessary. Temperature
changes and the chemical potential are encoded in the dimensionless
quantities $\delta \tau$ and $\delta \mu$ in units of the
temperature. Changes in the fluid velocity are encoded in $\delta u$
that fulfills $u^\mu \delta u_\mu \simeq 0$ in order to achieve the
correct normalization for the four-velocities $u_\mu$ and $u_\mu +
\delta u_\mu$. The space-independent part of each quantity is fixed at
its value far outside the bubble wall. In particular, the constant
part of the chemical potential can be neglected~\cite{MoAndPro}.

In the following we need two different types of averages
\be
\left< O \right> = \int \frac{d^3 k}{E} \, O \, f(k) \, ,
\quad\qquad
\left[ O \right] = \int \frac{d^3 k}{E} \, O \, \partial_X f (k) \, ,
\ee
and we define
\begin{align}
N &= \left< 1 \right>   \, ,   &   \bar N &= \left[ 1 \right]  \, ,           \\
J^\mu &= \left< p^\mu \right>  \,  ,\   &   \bar J^\mu &=\left[ p^\mu \right]  \, , \\
T^{\mu\nu} &= \left< p^\mu p^\nu \right>   \, , &   \bar T^{\mu\nu} &= \left[ p^\mu p^\nu \right]  \, , \\
M^{\mu\nu\lambda} &= \left< p^\mu p^\nu p^\lambda \right>  \, ,& \bar M^{\mu\nu\lambda} &= \left[ p^\mu p^\nu p^\lambda \right]  \, .
\end{align}
Using these definitions, the out-of-equilibrium densities can be
expressed in terms of fluctuations and equilibrium densities. For
example for the four-current one finds in leading order of the fluctuations
\be
\label{eq:J_expansion}
J^\mu = J_0^\mu + \bar J^\mu_0 \delta \mu
+ \beta \, \bar T_0^{\mu\nu} (\delta \tau u_\nu + \delta u_\nu)  \, ,
\ee
where the zero subscript denotes the equilibrium quantities with
a distribution function $f$ for fixed background values ($\delta u = \delta \tau = \delta \mu = 0$).
These functions are still space-time dependent due to their mass dependence.

The equations of motion of the system can in principle be obtained
from (\ref{eq:J_expansion}).  However, using the properties under
Lorentz transformations of the different functions and dimensional
analysis, they can be brought to a form that allows for a more
intuitive interpretation (details are given in appendix
\ref{app:relations}). Subsequently, the divergences of the
four-current and the energy-momentum tensor turn into
\bea
\frac12 \partial_\mu m^2 \, \beta \bar N \, ( u^\mu + u^\mu \delta\tau + \delta u^\mu )  
&+& \beta \bar T^{\mu\nu} \partial_\mu (u_\nu \delta\tau + \delta u_\nu) \nn \\
&+& \bar J^\mu \partial_\mu \delta\mu = \rm{coll} \label{eq:master1} \ ,
\eea
and
\bea
\frac12 \partial_\mu m^2 \, \beta\bar J^\lambda \,( u^\mu + u^\mu \delta\tau + \delta u^\mu )
&+& \beta \bar M^{\mu\nu\lambda} \partial_\mu (u_\nu \delta\tau + \delta u_\nu) \nn \\
&+& \bar T^{\mu\lambda} \partial_\mu \delta\mu = \rm{coll}~. \label{eq:master2}
\eea

We solve these equations in the planar wall approximation. We also
work in the wall frame (with the z-axis orthogonal to the wall and
oriented towards the broken phase) in which the plasma velocity and
its fluctuation are $u^\mu = \gamma (1, v_w)$ and $\delta u^\mu =
\delta v \, \bar u^\mu = \delta v \gamma (v_w , 1)$~\footnote{Note
  that the fluctuation $\delta v$ is equivalent to the fluctuation
  around $v_w$ up to a factor $\gamma^2$.}. As we focus on the steady
velocity regime, the substitutions $m^2(x)\to m^2(z)$, $u^\mu
\partial_\mu \to \gamma v_w \partial_z$ and $\bar u^\mu \partial_\mu
\to \gamma \partial_z$ apply. The linearized eqs.~\eqref{eq:master1}
and \eqref{eq:master2} can hence be expressed as~\footnote{Linearizing the equations is justified 
(among other conditions)
   when the change in mass is small compared to the temperature, $m^2 \lesssim T^2$.}
\be
\label{eq:linear_formal}
A \cdot \vec q\,' + \rm{coll} = S\, ,
\ee
where $\vec q =(\delta \mu, \delta \tau, \delta v)$ and the prime
denotes the dimensionless derivative $\vec{q}\,' = \gamma \beta
\partial_z \vec{q}$.
The matrix $A$ and the source $S$ have the form
\be
\label{eq:linear}
A \equiv
\begin{pmatrix}
 v_w c_2 & v_w  c_3 & \frac13 c_3 \\
 v_w  c_3 & v_w c_4 & \frac13 c_4 \\
 \frac13 c_3 & \frac13 c_4 & \frac13 v_w c_4 \\
\end{pmatrix} \, , \quad 
S \equiv \frac{m'm}{T^2}
\begin{pmatrix}
 v_w c_1 \\
 v_w c_2 \\
 0 \\
\end{pmatrix}~,
\ee
where the coefficients $c_i$ depend on the spin statistics of the
species we are dealing with. Working at lowest order in $m/T$ as in
M\&P, for bosons [having $p = \pi^2 T^4/90$ and $n = \zeta(3)
  T^3/\pi^2$] one finds~\footnote{One might wonder if it is feasible
  to neglect the mass dependence in those coefficients. After all, the
  leading coefficient only corresponds to the mean-field approximation
  while the phase transition relies on the interplay between the
  mean-field and higher contributions. However, these terms are only
  comparable because the zero-temperature contribution to the $\phi^2$
  operator almost cancels the mean-field contribution close to the critical
  temperature. In the matrix $A$ only the finite temperature
  contributions are relevant. No cancellation occurs and higher orders
  can be neglected.}
\be
c_1 = \frac{\log(2 T/m)}{2 \pi^2}\, , \quad
c_2 = \frac16 \, , \quad
c_3 = \frac{3\zeta(3)}{\pi^2} \, , \quad
c_4 = \frac{2\pi^2}{15}\, , \quad \label{eq:c_bos}
\ee
whereas for fermions [with $p = 7 \pi^2 T^4/720$ and $n = 3\zeta(3)
T^3/4\pi^2$]
\be
c_1 = \frac{\log(2)}{2 \pi^2}\, , \quad
c_2 = \frac1{12} \, , \quad
c_3 = \frac{9\zeta(3)}{4\pi^2} \, , \quad
c_4 = \frac{7\pi^2}{60}\, . \quad \label{eq:c_fer} 
\ee
The linearized fluid equation~\eqref{eq:linear_formal}
agree with the system obtained in M\&P in the limit of non-relativistic 
wall velocities.

Notice that in the limit $v_w \to 0$ the matrix $A$ has one vanishing
eigenvalue. On the other hand, the collision terms do not go to
zero. It is then ensured that in this limit, $\vec q = 0$ is the
unique solution.

\subsection{Standard Model-like plasma content}

In principle, in order to determine the friction, one has to solve a
system of differential equations \eqref{eq:linear_formal} (as many as
the number of species) coupled to each other via the collision
terms. Nevertheless, we are interested in theories in which the
particle content of the thermal bath resembles the SM
one. The species that are relevant during the phase transition are
therefore the electroweak gauge bosons (simply called W bosons
hereafter) and top quarks. The remaining particles are not driven out
of equilibrium and act as a background~\footnote{The Higgs itself also
  notices the phase transition, but it can be safely neglected since
  it constitutes only one degree of freedom. The same may hold for some fields
  involved in theories beyond the SM, as for instance (a
  small number of) gauge scalar singlets (see
  section~\ref{sec:singlet}).}.  Quarks and gluons are the large
portion of the background particles. They are strongly coupled and
hence share the same plasma fluctuations. Furthermore, their chemical
potential vanishes since the gluons quickly equilibrate. The equations
of motion of the background can be deduced from the two relations
arising from energy-momentum conservation.

After linearizing the collision terms~\cite{MoAndPro}, the fluid
equation~\eqref{eq:linear_formal} applied to the W and top fields
reads 
\bea
\label{eq:fluid_stuff}
A_W (\vec q_W + \vec q_{bg})' + \Gamma_W \vec q_W &=& S_W \, , \\
A_t (\vec q_t + \vec q_{bg})' + \Gamma_t \vec q_t &=& S_t \, ,
\label{sys0}
\eea
whereas for the background particles it leads to
\be
\label{eq:fluid_bg}
A_{bg} \vec q_{bg}\,\!\!\!\!' + \Gamma_{bg,W} \,\vec q_W + \Gamma_{bg,t} \vec q_t = 0 \, .
\ee
The quantities $A_W$ and $A_t$ are given by $A_W=A_b$ and $A_t=A_f$,
where $A_b$ and $A_f$ are defined as the matrix $A$ with the
coefficients~\eqref{eq:c_bos} and~\eqref{eq:c_fer},
respectively. Considering a SM-like background (with
decoupled right handed neutrinos), one has $A_{bg} = 19 A_b + 78
A_f$. The values of $\Gamma_W$ and $\Gamma_t$ are summarized in appendix
\ref{sec:coll}.  Since the total energy momentum is conserved, it
follows that $N_W \, \Gamma_W + \Gamma_{bg,W} \propto (1,0,0)$ and $N_t \,
\Gamma_t + \Gamma_{bg,t} \propto (1,0,0)$, with $N_W=9$ and $N_t=12$.

This system of differential equations can be more easily solved by
removing $\vec q_{bg}$ from \eqref{eq:fluid_stuff} and \eqref{sys0}
and then determining the background equation by integration. This
shows that the fluctuations $\vec q_W$ and $\vec q_t$ vanish by
construction far inside and far outside the bubble, where both sources
$S_W$ and $S_t$ vanish. On the other hand, the background fluctuations
$\vec q_{bg}$ cannot vanish on both sides. As previously mentioned, we
chose to match the solution in the symmetric phase in front of
the wall. Notice that due to energy-momentum conservation, the
absolute change of $\vec q_{bg}$ along the wall cannot depend on the
wall shape. In our approximation it can be determined by knowing the
change of the W-boson and top masses in units of the temperature.

\subsection{Higgs equation of motion}

\label{sec:eomHiggs}
The equation of motion of the Higgs~\eqref{eq:higgs2} can be
linearized as well. From the expansion of the fluid ansatz
\eqref{eq:ansatz} one obtains
\begin{align}
\label{eq:higgsfluid}
-\phi'' 
+ \frac{dV^T(\phi,T)}{d\phi}  &+ \frac{N_t T^2}{2} \frac{dm_t^2}{d\phi}
(c_{f1}
\delta\mu_{f}+c_{f2} \delta \tau_{f}+c_{f2} \delta \tau_{bg,f} ) \nonumber \\ 
&+ \frac{N_W T^2}{2} \frac{dm_W^2}{d\phi} (c_{b1} \delta\mu_{b}+c_{b2} \delta
\tau_{b}+c_{b2} \delta \tau_{bg,b} )=0 \ . 
\end{align}
The system of differential equations (\ref{eq:fluid_stuff}),
(\ref{eq:fluid_bg}) and (\ref{eq:higgsfluid}) constitutes the basis of
our numerical analysis.  We solve it by means of the two-parameters
wall-shape ansatz
\begin{align}
\label{eq:profile}
 \phi(z)= \frac{\phi_0}{2} \left( \tanh \frac{z}{L} + 1 \right) \ ,
\end{align}
where $\phi_0$ and $L$ are respectively the VEV of the Higgs in the
broken phase and the wall thickness, both during the bubble expansion.
This ansatz seems particularly appropriate for weak phase
transitions. In this case, the profile of the tunneling bounce
(i.e.~the instanton solution connecting the two phases
\cite{Quiros:1999jp}) is very similar to eq.~\eqref{eq:profile} and
such a shape is expected to be kept during the bubble
evolution. Instead for very strong phase transition the bounce profile
may qualitatively differ from eq.~\eqref{eq:profile}. Nevertheless, it
seems reasonable that the bubble wall acquires the above configuration
once it approaches the steady velocity regime. This is also seen in recent 
simulations~\cite{Hindmarsh:2013xza} and we assume this shape in our analysis.

In order to implement the constraint (\ref{eq:higgsfluid}), we take
the moments
\begin{align}
\label{eq:HiggsMomentum1}
& \int_{-\infty}^{\infty}\, dz \,  [\text{l.h.s. of eq. }  (\ref{eq:higgsfluid})] \times \phi'  = 0 \ ,\\
& \int_{-\infty}^{\infty}\, dz \,  [\text{l.h.s. of eq. }  (\ref{eq:higgsfluid})] \times (2 \phi-\phi_0) \, 
\phi'  = 0 \ . \label{eq:HiggsMomentum2}
\end{align}
These relations have a physical interpretation
(cf.~section~\ref{sec:momentum}). Eq.~\eqref{eq:HiggsMomentum1}
declares that in the steady velocity regime, the total pressure on the
wall vanishes. Its equilibrium
part is the potential difference $\Delta V^T = V(\phi_0,T)- V(0,T)$,
whereas the rest encodes the friction. Ultimately, requiring the
cancellation of their sum determines the wall velocity.  The second
equation corresponds to the pressures gradient in the
bubble wall. Its solution provides the wall
thickness. For weak transitions it mostly depends on equilibrium
physics and little on friction effects.

\subsection{The shock front in the deflagration mode\label{sec:shock}}

We have seen in the last section that by minor modifications, the
equations obtained in M\&P are also valid in the relativistic regime
as long as the phase transition is weak enough.  However, there is a
further reason why only slow walls are considered in M\&P. When $v_w$
is equal to the sound speed $c_s$, the sign of one eigenvalue in the
fluid system is flipped, and therefore the whole dynamics
changes. Additionally, the linearization for the background fields
ceases to be valid at $v_w\approx c_s$. 

In the deflagration mode, a shock wave builds up in front of the
expanding bubble \cite{Gyulassy:1983rq}. Accordingly, the fluid
velocity and temperature in front of the wall differ from those of the
symmetric phase. Indeed, the shock wave sets the fluid in motion and
heats the plasma. This decreases the pressure difference experienced
by the Higgs and reduces the latent heat released in the
plasma. This effect can be strong enough to dominate the dynamics of the
bubble expansion, whose description can be inferred only from
hydrodynamic considerations~\cite{Konstandin:2010dm}.  To deal with
this issue, we follow the procedure of ref.~\cite{Ignatius:1993qn, energybudget} which
matches the plasma velocity and enthalpy in front and behind the
wall. Unlike the analysis in M\&P, we solve the non-linear equations
and do not rely on small fluid/wall velocities in this step.


\section{Phenomenological approaches\label{sec:beyond}}

In this section we discuss phenomenological approaches to the bubble
wall friction. In this kind of study, the Higgs
equation \eqref{eq:higgs2} is assumed to be effectively described
as 
\be
\label{eq:phenoOld}
 -\phi'' + \frac{dV_T(\phi,T)}{d\phi}  =  \eta(\phi, v_w) u^\mu \partial_\mu \phi \, .
\ee
The effective friction $\eta$ may involve an explicit dependence on
$\phi$ and/or $v_w$. Typically, it is deduced either by a matching to
the existing results in Boltzmann treatments~\cite{Huber:2013kj} or by
the relaxation time approximation~\cite{Megevand:2012rt}. Often, it is
supplemented by a further equation that sets the temperature variation
in the wall and that may be derived from the energy-momentum
conservation of the plasma (assumed to be in local
equilibrium~\cite{Ignatius:1993qn, Huber:2013kj}).

Depending on the parameter region and the level of sophistication,
eq.~\eqref{eq:phenoOld} can reproduce almost all features of the full
Boltzmann treatment. It has however its limitation. The most striking
is that the friction force scales with the wall thickness as $1/L$.
Moreover, the dependence on the wall velocity can be quite involved. For
instance, in the highly relativistic regime, $v_w
\to 1$, the term $u^\mu \partial_\mu \phi$ in eq.~\eqref{eq:phenoOld} is enhanced by a Lorentz
factor, and leads to finite wall velocities even for extremely strong
phase transitions. It is however known that bubble walls can enter a
runaway regime~\cite{BodekerMoore}. Phenomenologically, this can be
cured by introducing a $1/\gamma_w$ factor in $\eta(\phi,
v_w)$~\cite{Huber:2013kj} (or an even more complicated
dependence~\cite{Megevand:2013hwa}). This seems quite {\it ad hoc} and
one can instead wonder whether the discrepancy has deeper origins. As we will see, 
the dependence on the wall velocity already starts to become
quite non-trivial nearby the speed of sound.

In the following we demonstrate in which cases the Boltzmann treatment agrees with the 
phenomenological approach. We consider a system of equations obtained by integrating
the moments (\ref{eq:HiggsMomentum1}) and \eqref{eq:HiggsMomentum2}.
The non-equilibrium part contains a first contribution coming from the
fluctuations $\delta \mu$ and $\delta \tau$, and a second contribution
coming from the background fields $\delta \tau_{bg}$. These scale
differently in terms of $v_w$, $\phi_0/T$ and $L$, and we hence treat
them differently.  Our parametrization then reads
\bea
\label{eq:pheno1}
\frac{\Delta V}{T ^4} &=& 
f_{fl} + f_{bg} \ , \nn \\
\label{eq:pheno2}
- \frac{2}{15 (T L)^2} \left(\frac{\phi_0}{T}\right)^{3} + \frac{W}{T^5} &=& 
g_{fl} + g_{bg}  \ ,
\eea
where the quantities $f_{fl}, g_{fl}$ and $f_{bg}, g_{bg}$ are the
fluid and background functions (depending on $v_w$, $\phi_0/T$ and
$L \, T$) and $W$ is given by
\be
W = \int_0^{\phi_0} \frac{dV(\phi,T)}{d\phi} (2 \phi - \phi_0) \,  d\phi \ . 
\ee

\begin{figure}
    \includegraphics[width=0.45\textwidth]{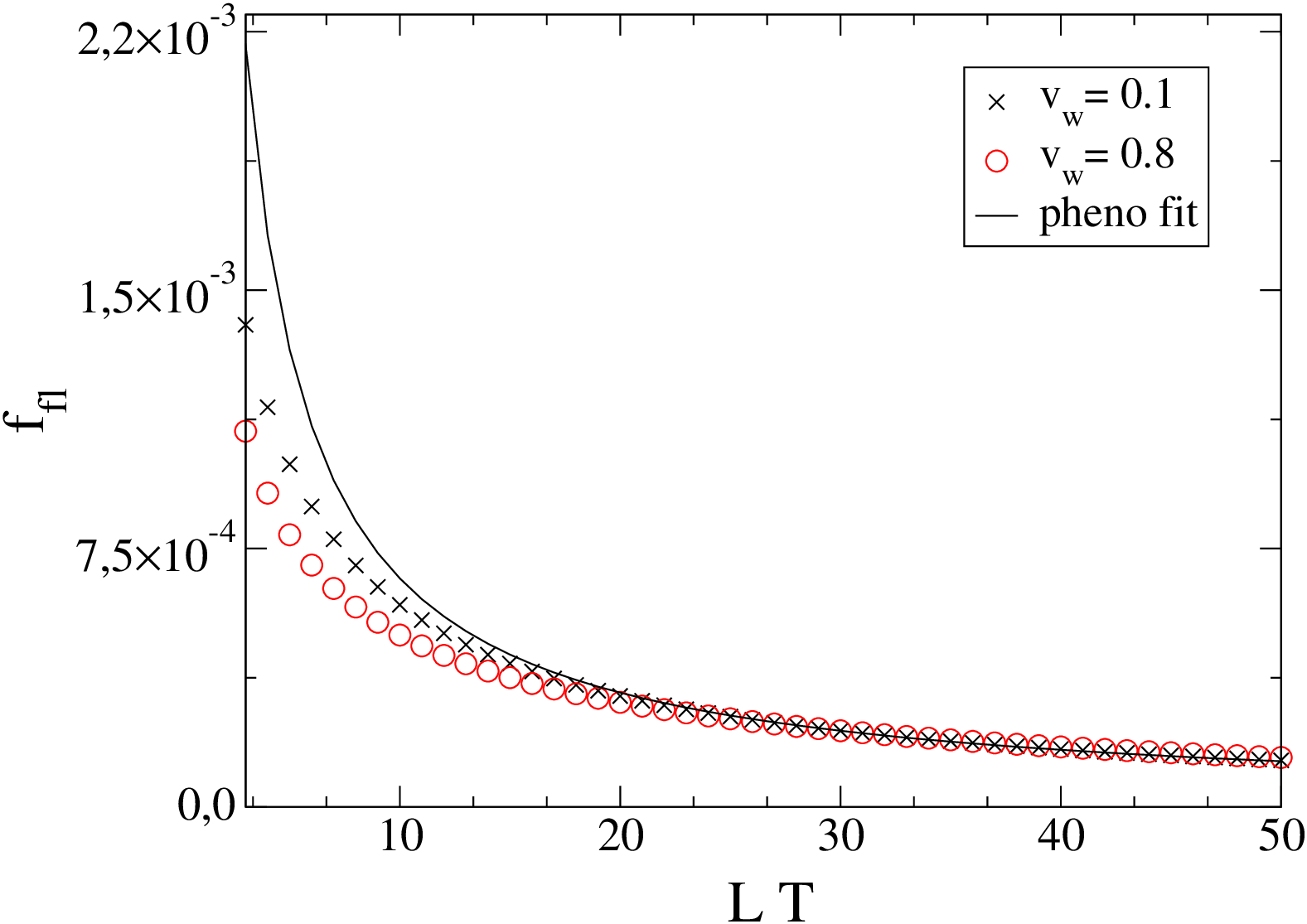} \quad \qquad
 \hspace{-2mm}   
    \includegraphics[width=0.45\textwidth]{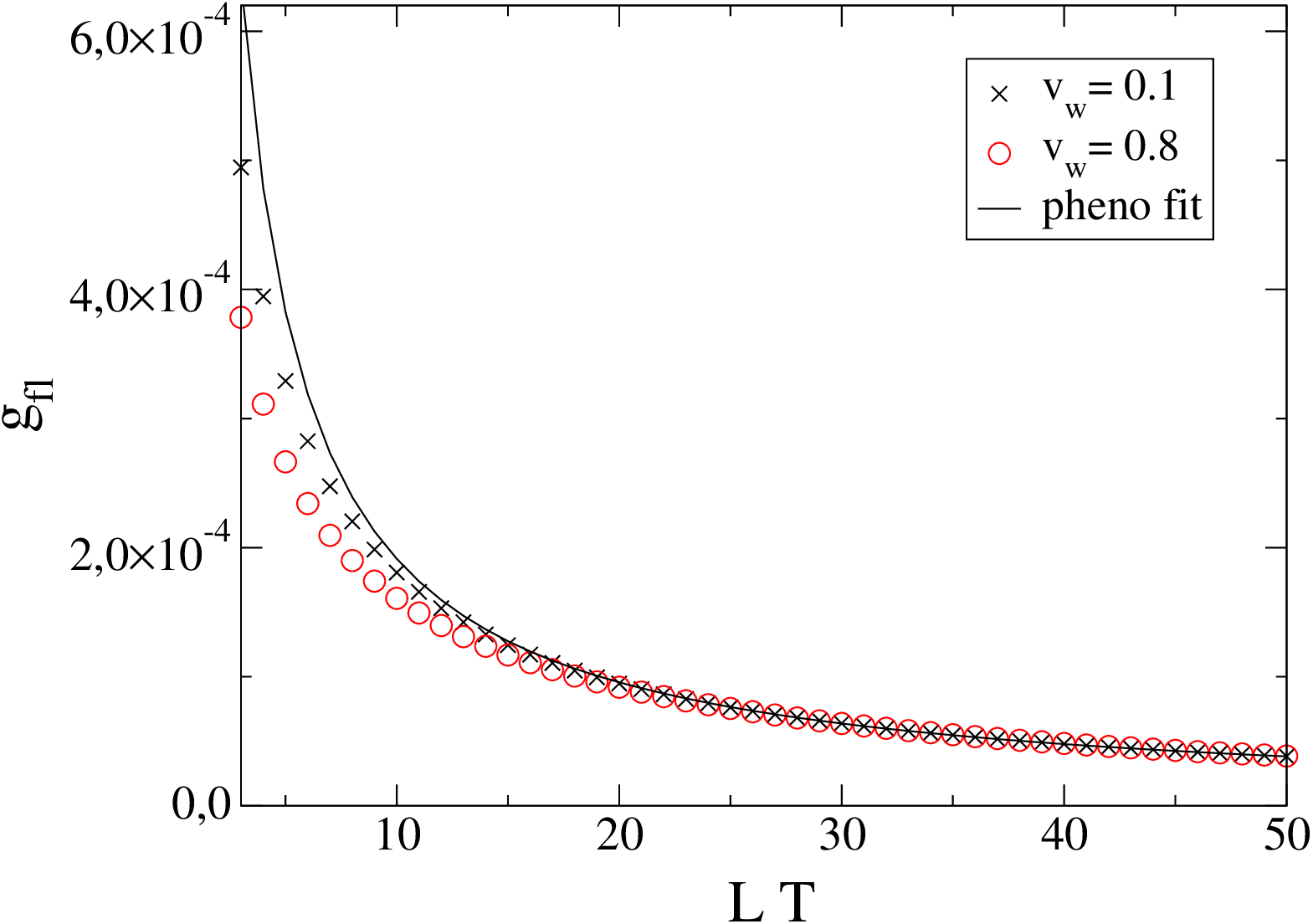}
\\[25pt]
{\color{white}.}\hspace{-3mm}
    \includegraphics[width=0.45\textwidth]{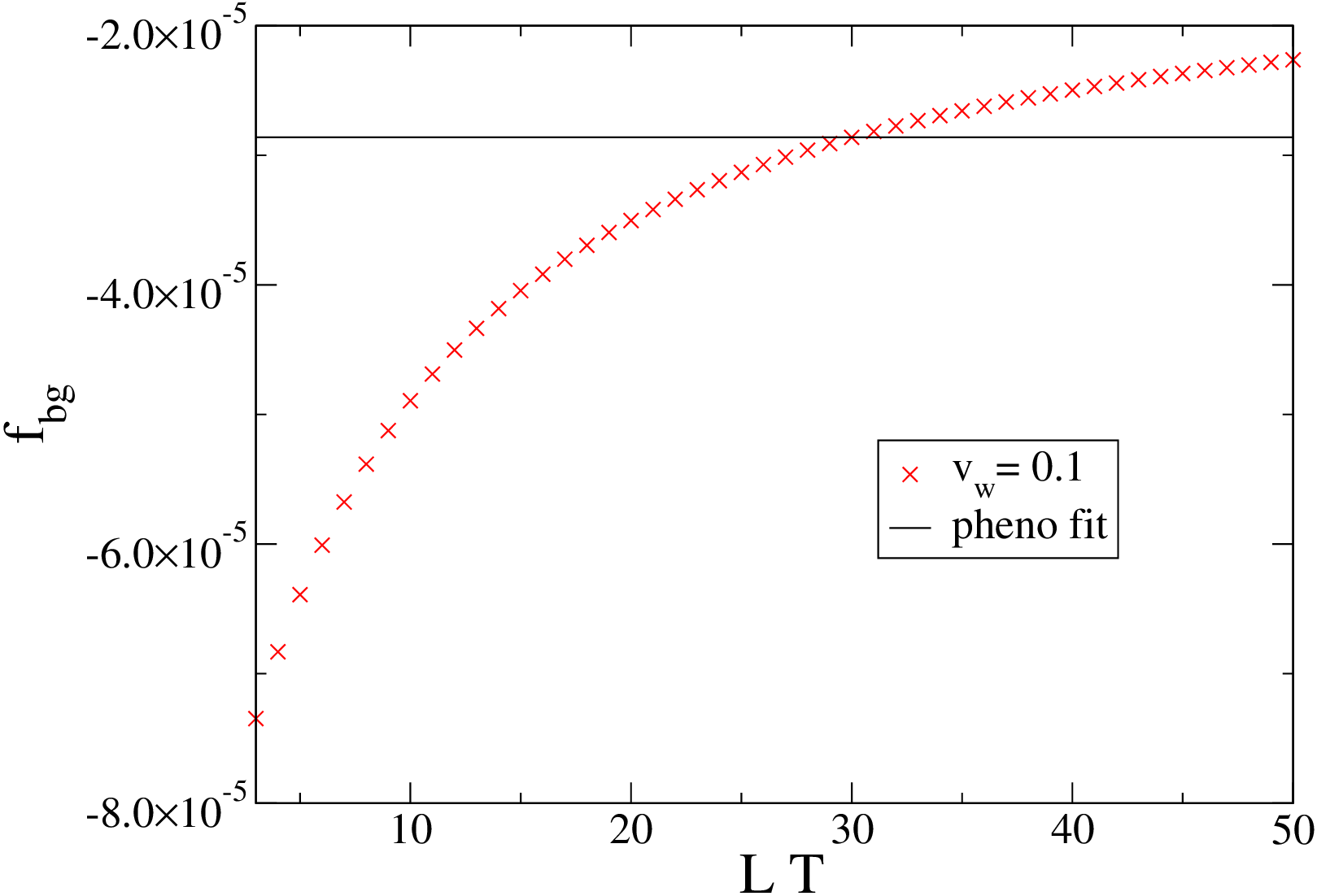}\quad \qquad
\hspace{-1mm}
      \includegraphics[width=0.45\textwidth]{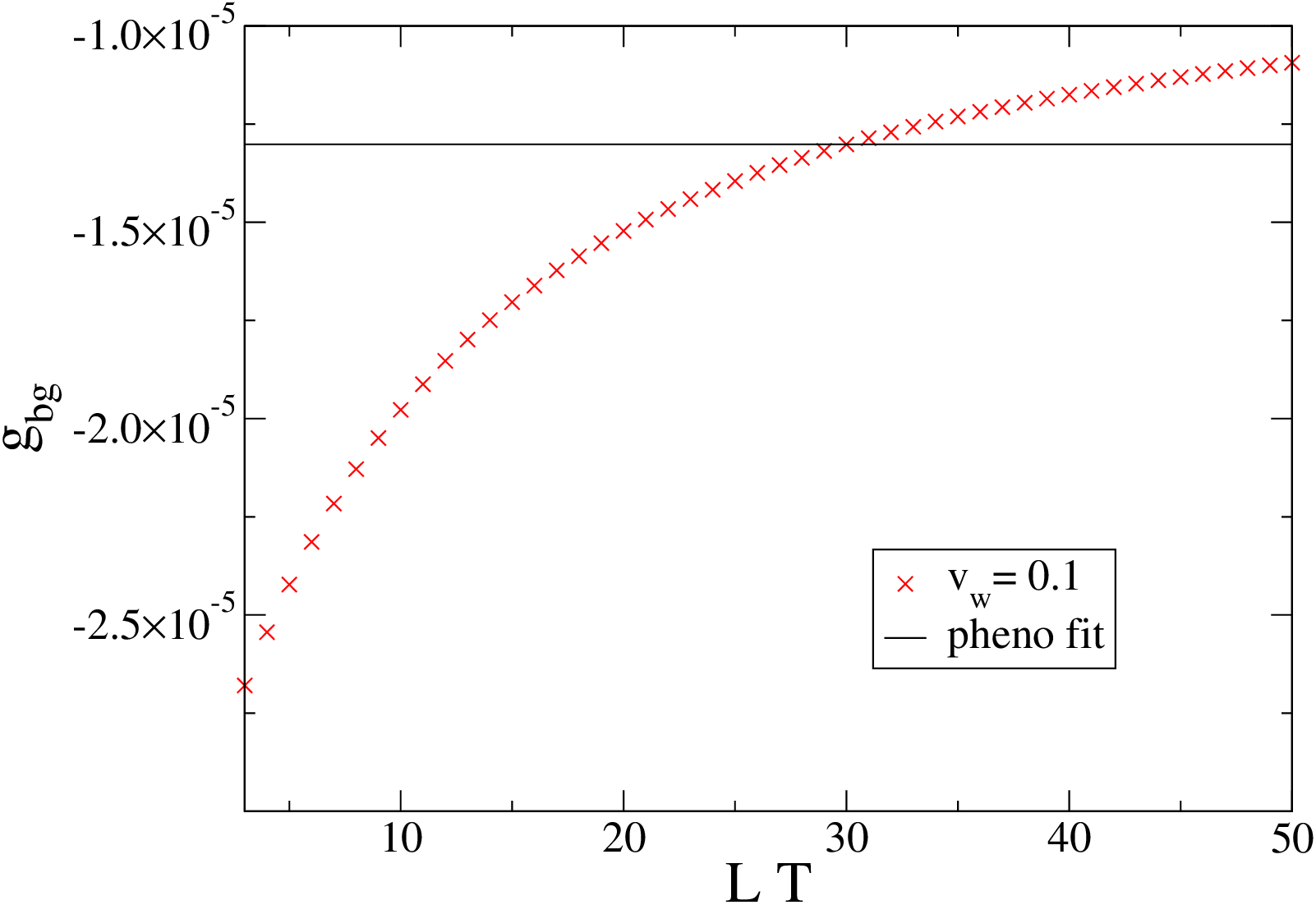}
\\[25pt]
{\color{white}.}\hspace{-4mm}
      \includegraphics[width=0.46\textwidth]{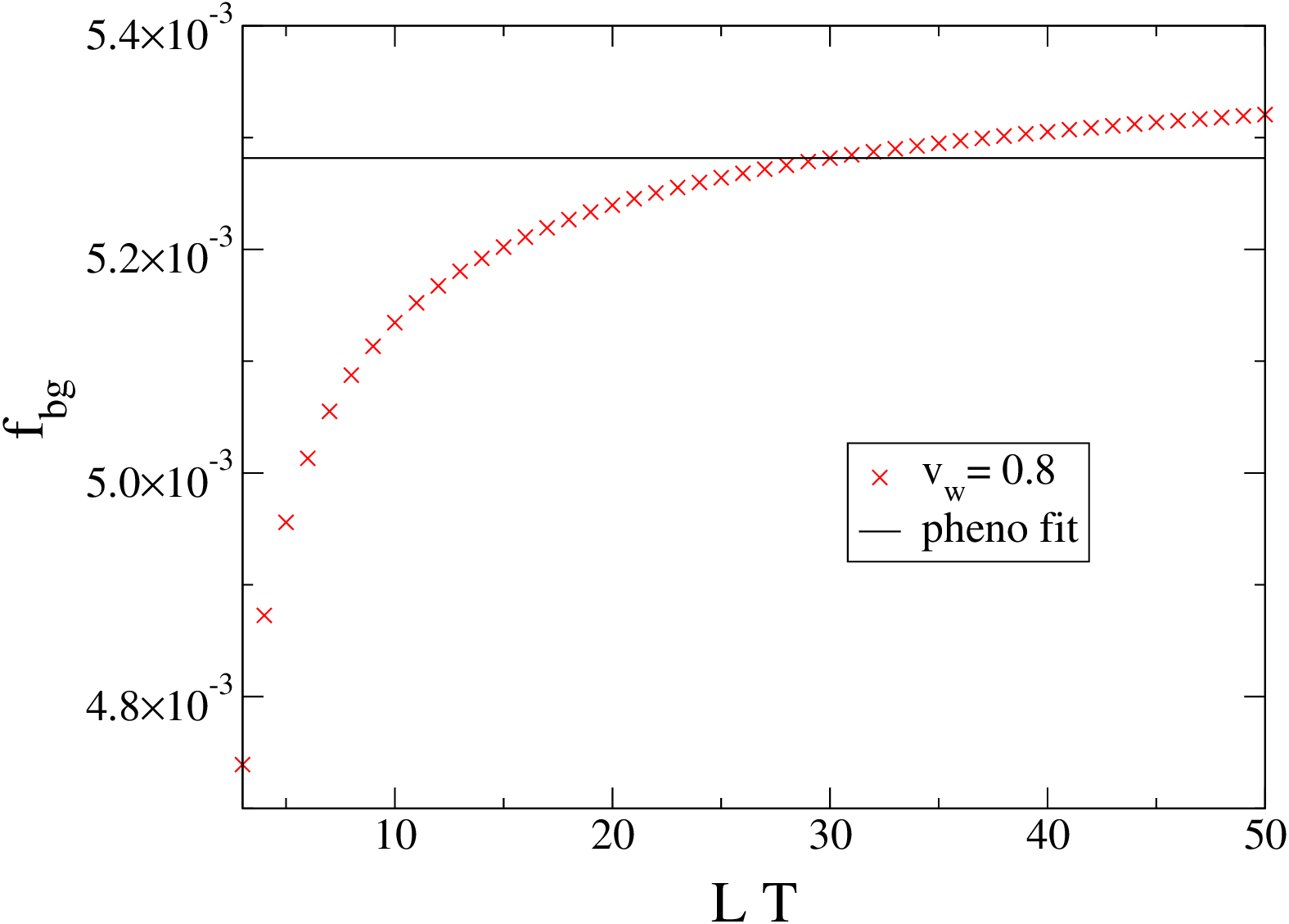} \qquad \hspace{2mm}
      \includegraphics[width=0.45\textwidth]{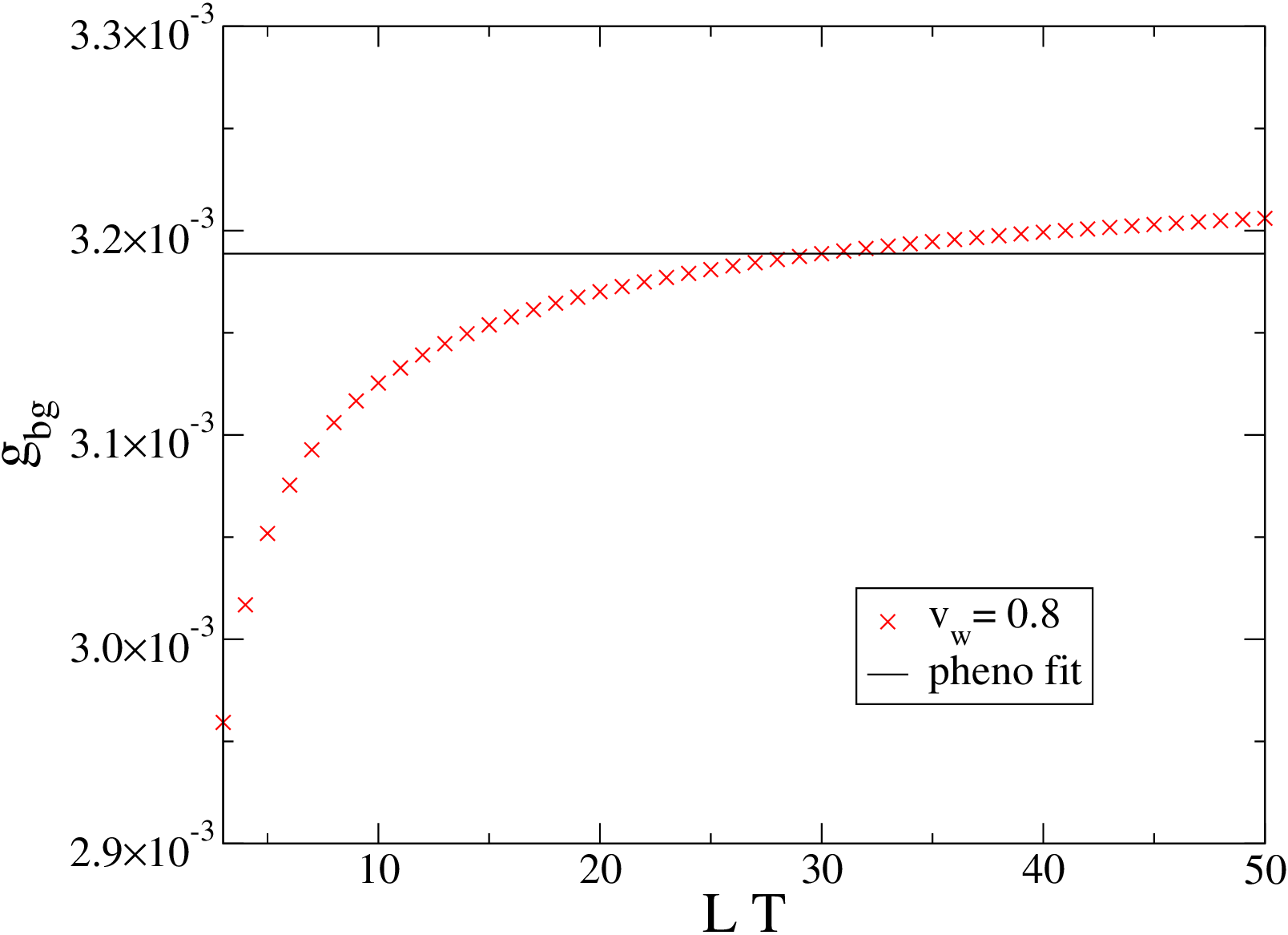} 
\caption{\small The  friction components as functions of the wall thickness $L \, T$ 
for $\phi_0/T= 1$ and two different wall velocities $v_w= 0.1$ and $v_w=0.8$. 
The friction components of the background for subsonic and supersonic 
wall velocities are shown in separate plots since they differ greatly.
\label{fig:lsc}}
\end{figure}

\begin{figure}
  \centering
                  \includegraphics[width=0.45\textwidth]{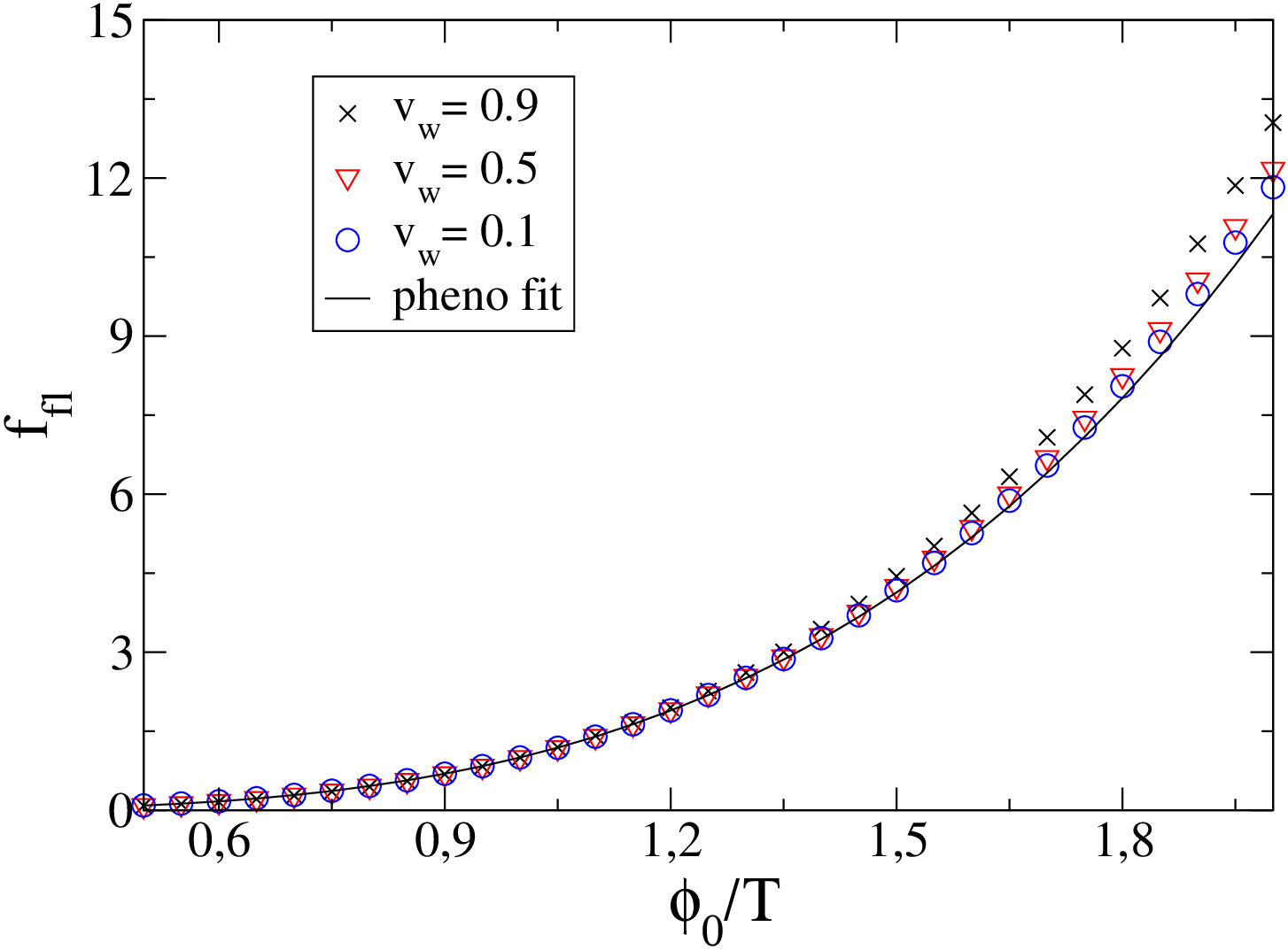}
\qquad \quad      \includegraphics[width=0.45\textwidth]{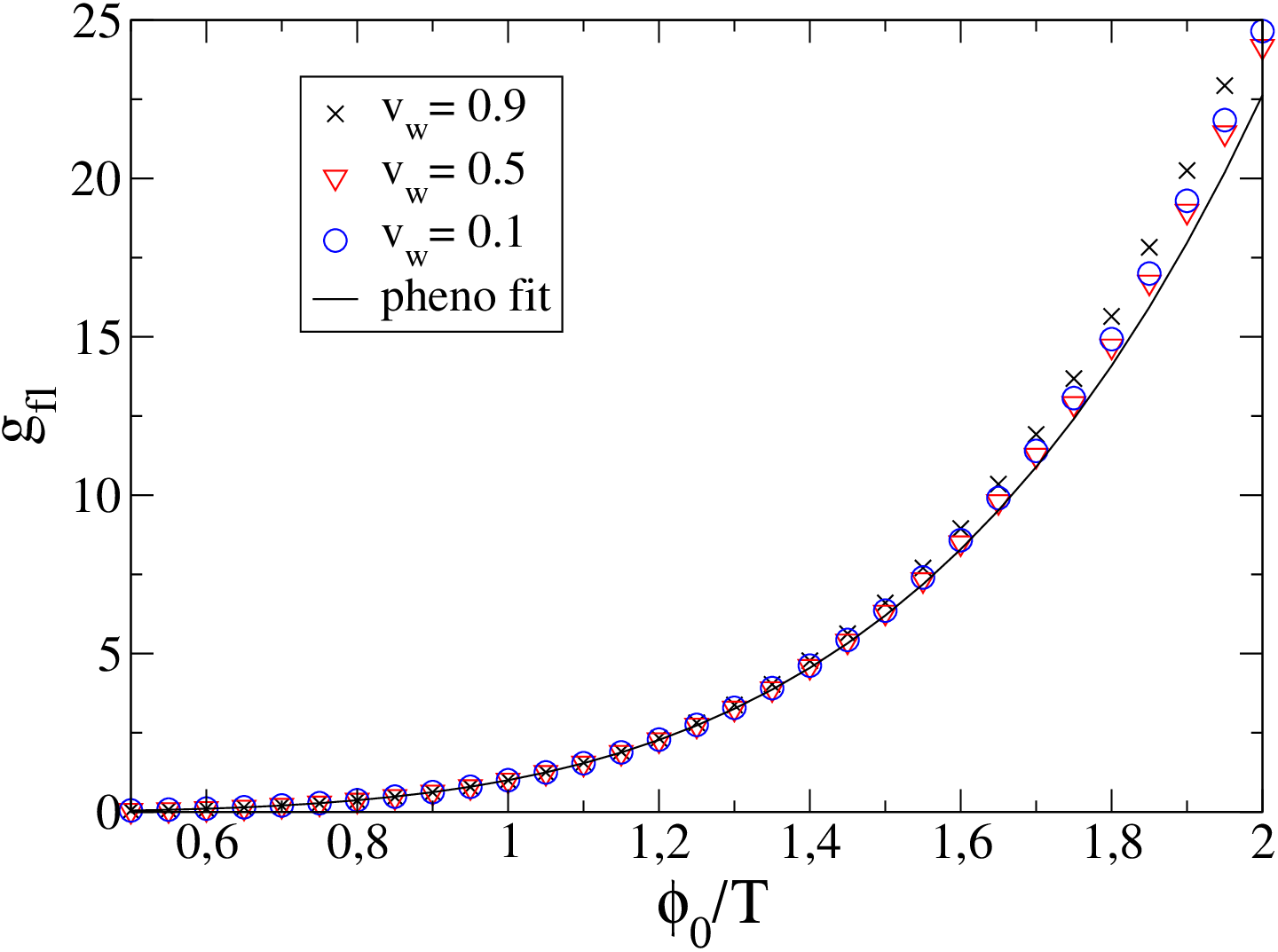}
        \caption{\small Dependence of the friction components on the strength of the phase transition $\phi_0 / T$. 
The lines for the different velocities are normalized to unity at $\phi_0/T = 1$. The wall
thickness is $L \, T = 30$.}
    \label{fig:flpT}
\end{figure}

\begin{figure}[h!]
                 \includegraphics[width=0.45\textwidth]{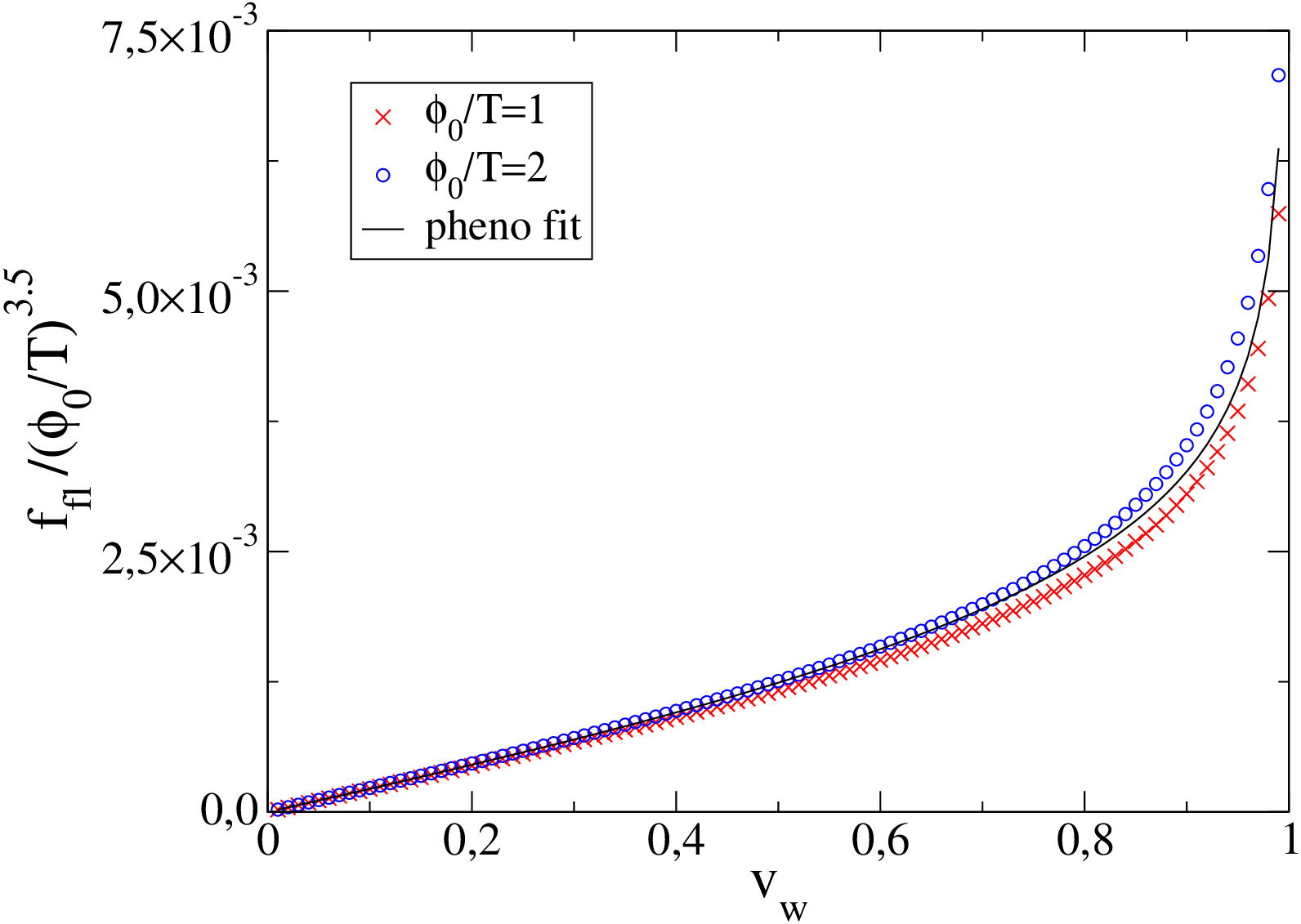}
\qquad \quad     \includegraphics[width=0.45\textwidth]{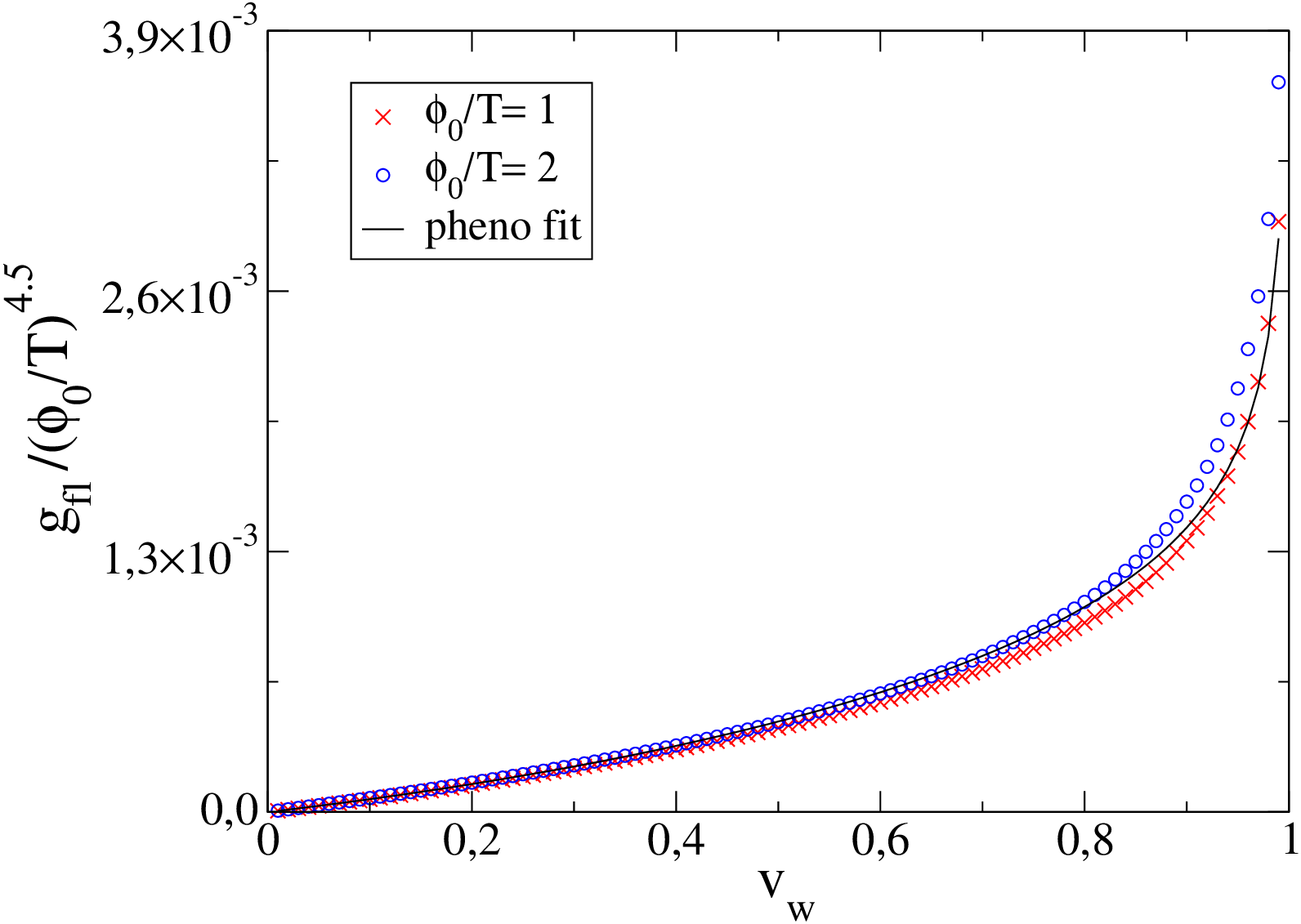}
\\[25pt]
 {\color{white}.}\!  \includegraphics[width=0.435\textwidth]{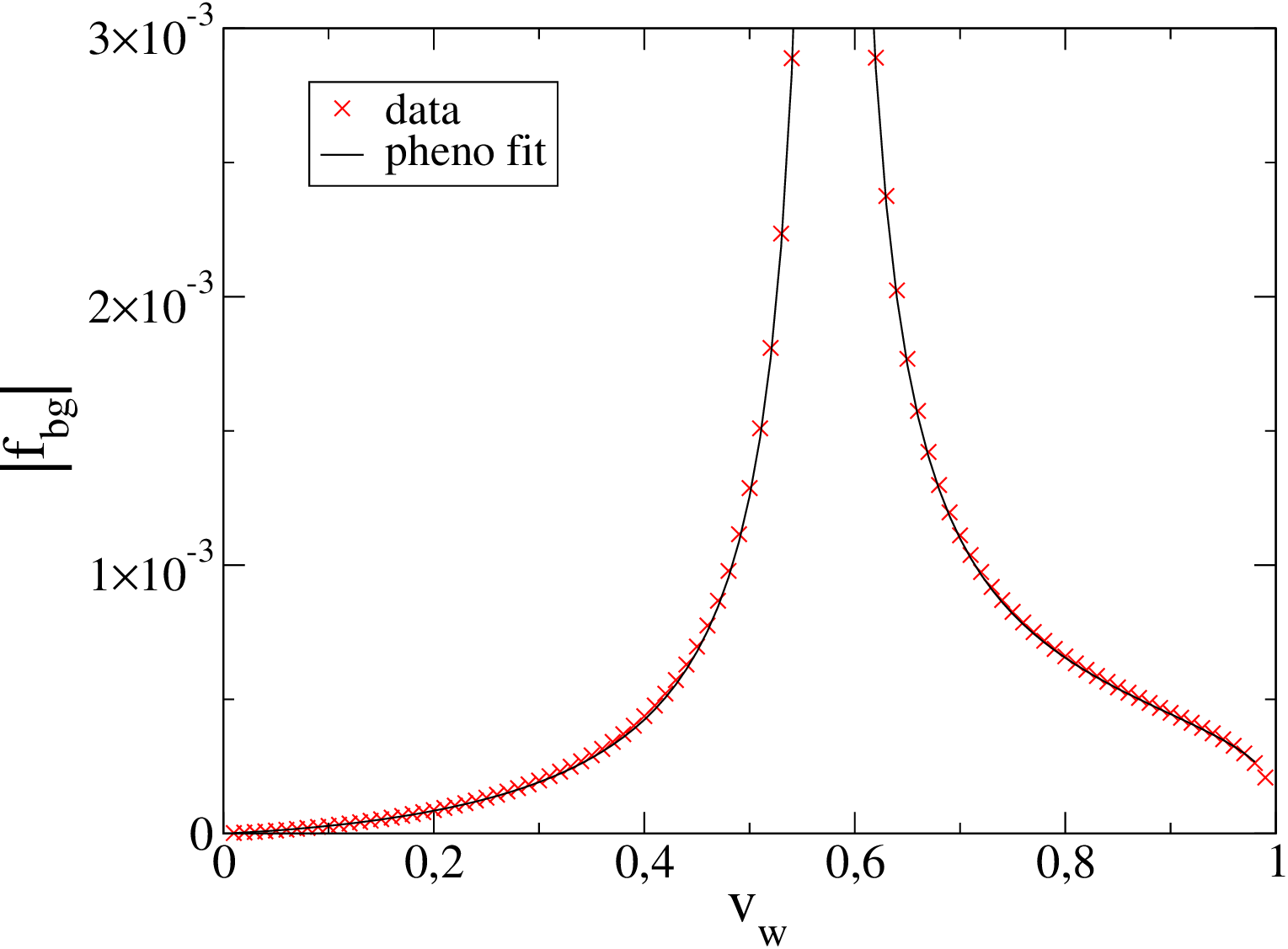}
\qquad \quad \hspace{0pt}   \includegraphics[width=0.435\textwidth]{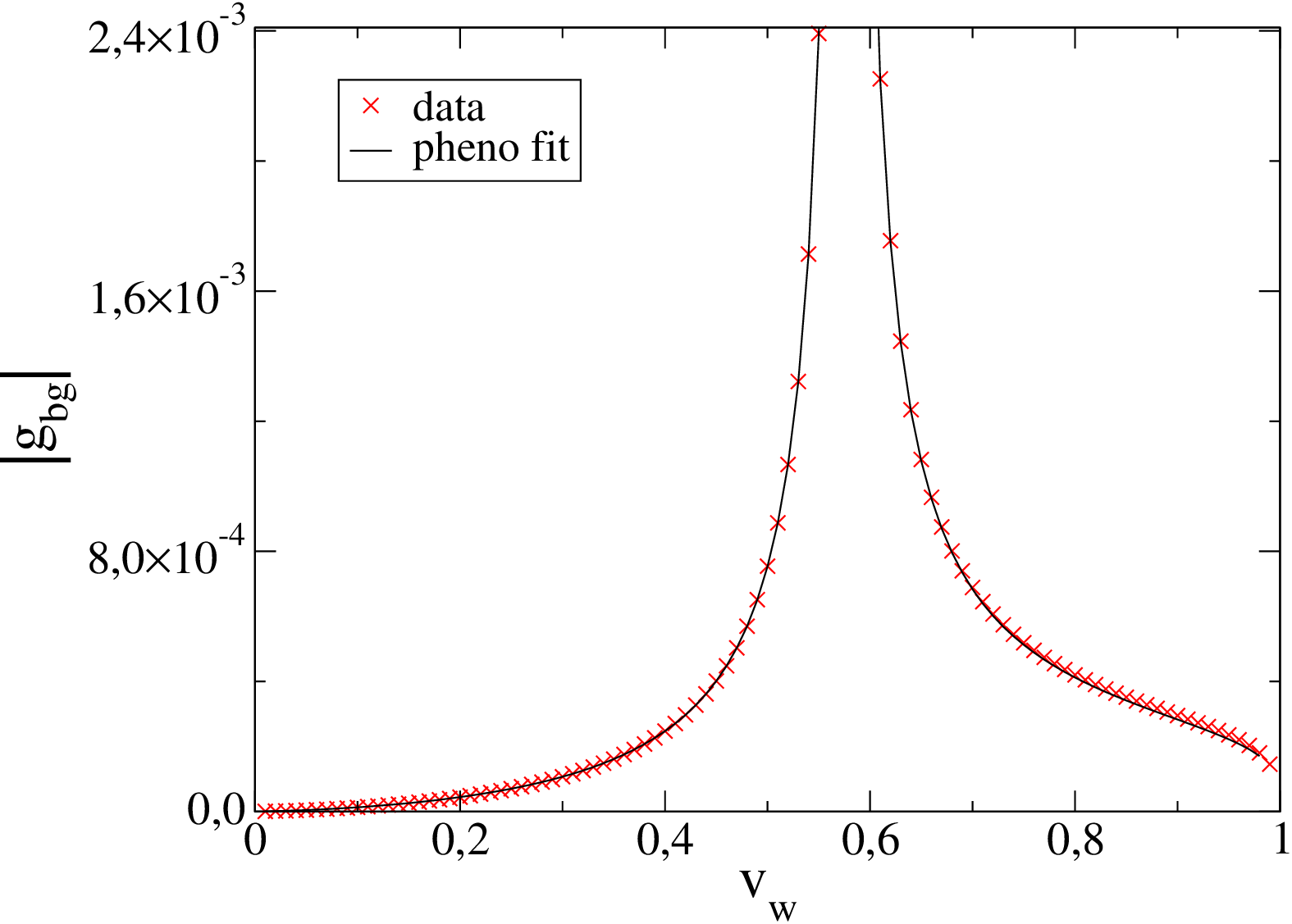}
 \caption{\small The velocity dependence of the friction components in comparison to the fit (\ref{fit1}) - (\ref{fit4}) 
(solid lines). Different colors represent different strengths of phase transition 
for the fluid parts ($\phi_0/T= \{ 1, 2 \}$), while the background components scale as $(\phi_0 / T)^4$. The wall thickness in all plots is $L \, T=30$.
\label{fig:flvw} 
}
\end{figure}

The functions $f_{fl}, f_{bg}, g_{fl}, g_{bg}$ heavily
simplify for thick bubble walls.  More specifically, under the
condition
\be
\label{eq:GammaL}
A^{-1} \Gamma L \gg \gamma \, ,
\ee
the kinetic term in the fluid equations (\ref{eq:fluid_stuff}) and (\ref{sys0}) can be neglected 
and the background equation \eqref{eq:fluid_bg} yields 
\bea
\label{eq:relaxation_time}
q_{W,t} &\simeq& (\Gamma_{W,t} - A_{W,t} \, A_{bg}^{-1} \, \Gamma_{bg})^{-1} S \propto m^\prime m \, , \\
q_{bg} &=& \int dz \, A_{bg}^{-1} \, \Gamma_{bg} \, q \propto m^2 
\eea
(in the last relation we neglected the $z$-dependence in $c_1$). 
Hence, $f_{fl}$ and $g_{fl}$ scale as $1/L \, T$, whereas $f_{bg}$ and
$g_{bg}$ are independent of it.  Their dependence on $L \, T$, $v_w$
and $\phi_0/T$ are shown in figures \ref{fig:lsc}--\ref{fig:flvw} for
several phase transitions parameters.
By inspecting the eigenvalues of the scattering terms
$\Gamma$, one can observe that the criterion (\ref{eq:GammaL}) amounts
to $L\,T \gg 20$. This is well reproduced by the numerical results in figure \ref{fig:lsc}
that show how the friction terms approach above scaling.

Generally, the friction terms of the background fields display the
proportionalities $f_{bg} \propto m^4 \propto \phi^4$ and
$g_{bg} \propto \phi m^4 \propto \phi^5$.  The friction terms of the
fluid have an additional dependence on $\phi$ via the mass-dependent
coefficient $c_1$ in (\ref{eq:linear}), which is absent in the
background due to its vanishing chemical potential $\delta \mu_{bg}$ (see
the discussion on the background in section \ref{sec:fluid}). A fit to
the numerical data yields $f_{fl} \propto \phi^{7/2}$ and
$g_{fl} \propto \phi^{9/2}$. This behavior is shown in figure \ref{fig:flpT}.

Using the above
proportionalities, a reasonable fit for the parameterizations
(\ref{eq:pheno2}) turns out to be
(for $L \, T \gg 20$)
\bea
f_{fl} &=& \frac{6.6 \times 10^{-2}}{ T L}  (v_w + 0.1 v_w^2) \sqrt \gamma
  \left( \frac {\phi_0}{T}\right)^{7/2}  \ ,
\label{fit1}\\
g_{fl} &=& \frac{1.8 \times 10^{-2}}{ T L} (v_w + 0.85 v_w^2) \sqrt \gamma
  \left( \frac {\phi_0}{T}\right)^{9/2}   \ ,\label{fit2}\\
f_{bg} &=& -1.8 \cdot 10^{-3} \times \frac{(v_w + 5.5 v_w^2)}{(c_s^2 -v_w^2)} \times \frac{1}{\sqrt \gamma} \ , \label{fit3}\\
g_{bg} &=& -6 \cdot 10^{-4} \times  \frac{(v_w + 11.5 v_w^2)}{(c_s^2 -v_w^2)} \times \frac{1}{\sqrt \gamma} \ .
\label{fit4}
\eea
The rather complicated dependence on the wall velocity can be
disentangled and traced back to different origins. The quadratic
corrections of the friction components in terms of wall velocity come from the
dependence of the eigenvalues of the system. Also the factor $1/(c_s^2 - v_w^2)$ 
in the background field arises from the eigenvalues of the matrix $A$ in (\ref{eq:relaxation_time}).
On the other hand, the $\sqrt{\gamma}$ enhancement in the fluid functions is
due to the suppression of the collision terms. This enhancement
suggests a divergent friction, but for very fast walls, $ \gamma \gg
10 $, the friction approaches a constant value, which is just given
by the ``free fluid solution" (i.e.~$\Gamma \to 0$). In contrast,
the background functions are suppressed by an additional factor $1/\gamma$
compared to the fluid functions. 
This can be deduced from the fluid equations, which imply that in the
ultra-relativistic limit the background fields have to be
space-independent due to vanishing source and collision terms.
In this case equilibration to the true temperature and fluid velocity only happens 
far behind the bubble wall.

Interestingly, the background contribution to the friction is negative
for subsonic wall velocities.  In fact, this term encodes the impact
of the temperature variation on the Higgs field,
namely
\be
\label{eq:bg_temp_term}
f_{bg} = \int dz \, \partial_z \phi \, \delta \tau \, \frac{d^2V_T}{d\tau d\phi} \ . 
\ee
The leading contribution to $d^2V_T/d\tau d\phi$ comes from the mean-field
term $V_T \propto m^2 T^2$ and is positive. For the deflagration mode,
the temperature drops across the wall and makes this term (\ref{eq:bg_temp_term})
negative. For supersonic wall velocities, the sign changes and this
term acts as an additional friction, hindering the wall expansion.

For small velocities, the usual friction dominates, but the
contribution from the background grows with an additional factor
$1/(c_s^2 - v_w^2)$. One curious consequence of this behavior is that
there is a wall velocity with maximal friction and hence a maximal
velocity in the deflagration mode. This is a dynamically effect and
not related to the considerations about entropy increase
in~\cite{Ignatius:1993qn}. For example, we find the numerical values
\bea
\label{eq:vw_bounds}
v_w < 0.37 \quad \textrm{or} \quad v_w > 0.74  
&\qquad& {\rm for}\quad \phi_0/T = 1, \quad LT \simeq 30 \, , \nn \\
v_w < 0.33 \quad \textrm{or} \quad v_w > 0.76 
&\qquad&  {\rm for} \quad \phi_0/T = 2, \quad LT \simeq 30 \, .
\eea
This effect seems not be present in the phenomenological approach
(\ref{eq:phenoOld}). In this case, even if the change in temperature
is accounted for, the terminal velocity in the deflagration mode seems
to be the speed of sound and not significantly below
it~\cite{Huber:2013kj}. Likewise, there is a minimal velocity for the
detonation mode. Hence, there results a gap (in terms of pressure
difference $\Delta V$) for which no solution exists to the linearized
Boltzmann equations. So, even if the friction is well represented
in the phenomenological approach eq.~(\ref{eq:pheno2}), 
the contribution from the background is quite different.

\begin{figure}
 \hspace{1mm}     \includegraphics[width=0.435 \textwidth]{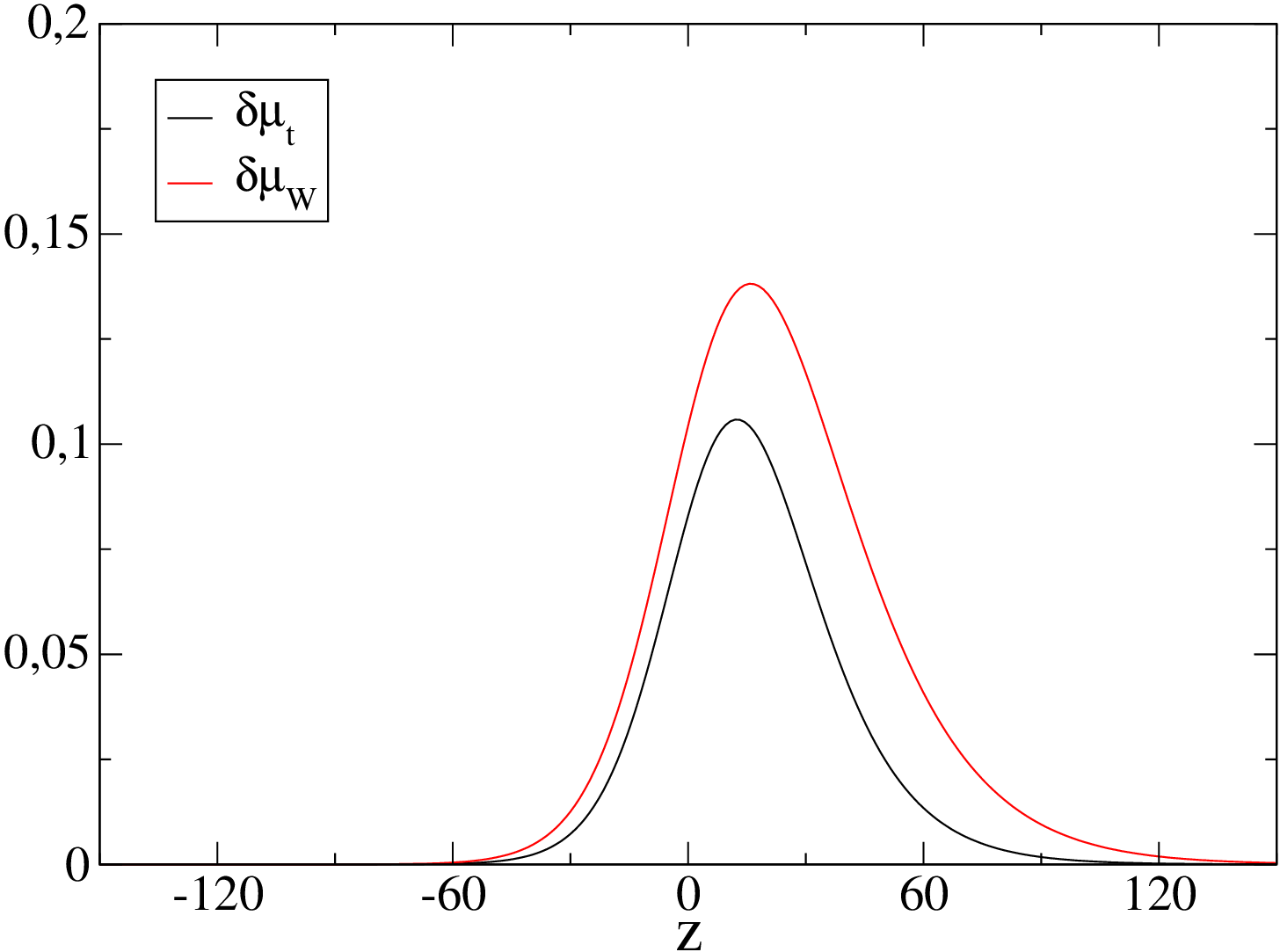}\qquad \quad
      \includegraphics[width=0.445\textwidth]{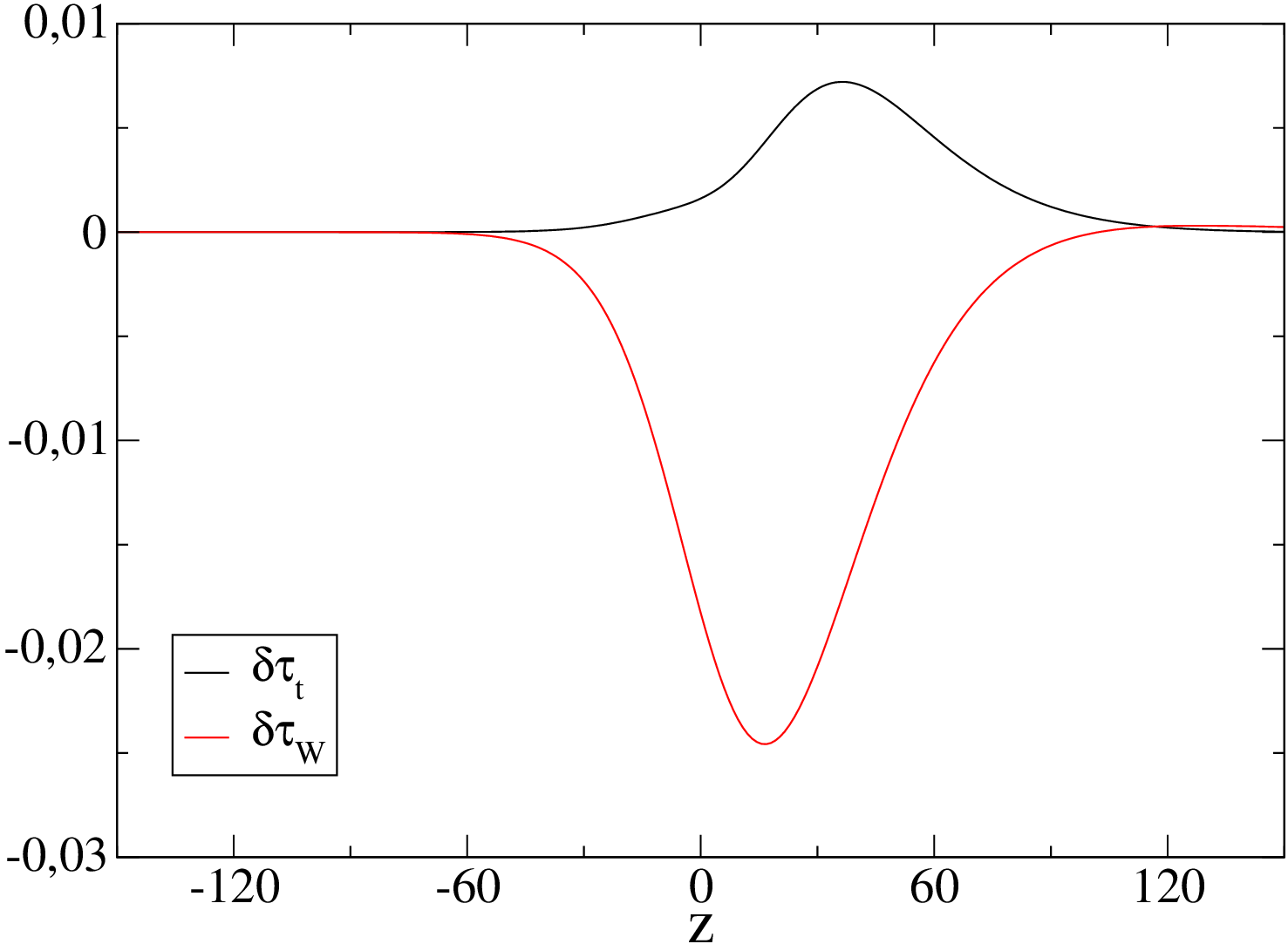}
\\[25pt]
\centering      \includegraphics[width=0.45\textwidth]{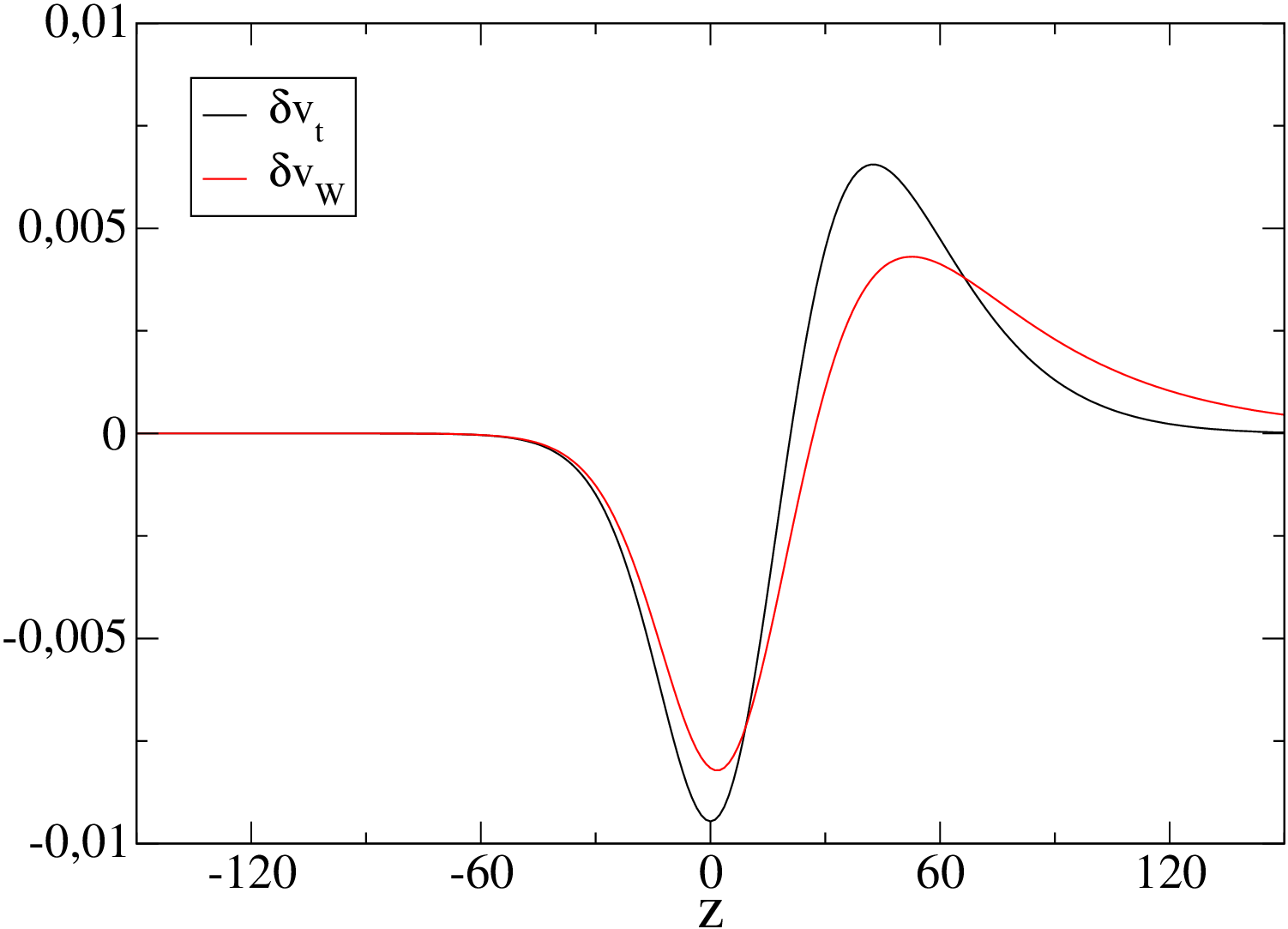}\qquad \quad
      \includegraphics[width=0.45\textwidth]{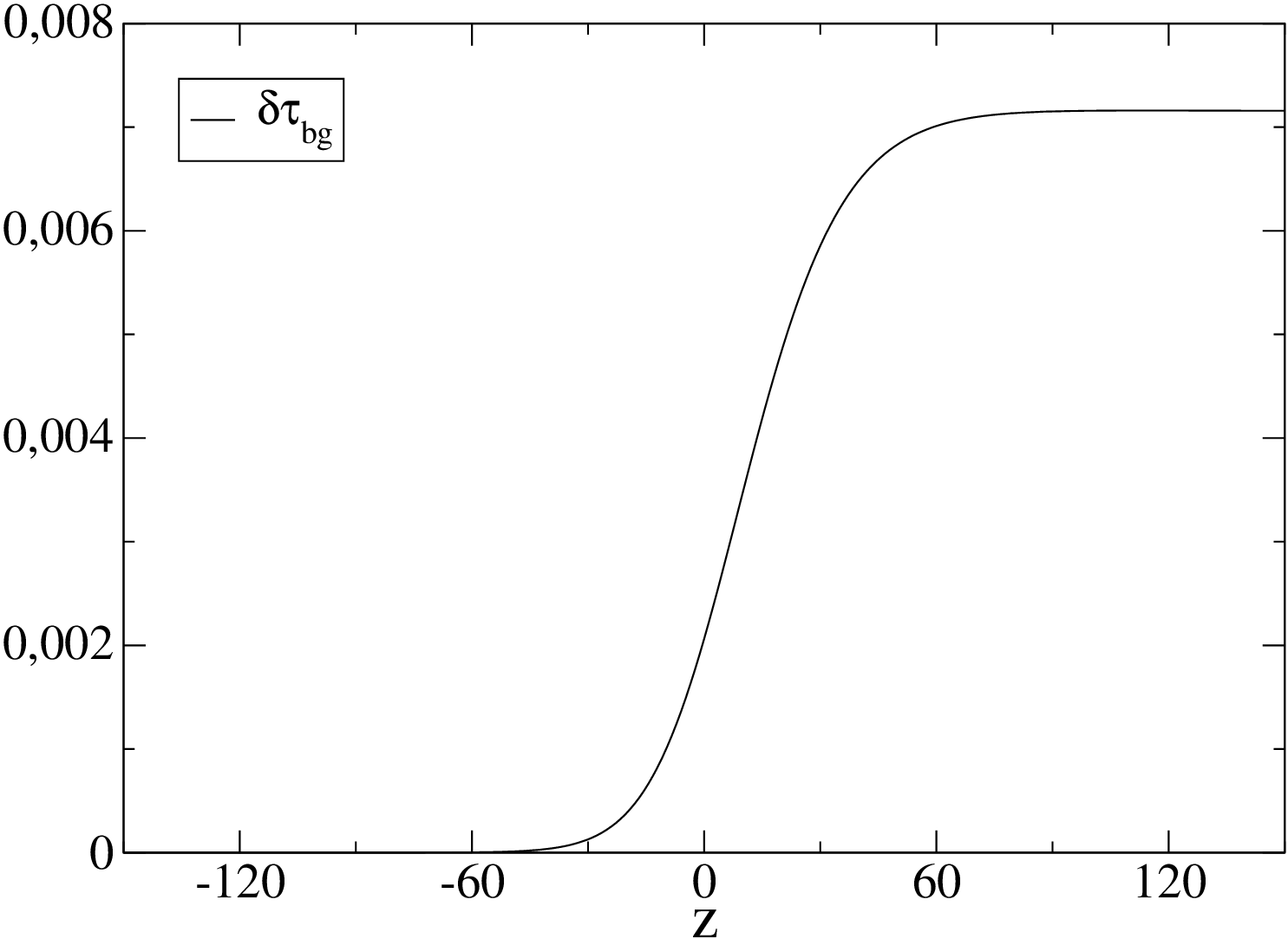}
\\[25pt]
\includegraphics[width=0.45\textwidth]{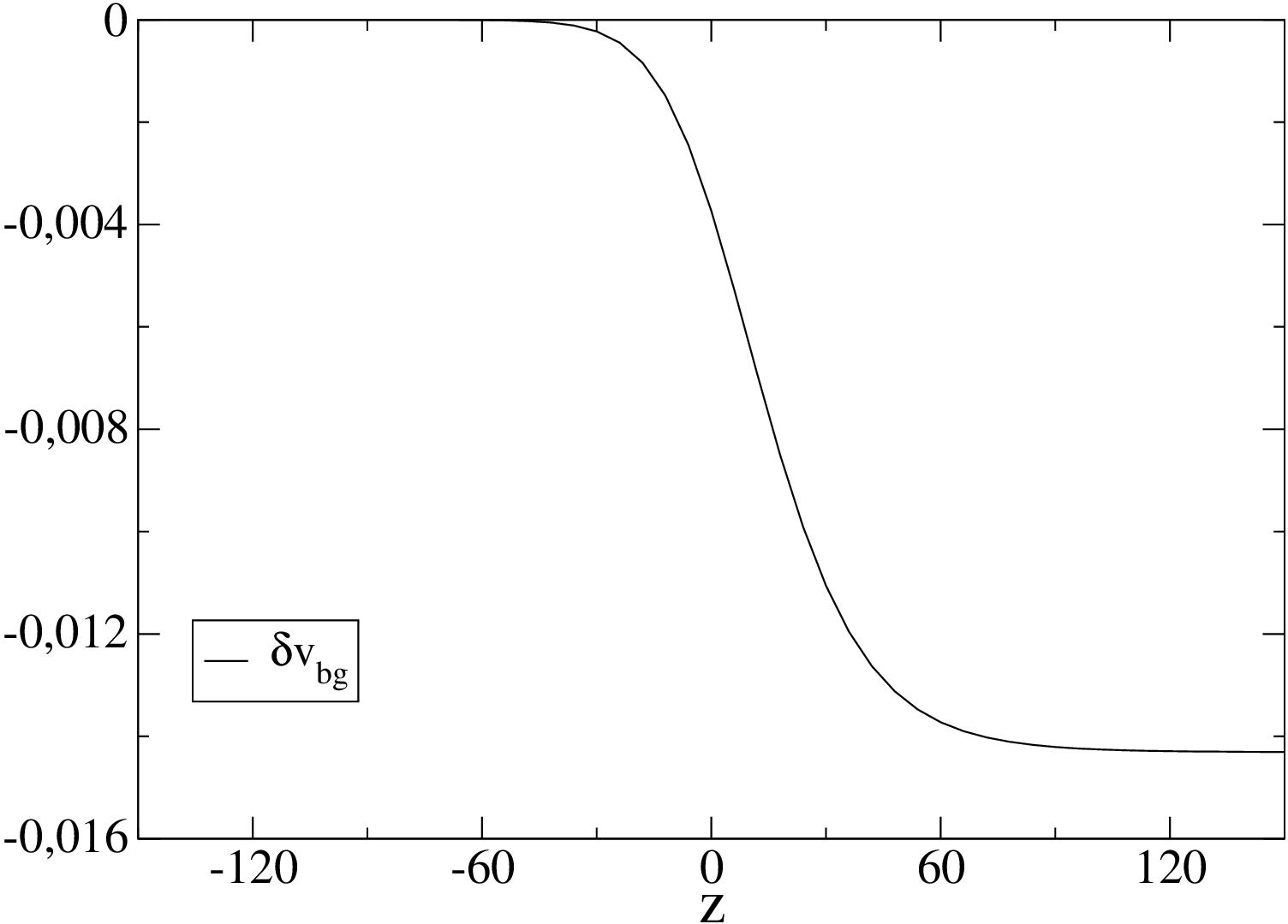}
 \caption{\small Example for the fluctuations in the fluid and background fields; $v_w= 0.5,LT= 30, \phi_0/T=1 $.}
    \label{fig:flbgsol1}
\end{figure}

\begin{figure}
  \centering
      \includegraphics[width=0.45\textwidth]{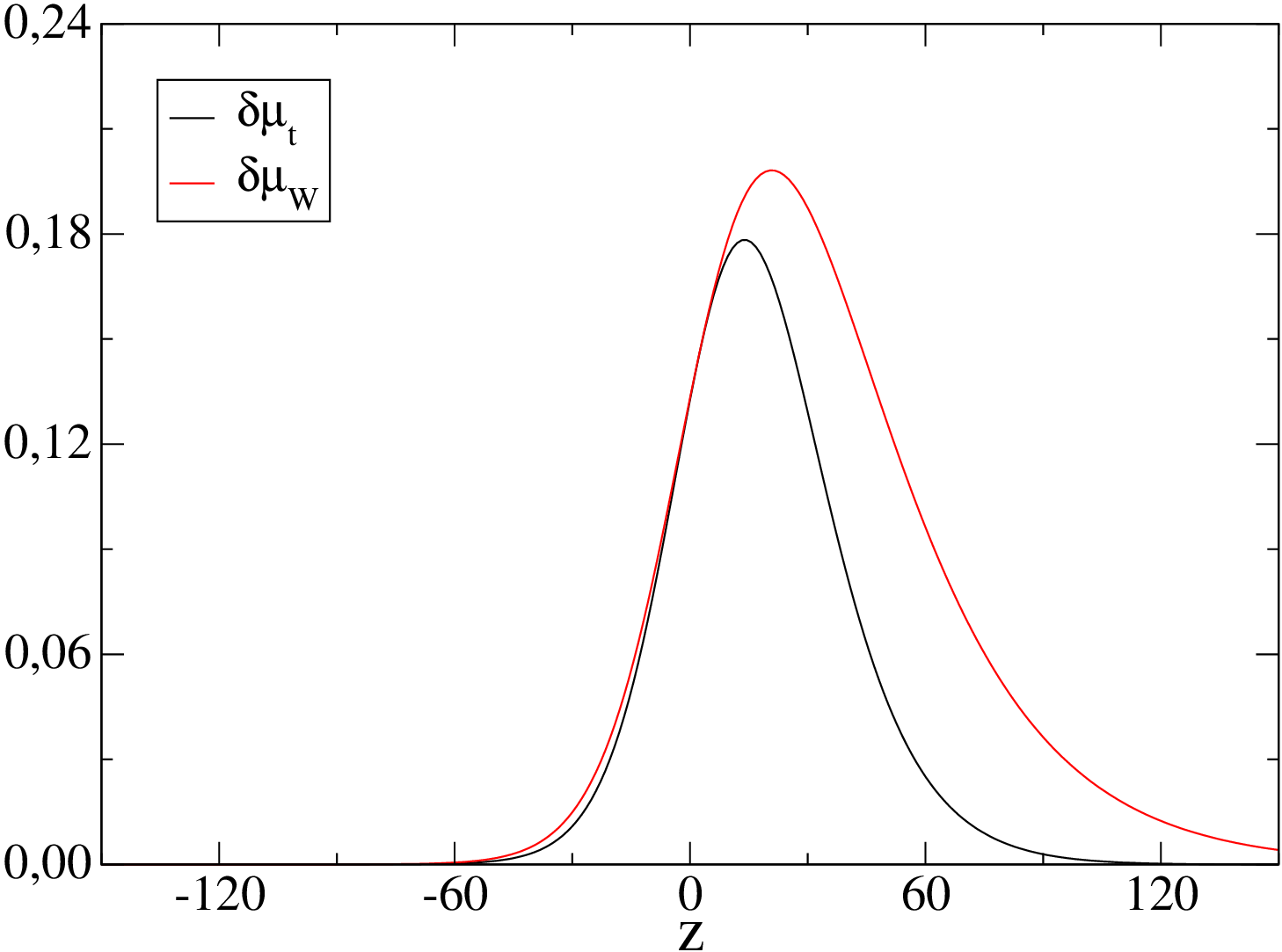}\qquad \quad
      \includegraphics[width=0.45\textwidth]{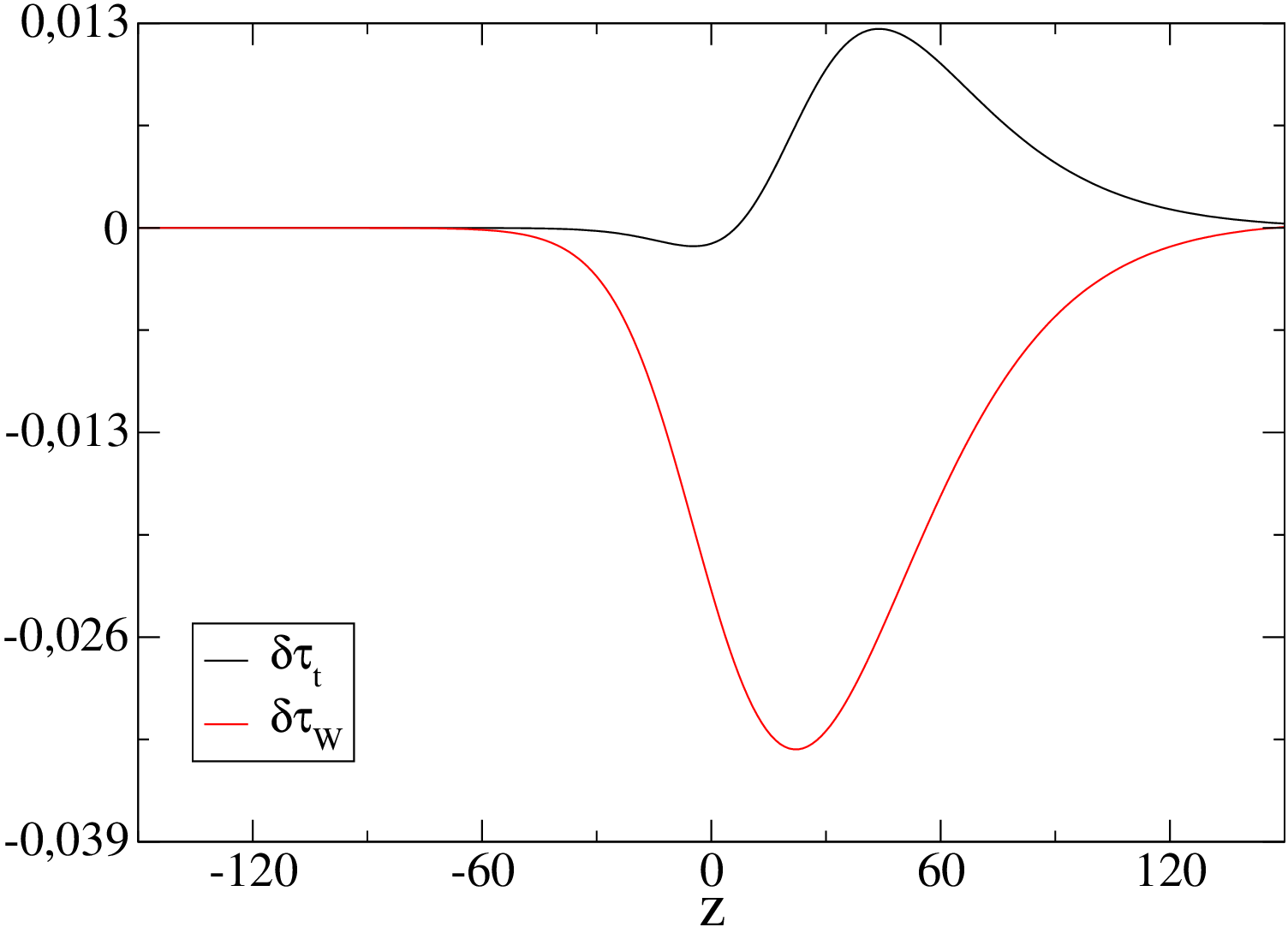}
\\[25pt]
      \includegraphics[width=0.45\textwidth]{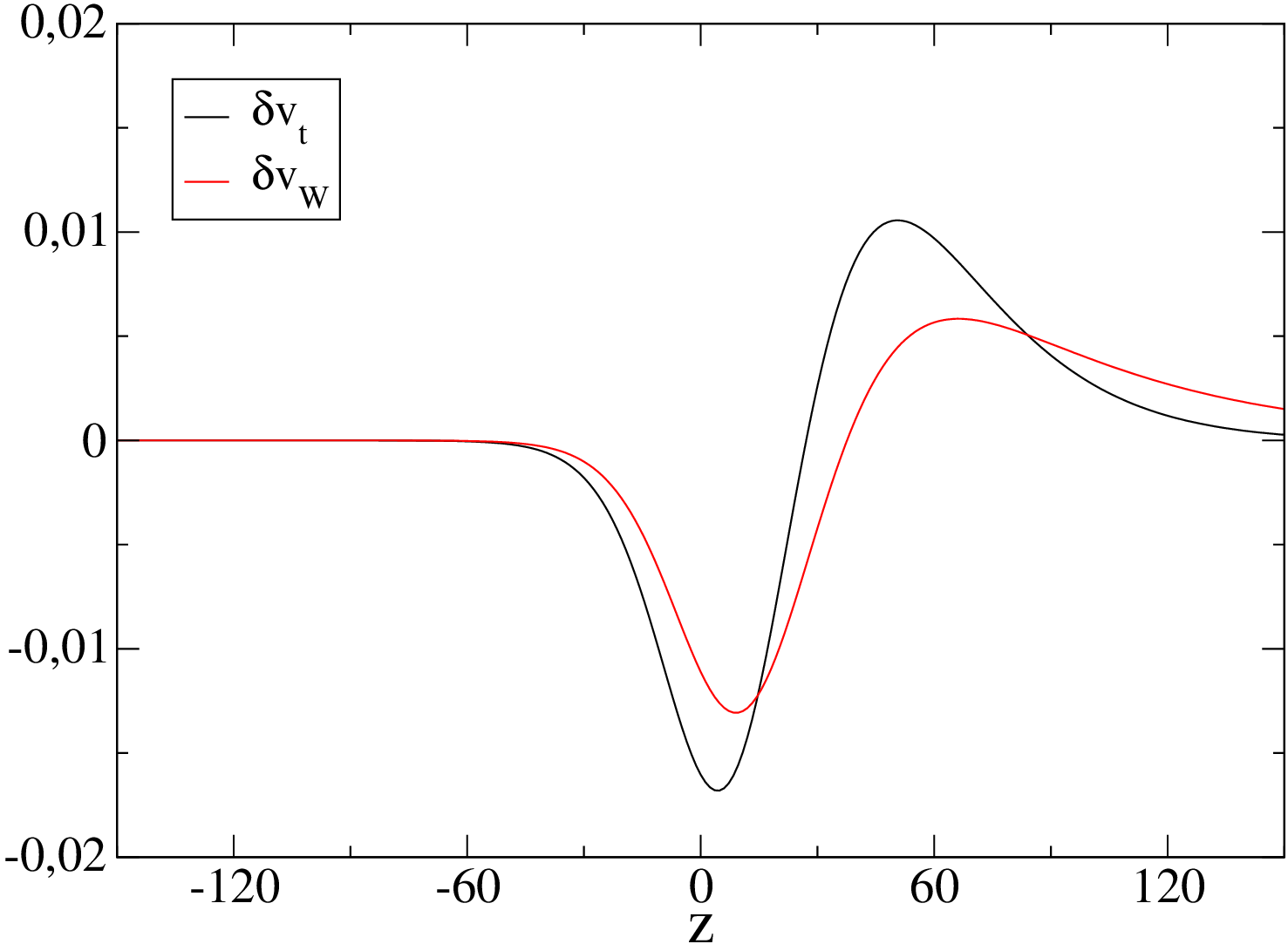}\qquad \quad
      \includegraphics[width=0.45\textwidth]{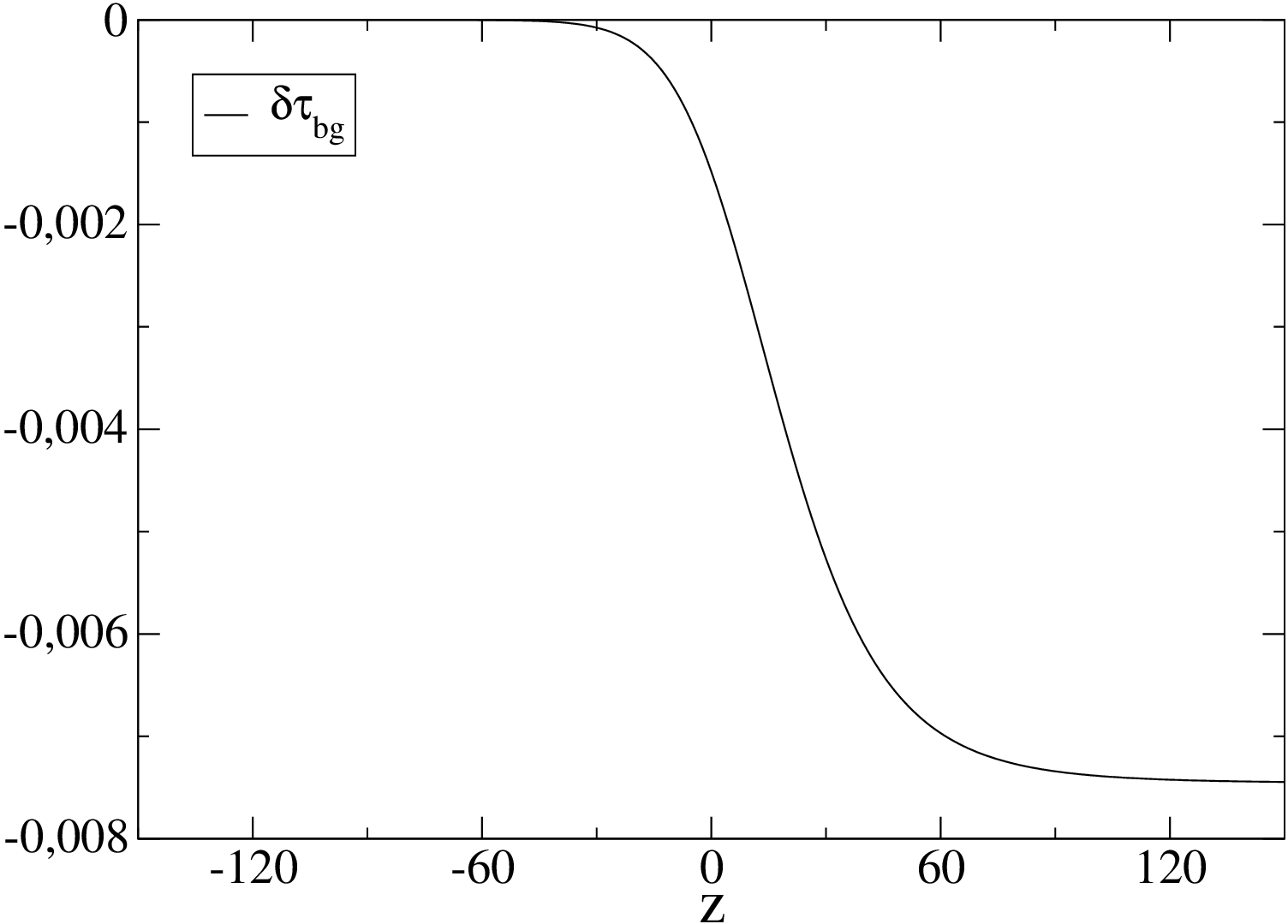}
\\[25pt]
\includegraphics[width=0.45\textwidth]{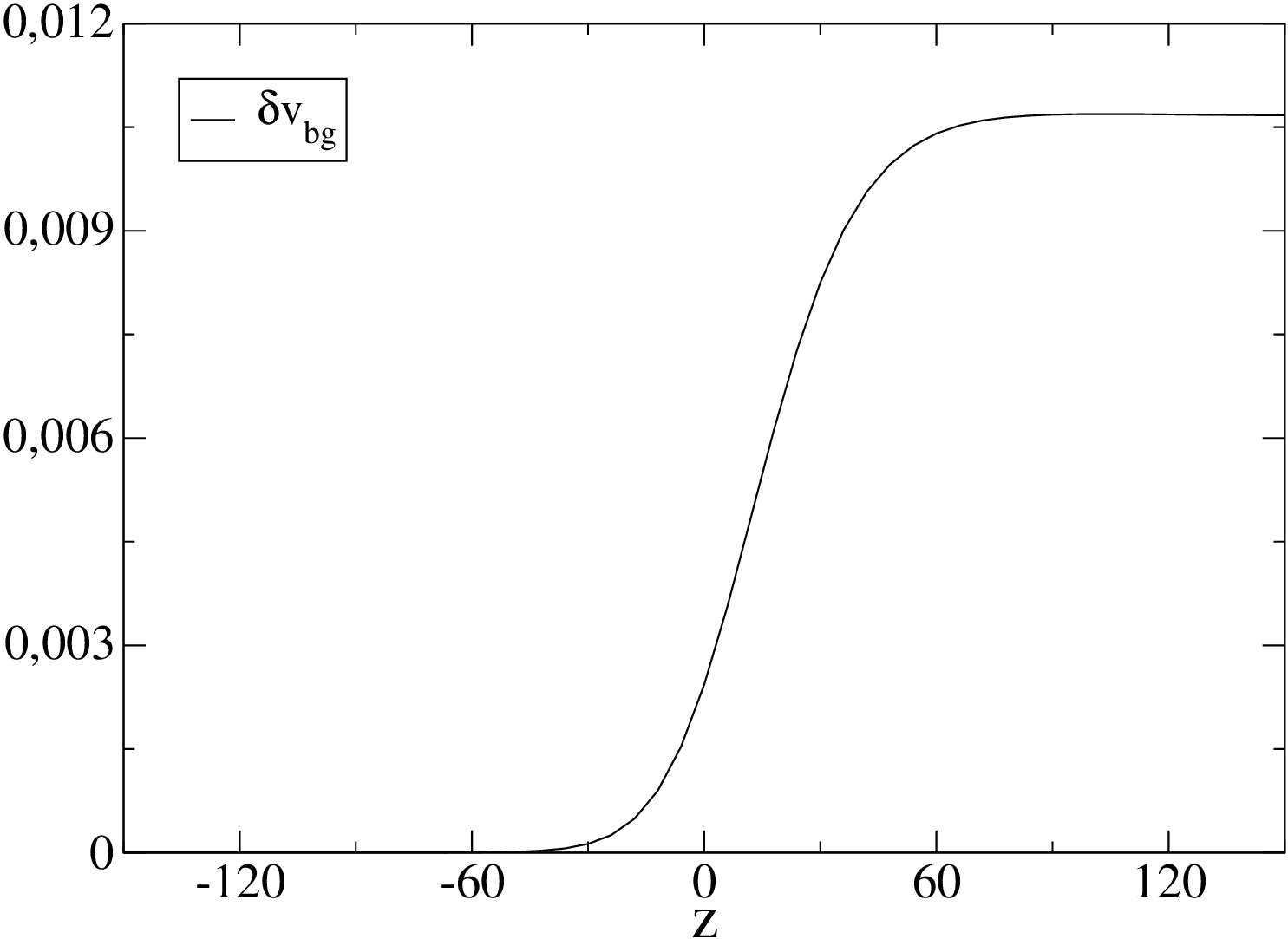}
 \caption{\small Example for the fluctuations in the fluid and background fields; $v_w= 0.7,LT= 30, \phi_0/T=1 $.}
    \label{fig:flbgsol2}
\end{figure}

\begin{figure}
  \centering
{\color{white}\,}      \includegraphics[width=0.435\textwidth]{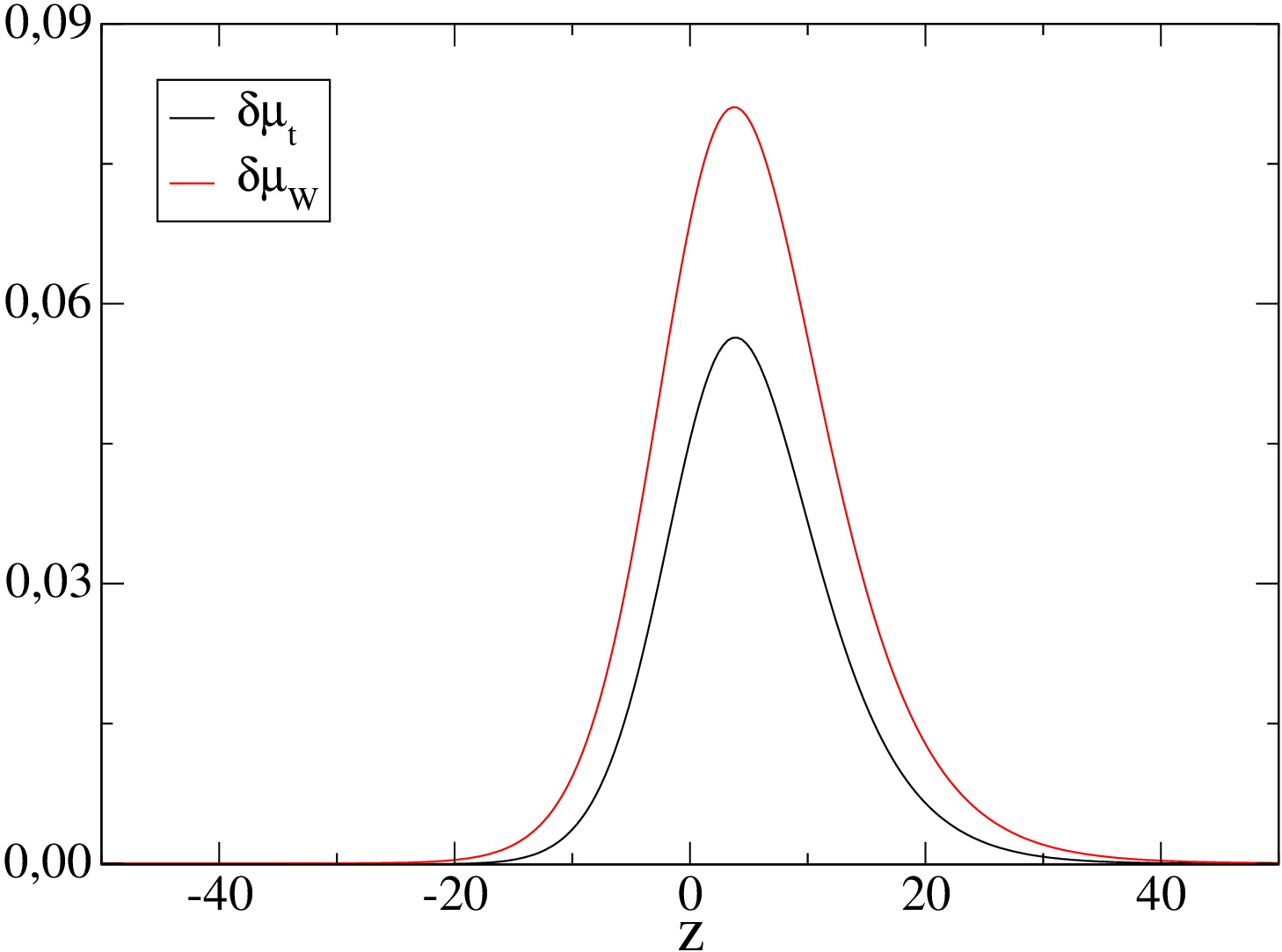} \qquad \quad
      \includegraphics[width=0.45\textwidth]{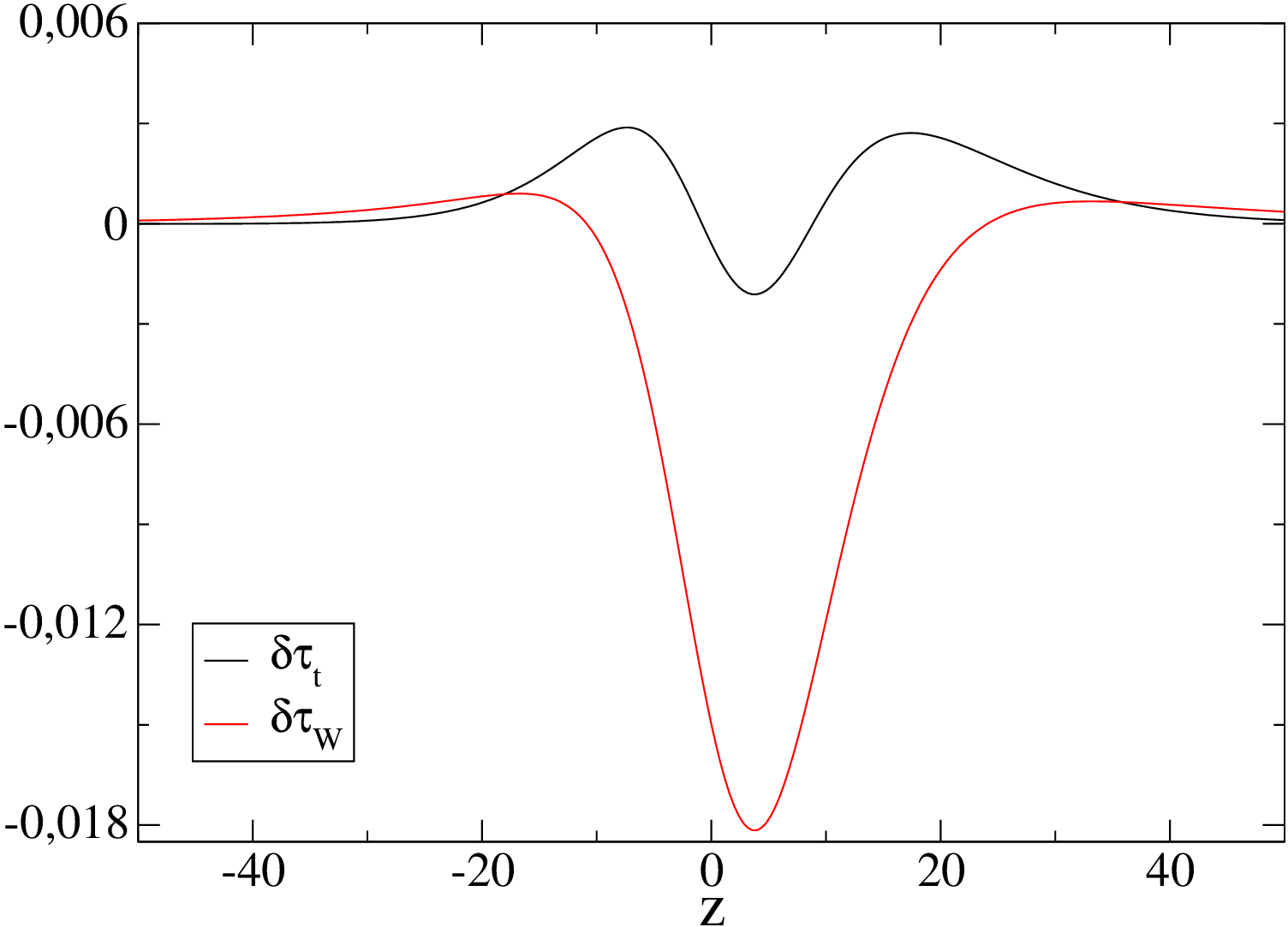}
\\[25pt]
{\color{white}.} \hspace{-7mm}     \includegraphics[width=0.447\textwidth]{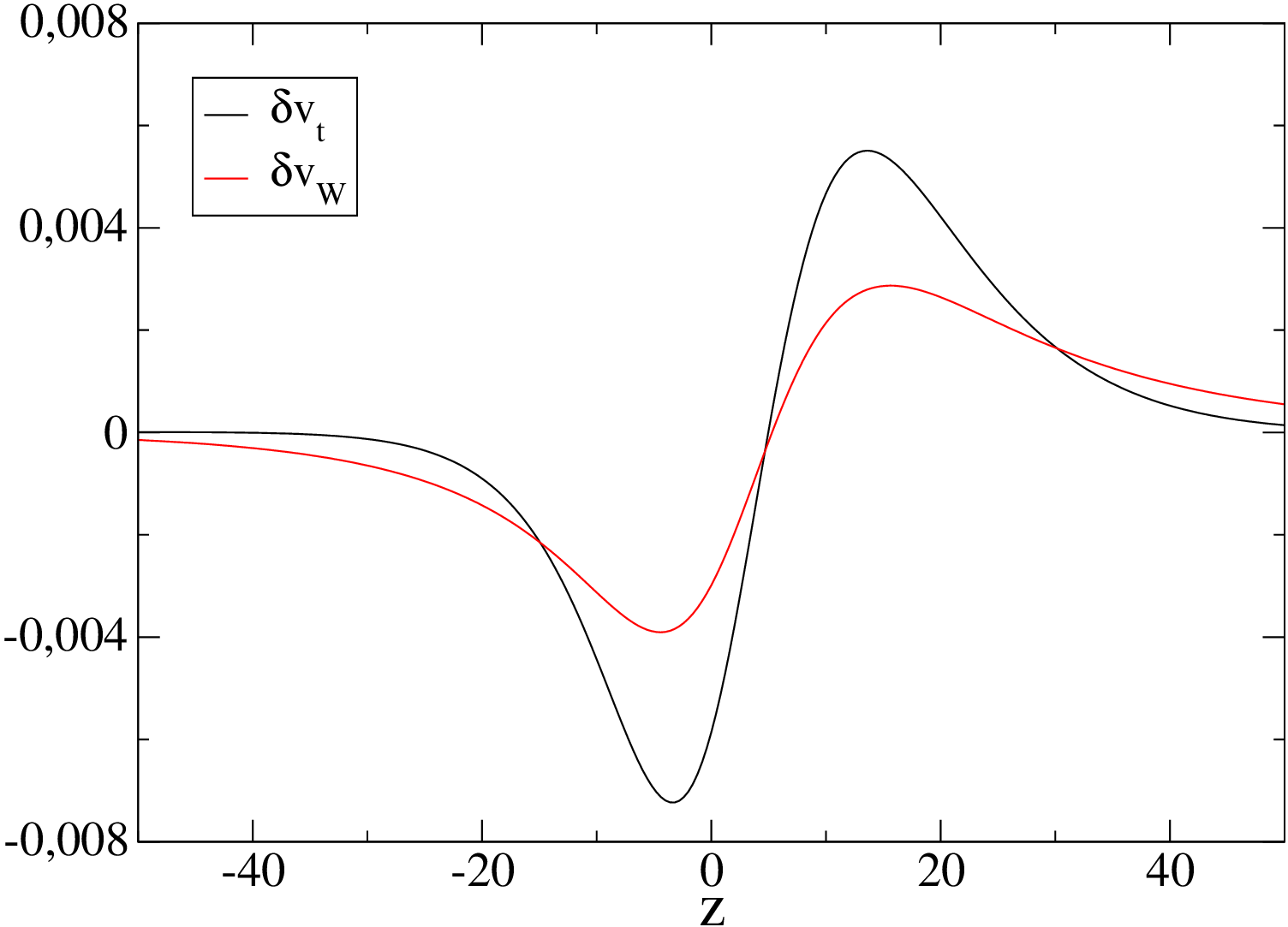}  \qquad 
    \hspace{1mm}  \includegraphics[width=0.465\textwidth]{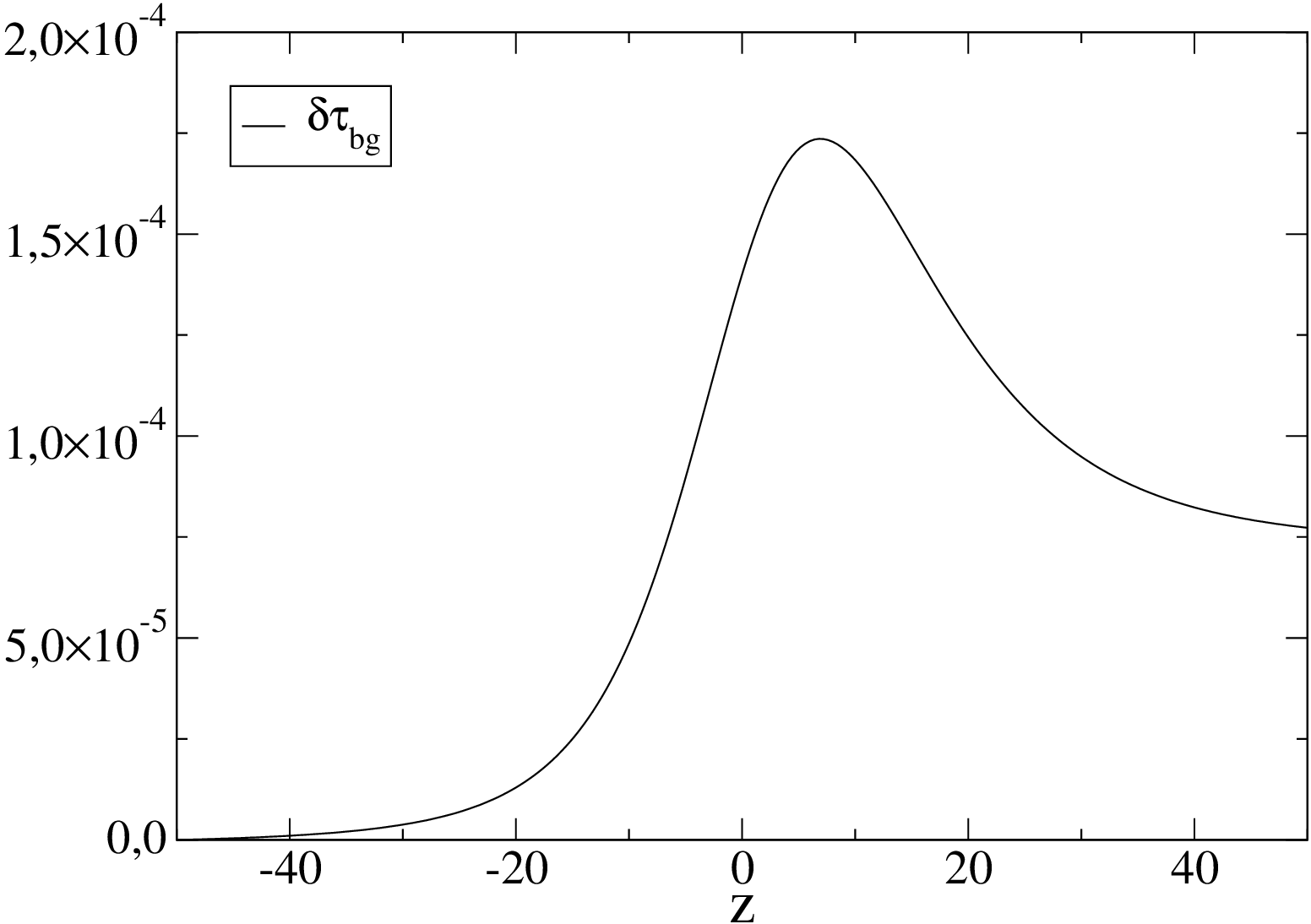}
\\[25pt]
\includegraphics[width=0.45\textwidth]{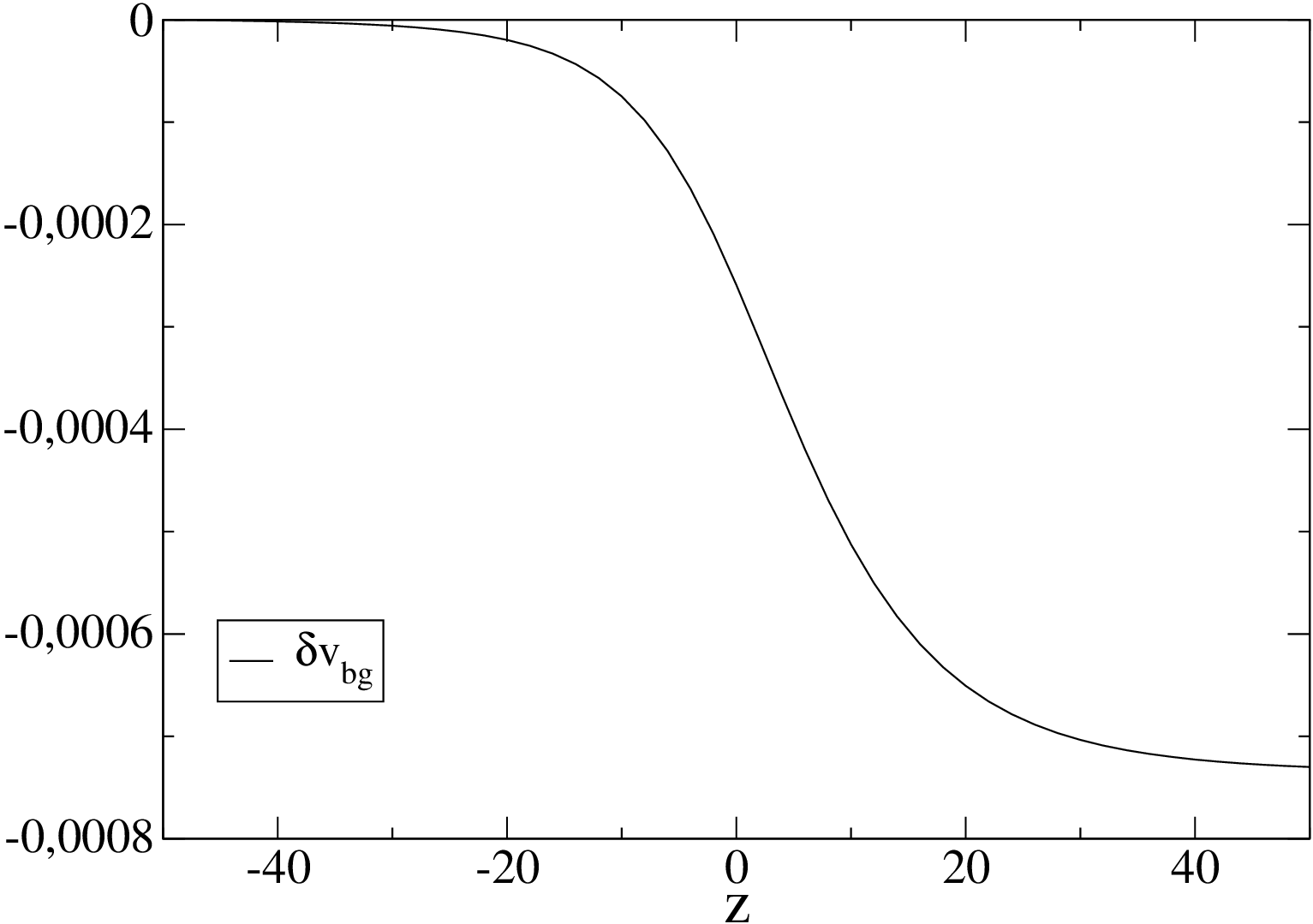}
 \caption{\small Example for the fluctuations in the fluid and background fields; $v_w= 0.1,LT= 10, \phi_0/T=1$.}
    \label{fig:flbgsol3}
\end{figure}

Some examples for the fluctuations are shown in the 
figures \ref{fig:flbgsol1} to \ref{fig:flbgsol3}. 
The first two plots show a deflagration and a detonation in the thick wall regime.
The last example is a detonation with a relatively thin wall.
The first important 
point is that the fluid and background fluctuations
are small in all cases, what justifies the linearization of the equations in (\ref{eq:fluid_stuff}).
This is even true for wall velocities close to the speed of sound or supersonic wall velocities.
Next, we see that depending on the parameters, the profiles in the wall can be quite different than
in the relaxation time approximation (\ref{eq:relaxation_time}). In particular, the background 
fields do not need to be monotonic. 

In some baryogenesis analyses, the wall
thickness is not derived from the Higgs equation (\ref{eq:phenoOld})
but taken from the tunneling bounce profile. This amounts to
neglecting the friction term in eq.~(\ref{eq:pheno2}). It is clear
that this approximation breaks down for very thick walls, since in
this regime the friction dominates over the Higgs kinetic terms.
Depending on the velocity of the bubble wall, this procedure can lead
to thicker or thinner walls than in dynamical treatments including
friction.

In conclusion, the phenomenological model (\ref{eq:phenoOld}) is only
a good description in the regime where the bubble walls are thick 
[cf.~eq.~(\ref{eq:GammaL})] and
the wall velocity is much below the speed of sound.
One can use eqs.~\eqref{eq:pheno2} and
\eqref{fit1}--\eqref{fit4} as an improved phenomenological
model. To this aim, one can proceed by: {\texttt{i)}} computing the
nucleation temperature of the phase transition, e.g.~via the bounce
analysis; {\texttt{ii)}} guessing the value of $v_w$; {\texttt{iii)}}
determining the shock front and the temperature in front of the wall;
{\texttt{iv})} calculating the wall thickness from the second
constraint in eq.~\eqref{eq:pheno2} (the result is rather insensitive
to $v_w$); {\texttt{v)}} checking whether the first equality in
eq.~\eqref{eq:pheno2} is satisfied and, if not, repeating the
procedure from {\texttt{ii} }.

\subsection{Runaway regime}

In \cite{Huber:2013kj} it has been argued that the analysis of runaway
walls can be used to deduce the friction coefficient $\eta$ and that
this procedure leads to very similar results as the matching to the
solutions of the Boltzmann equations. In order to find finite friction
in the limit $v \to 1$, this paper assumed an additional explicit
factor $1/\gamma$ in $\eta$ that cancels the factor $\gamma$ present
in the four-velocity $u^\mu$.

The runaway regime results when the pressure difference from the fluid
is too low to compensate for the pressure difference from the Higgs
field in the wall. In the highly-relativistic regime, the pressure
difference from the fluid can be readily
evaluated~\cite{BodekerMoore}. It is equal to the free energy
difference in the mean-field approximation (evaluated using the
temperature in front of the wall). Hence, the friction approach
(\ref{eq:phenoOld}) can lead to the same runaway criterion only if the
contributions to the finite temperature potential {\em beyond}
mean-field equals the friction term.

Interestingly, the Boltzmann approach leads to a finite friction in
the $v_w \to 1$ limit. This can be seen by inspecting
eqs.~(\ref{eq:linear_formal}) and (\ref{eq:linear}). For
$\gamma_w \to \infty$, their kinetic and source terms are linear in
$\gamma_w$, whereas the collision terms are not. Thus, one can neglect
the collision terms in the Boltzmann equations. Naively, one may
wonder whether this depends on our notation since we absorbed a factor
$\gamma_w$ in the definition of $\bar u^\mu$. It has instead to be
noticed that the velocity fluctuations do not enter the Higgs
equations, and this is why the friction is finite in this
limit. Nevertheless, our fluid ansatz is not justified for such a regime. It is not
guaranteed that the friction calculation actually leads to the
results in~\cite{BodekerMoore} that uses the proper particle distribution
functions of the highly-relativistic limit.  In fact, the two
results scale quite differently. For example, in the SM the leading
terms beyond mean-field are the thermal cubic contributions. Only
bosonic degrees of freedom contribute to them which is quite opposite
to the friction terms, where fermions yield numerically even larger
contributions. In this light, it seems plausible that the agreement
found in reference~\cite{Huber:2013kj} is specific to the considered
model, namely the SM with low cutoff (see section \ref{sec:lowCO}).


\section{Applications to models\label{sec:resu}}
In this section we apply the method we have previously discussed, to
calculate the wall dynamics in simple models providing first-order
phase transitions.  The first model we consider is the SM with a small
(and experimentally excluded) Higgs mass. The analysis of this
scenario allows to compare our approach with the original
calculation in M\&P. It also permits to point out some peculiarities
of those first-order phase transitions that rely solely on a
temperature-induced cubic term in the free energy.

The second model we analyze is the SM with a low cutoff. In this
framework we will check the consistency of our results with those of
ref.~\cite{Huber:2013kj} where a phenomenological approach is
employed.

Finally, we discuss the phase transition in the gauge-singlet scalar
extension of the SM.  This framework has enough free parameters to
disentangle the effect of the pressure due to the finite-temperature
potential from that of the phase transition strength. Qualitative
behaviors not emerging in the previous two models will be
highlighted.

The numerical results we present are based on the following
procedure. We use the bounce method to determine the bubble action
$S(T)$ and we define the nucleation temperature $T_n$ such that
$S_3(T_n)/T_n=140$~\cite{Quiros:1999jp}. This also provides $L_{nucl}$, the thickness of
the wall at the nucleation time. ($L_{nucl}$ is defined such that the
integrations of the bounce profile and of the
function \eqref{eq:profile} with $L=L_{nucl}$, are equal).
The wall thickness satisfying the fluid constraints is dubbed $L_{dyna}$ in the following.
Depending on the wall velocity, the temperature in front of the wall varies. 
We calculate this temperature using the methods discussed in section \ref{sec:shock} and denote it 
by $T_w$.

\subsection{The Standard Model with light Higgs}

For a first model, we consider the SM with a Higgs mass
$m_h \leq 70\, $GeV and compare with the findings of M\&P. 
The comparison has of course only illustrative
purposes: for a Higgs mass in agreement with the LHC
measurements~\cite{Aad:2012tfa,Chatrchyan:2012ufa} the electroweak
phase transition in the SM is a
crossover~\cite{Kajantie:1996mn,Aoki:1999fi}.

We implement the high-temperature expansion of the one-loop effective
potential \cite{Quiros:1999jp}
\begin{align}
 V_{\rm SM} = - D (T_D^2-T^2) \phi^2 - E \phi^3 T
 + \frac{\lambda_T}{4} \phi^4 \ ,
\end{align}
where the Coleman-Weinberg corrections are included as follows:
\begin{align}
\lambda_T &= \frac{m_h^2}{2 v_{0}} - \frac{3}{16 \pi^2 v_{0}^4} \left( 2m_W^4 \ln \frac{m_W^2}{a_b T^2} + m_Z^4 \ln \frac{m_Z^2}{a_b T^2} - m_t^4 \ln \frac{m_t^2}{a_f T^2} \right) \ ,\\
D&= \frac{1}{8 v_{0}^2} (2 m_W^2 + m_Z + 2 m_t^2) \ ,\\
E&= \frac{1}{4 \pi v_{0}^3} \left(2m_W^3 + m_Z^3 \right) \ ,
\end{align}
\begin{align}
T_D^2&= \frac{1}{4 D} \left(m_h^2- \frac{3}{8 \pi^2 v_0^2} (2 m_W^4 + m_Z^4 -4 m_t^4) \right) \ ,
\end{align}
with $v_{0}= 246\,$GeV, $a_f\simeq 14$ and $a_b\simeq 223$. 

\begin{figure}
        \includegraphics[width=0.46\textwidth]{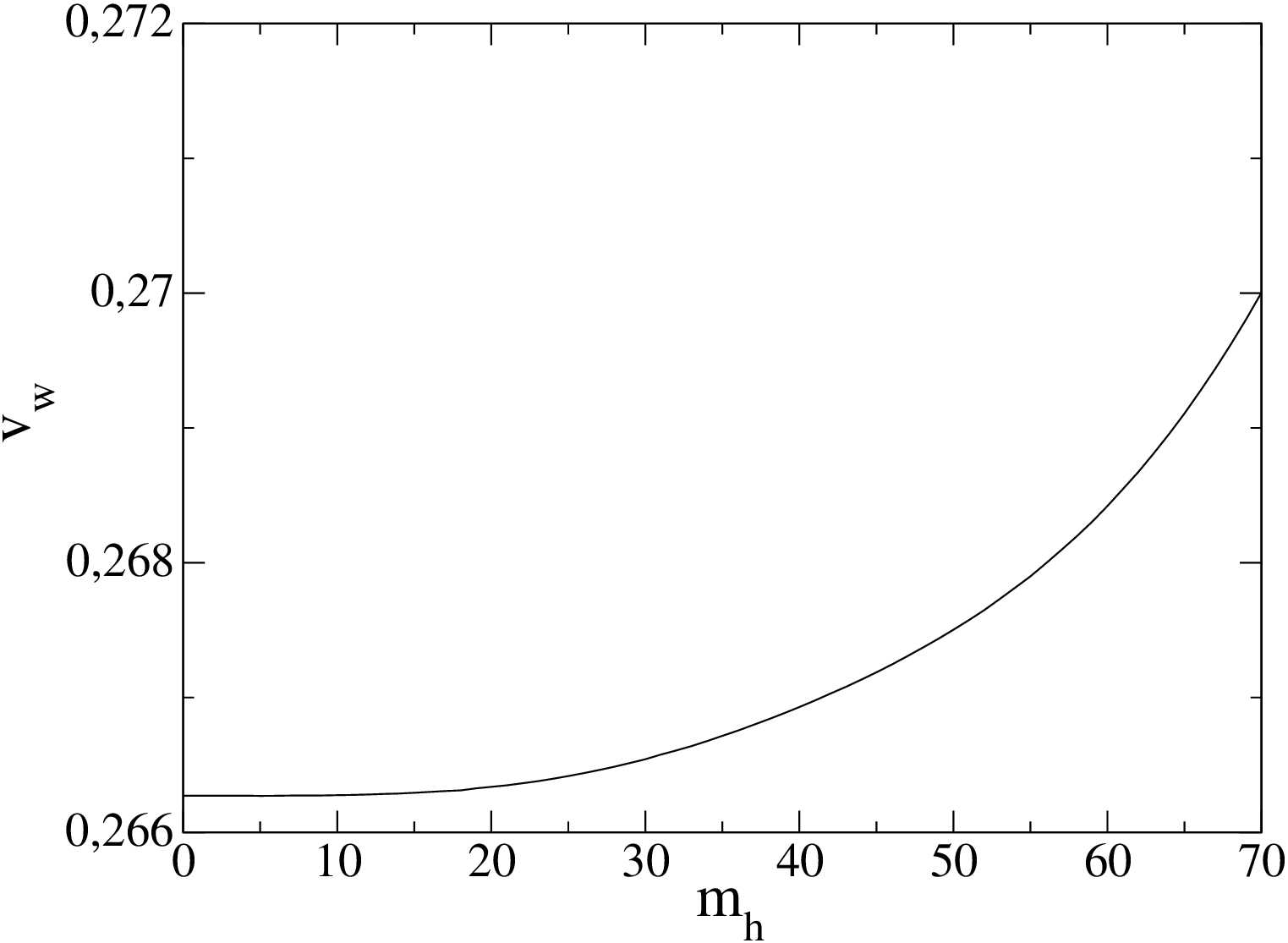}\qquad \quad
   \hspace{1mm}     \includegraphics[width=0.432\textwidth]{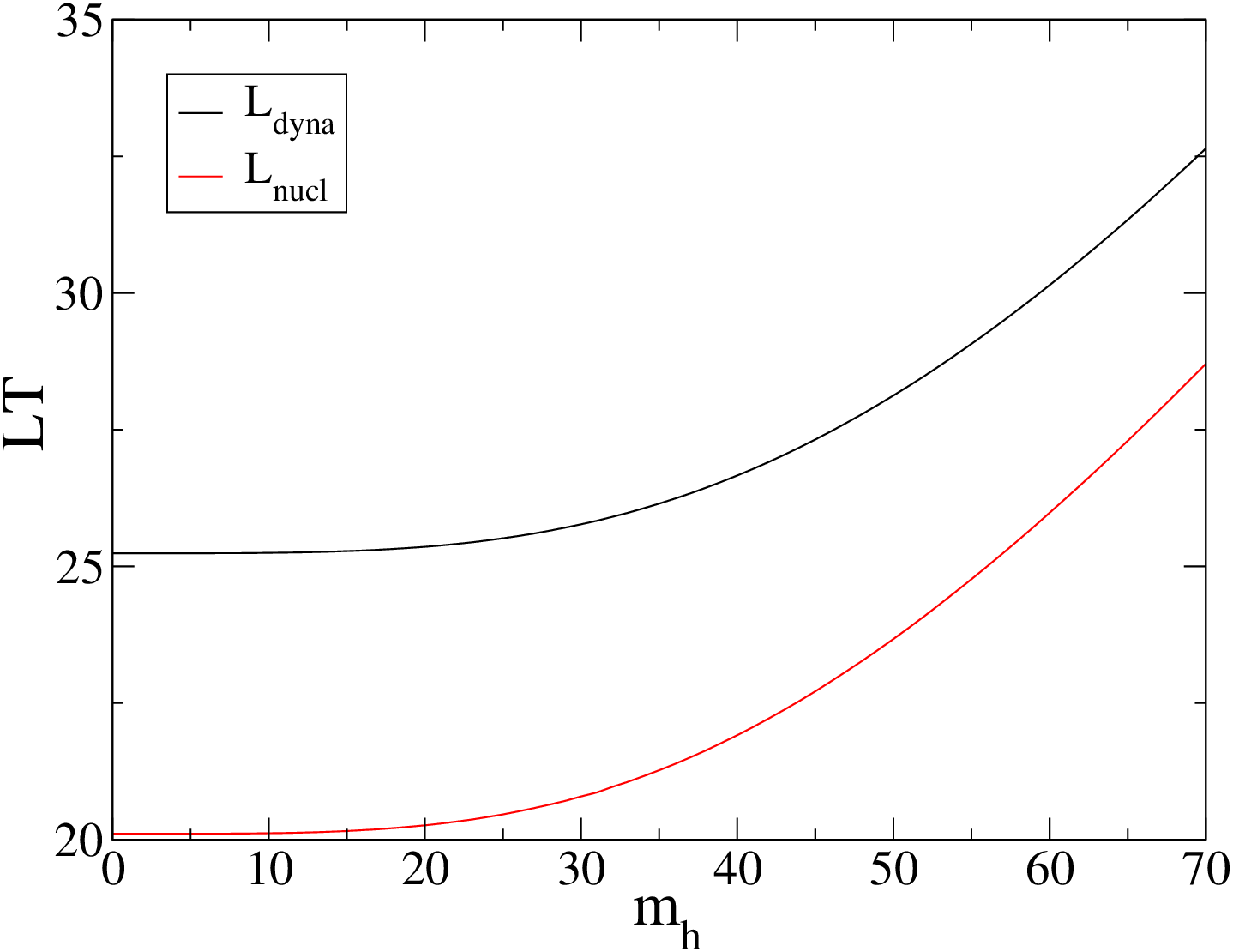}
\\[20pt]
{color{white}.}\hspace{-20mm}        \includegraphics[width=0.45\textwidth]{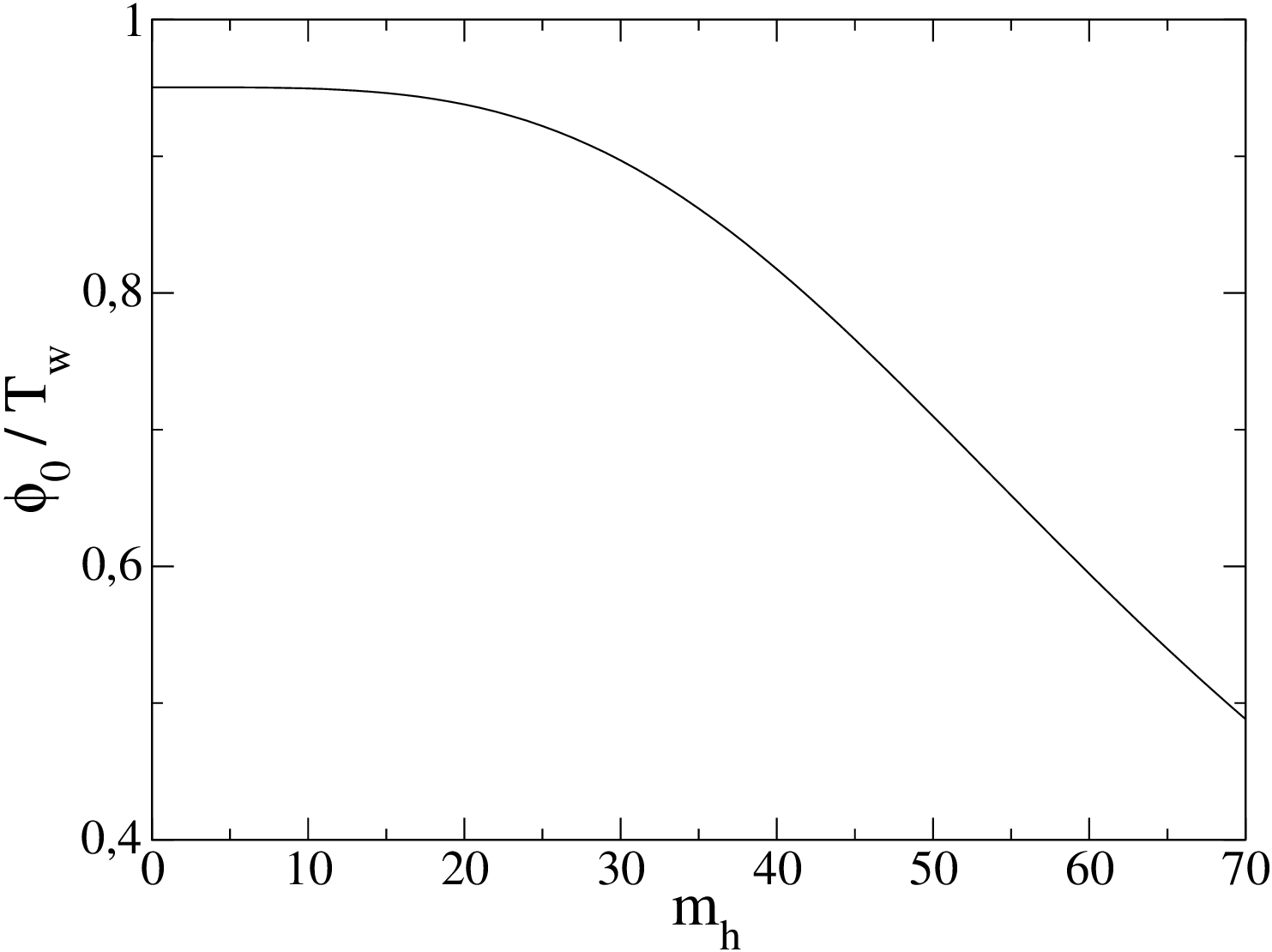}\qquad \hspace{-2mm}
\includegraphics[width=0.48\textwidth]{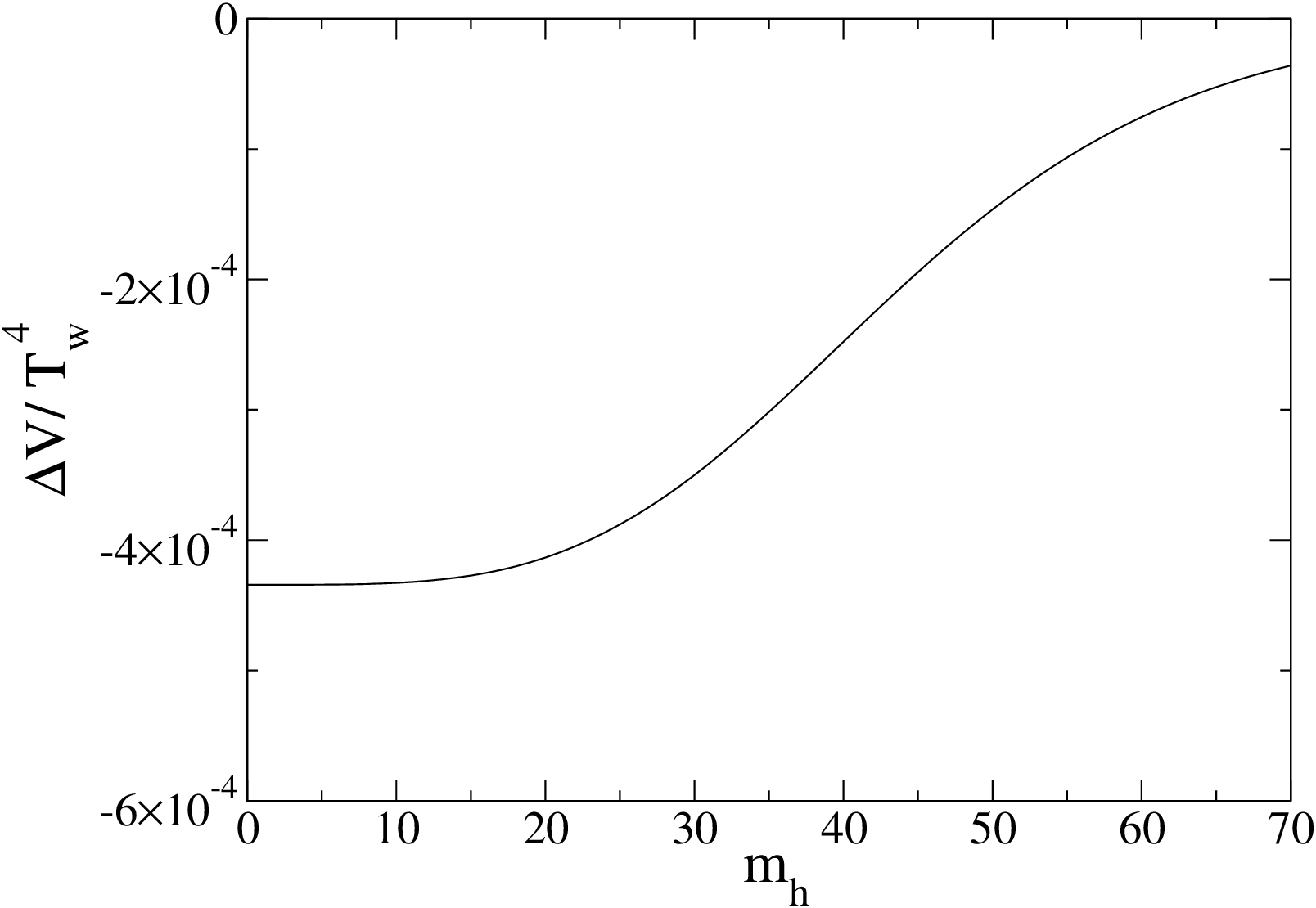}
\\[20pt]
\includegraphics[width=0.46\textwidth]{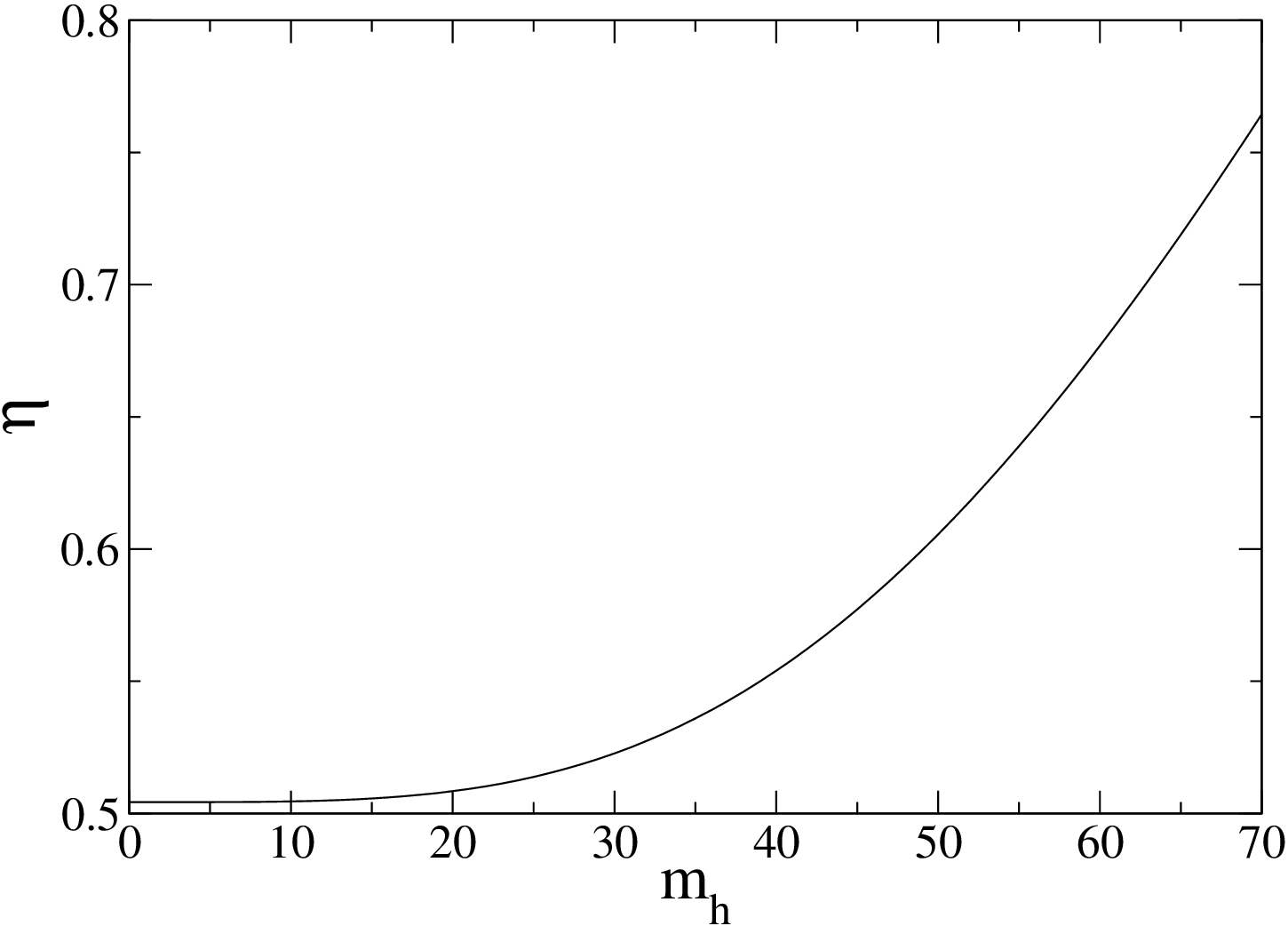}\qquad \quad
   \hspace{-1mm}    \includegraphics[width=0.45\textwidth]{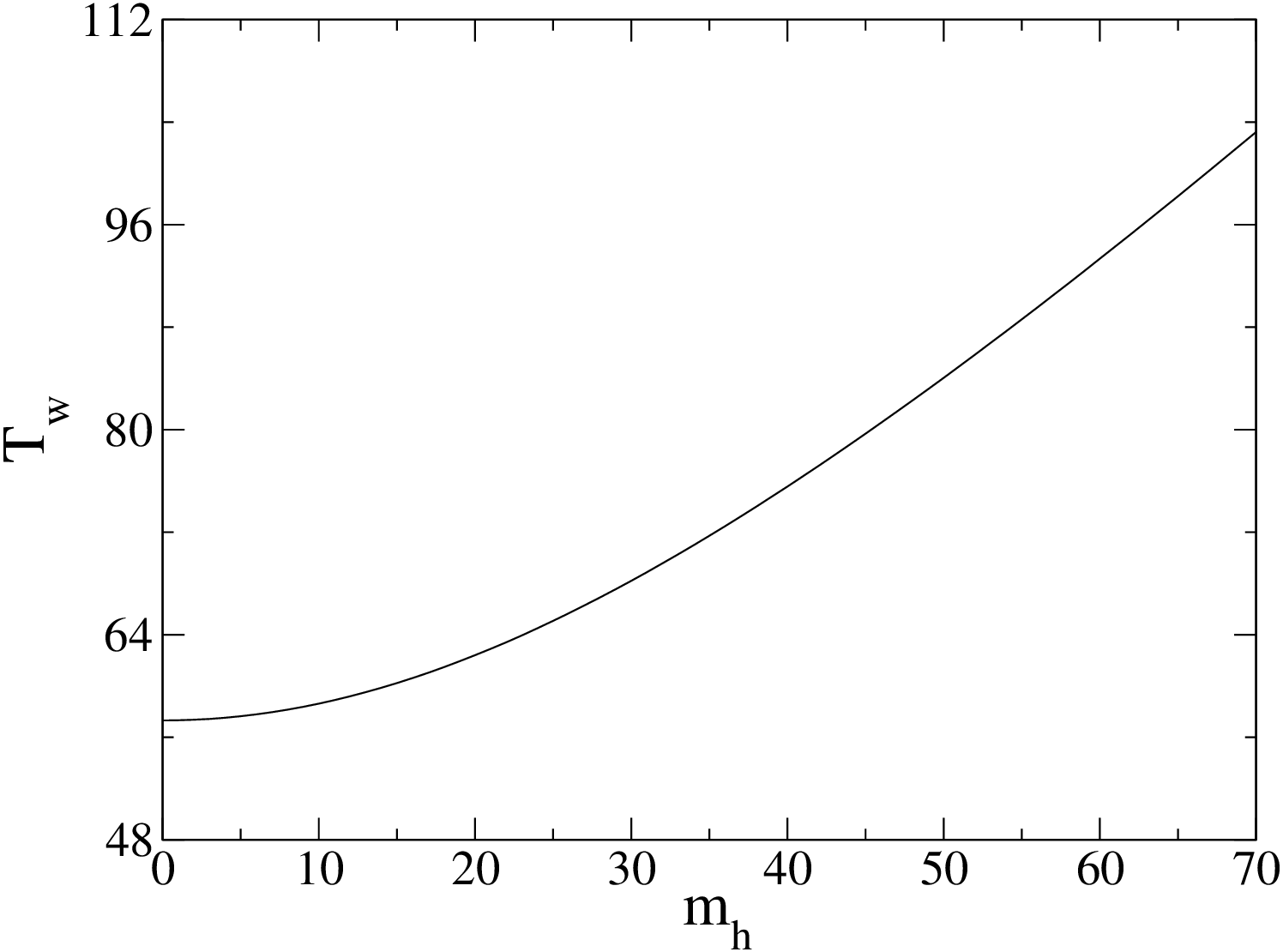}\\
 \caption{
\small 
Characteristics of the phase transition in the SM with a light Higgs.
\label{fig:SM}}  
\end{figure}

For such a potential, we obtain the numerical results presented in
figure \ref{fig:SM}. As expected, the thicknesses $L_{nucl}$ and
$L_{dyna}$ (upper right panel) are closer for weak phase transitions
(cf.~central left panel). Moreover, the wall velocity (upper left
panel) is rather constant, but the thickness shrinks with stronger
phase transitions.
This behavior of $v_w$ is due to the fact that the dependence of the
friction on $\phi_0/T$ and the wall thickness $L T$ is almost
identical to the one of the pressure difference along the wall. 
Concerning the phenomenological approach in the SM, the wall 
thickness is sufficiently large and the wall velocity is
sufficiently small to make the phenomenological approaches
(\ref{eq:phenoOld}) or (\ref{eq:pheno2}) feasible.

Finally, our numerical findings are close to the results found by M\&P
but are not identical. The reasons for this discrepancy can be traced
back to the determination of the shock front (which we treat
non-linearly unlike M\&P), the mass dependence in $c_1$ (that is neglected in M\&P) 
and the slightly different potential.

\subsection{Standard Model with a low cutoff\label{sec:lowCO}}

As a second example we chose a simple extension of the SM. It contains
the SM supplemented by new physics coming into play at a scale $M$ and
producing an effective $\phi^6$ operator at low energy. This framework
allows for strong first-order phase transitions with a Higgs mass
compatible with present LHC data~\cite{Grojean:2004xa}. The additional
content is chosen such that it affects the Higgs potential but does
not contribute to the friction.  

In order to compare our results to those obtained via a purely
hydrodynamic approach to wall velocities, we consider a similar
framework as analyzed in ref.~\cite{Huber:2013kj}. In this case the high
temperature expansion of the Higgs effective potential is given by
\begin{align}
V_{eff}(\phi,T)=&\frac{1}{2}\left[-\mu^2 + \left(\frac 1 2 \lambda+ \frac{3}{16} g_1^2 + \frac{1}{16}g_2^2 + \frac{1}{4} y_t^2\right)T^2\right]\phi^2 \nn \\ 
&- \frac{g_2^3}{16\pi} T \phi^3 + \frac{\lambda}{4} \phi^4 + \frac{3}{64\pi} y_t^4 \phi^4 \ln \left(\frac{Q^2}{c_f T^2}\right)\nn\\ &+ \frac{1}{8 M^2}(\phi^6 + 2 \phi^4 T^2 + \phi^2 T^4 ) \ ,
\end{align}
where $g_1,g_2$ are the electroweak gauge couplings, $h_t$ is the
top-Yukawa coupling, and $Q$ is the renormalization scale fixed at
$Q=m_t$.  The zero-temperature part
\begin{align}
V_{eff}(\phi,0)=& -\frac{\mu^2}{2} \phi^2 + \frac{\lambda}{4}\phi^4 + \frac{1}{8M^2} \phi^6 - \frac{3}{64\pi^2} y_t^4 \phi^4 \left[\ln \left(\frac{y_t^2 \phi^2}{2Q^2}\right)-\frac 3 2 \right] \nn \\
&+ \frac{3}{512 \pi^2} g_2^4 \phi^4 \ln \left[\ln \left(\frac{g_2^2 \phi^2}{4 Q^2} \right)-\frac 3 2 \right] \nn\\
&+ \frac{3}{64 \pi^2}(\frac{g_1^2}{4} + \frac{g_2^2}{4})^2 \left[\left(\ln \frac{(g_1^2 + g_2^2)\phi^2}{4Q^2}\right)-\frac 3 2\right] \ ,
\end{align}
together with the renormalization conditions
\begin{align}
\left. \frac{\partial V_{eff} (\phi,0)}{\partial \phi}\right|_{\phi=v_0}= 0, \qquad \left. \frac{\partial^2 V_{eff} (\phi,0)}{\partial \phi^2}\right|_{\phi=v_0}= m_h^2=(125\,{\rm GeV})^2 \ ,
\end{align}
is used to determine the $\mu$ and $\lambda$ parameters.  

\begin{figure}
\includegraphics[width=0.455\textwidth]{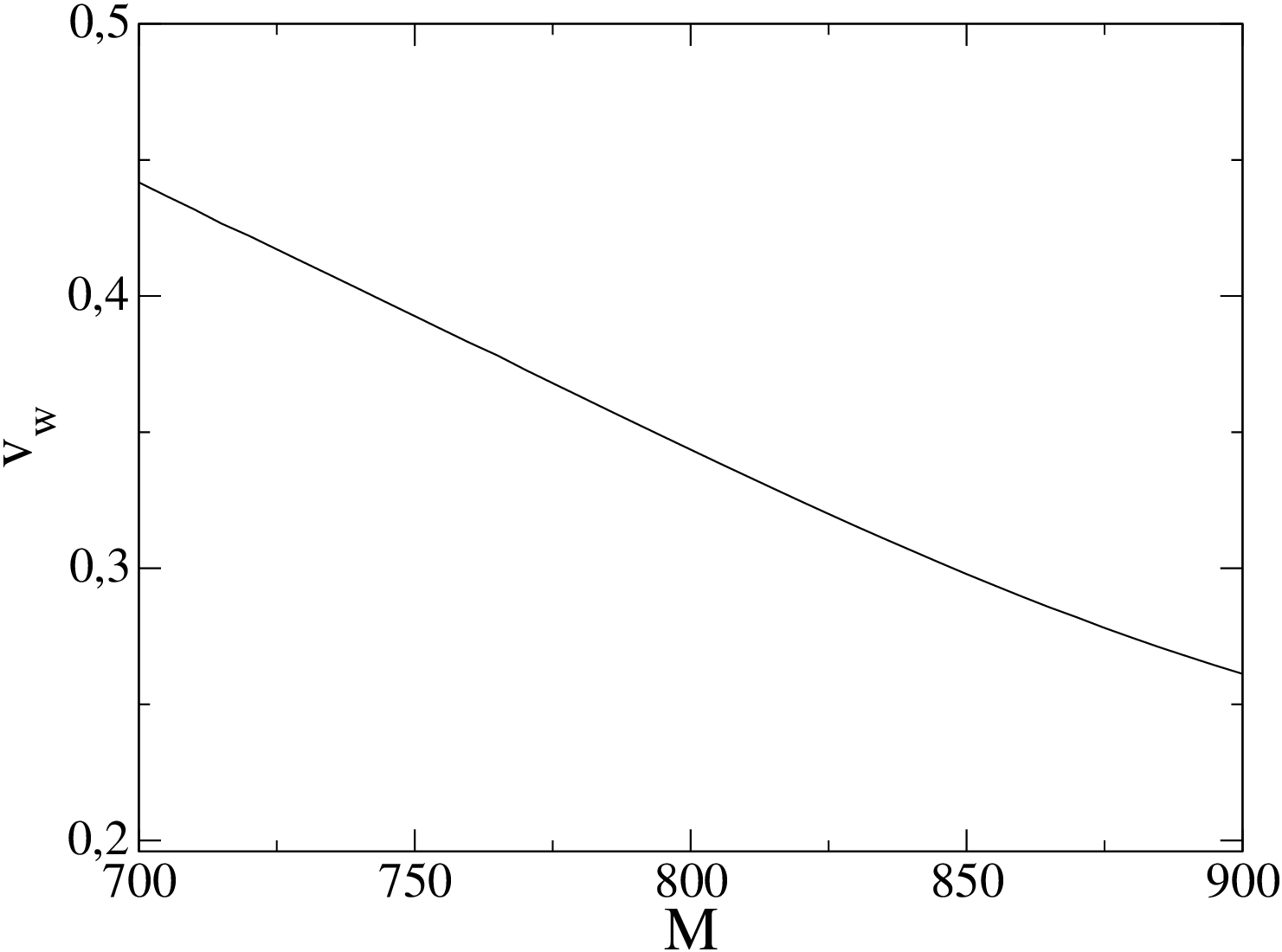}\qquad\quad
\hspace{4.5mm} 
\includegraphics[width=0.445\textwidth]{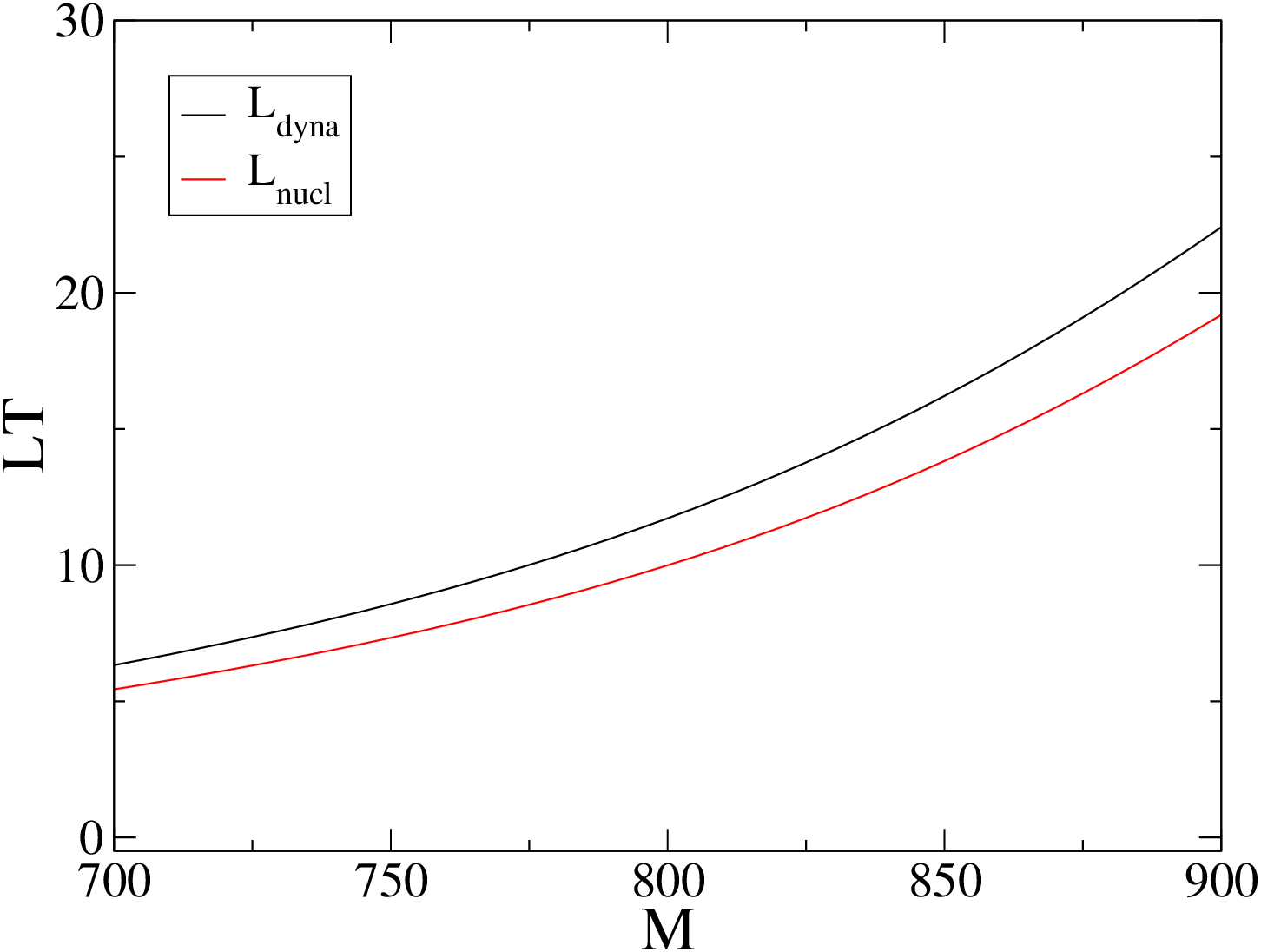}
\\[20pt]
{color{white}.}\hspace{-22mm}  
\includegraphics[width=0.46\textwidth]{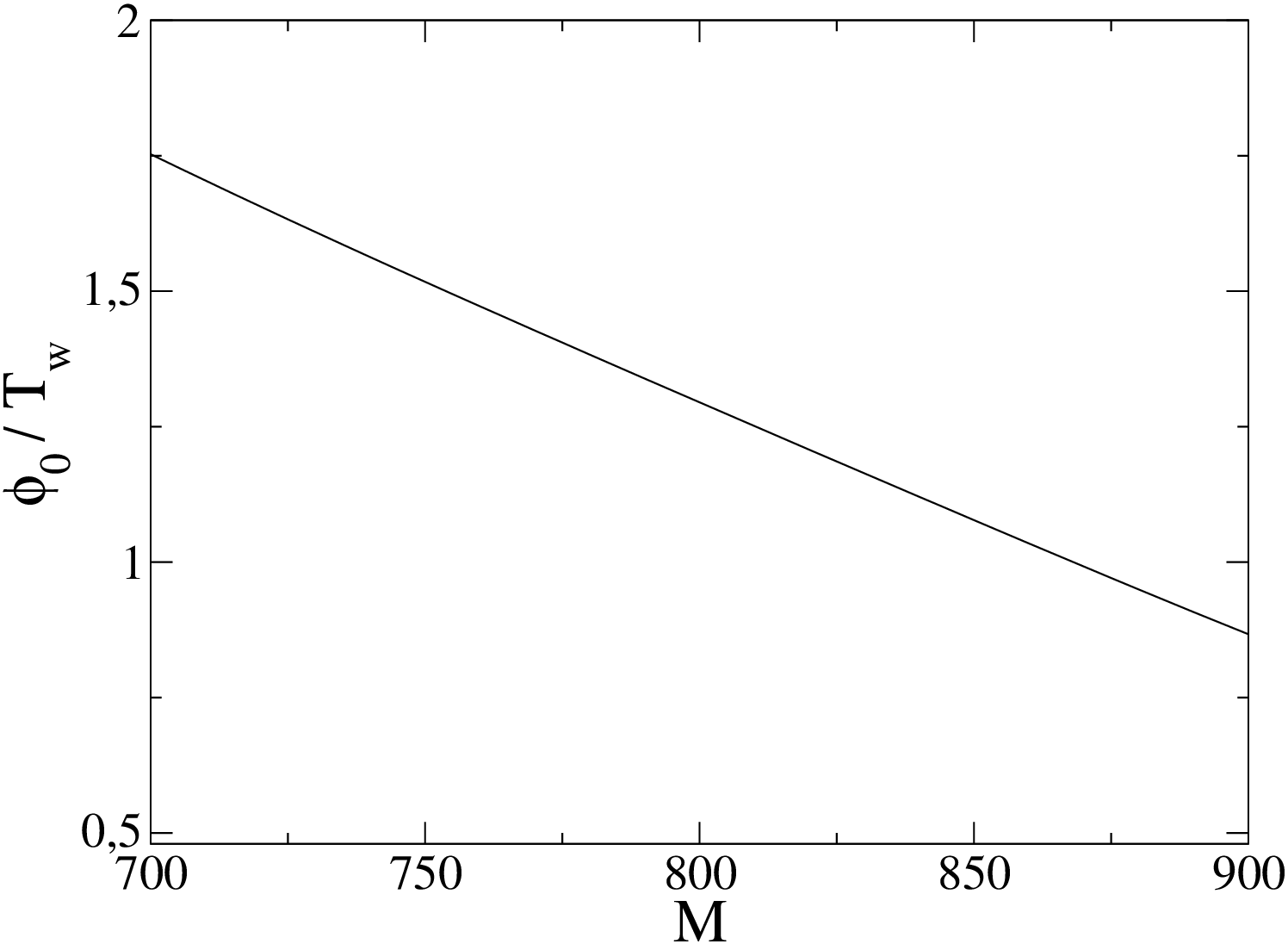}\qquad\quad\hspace{-1.5mm}
\includegraphics[width=0.487\textwidth]{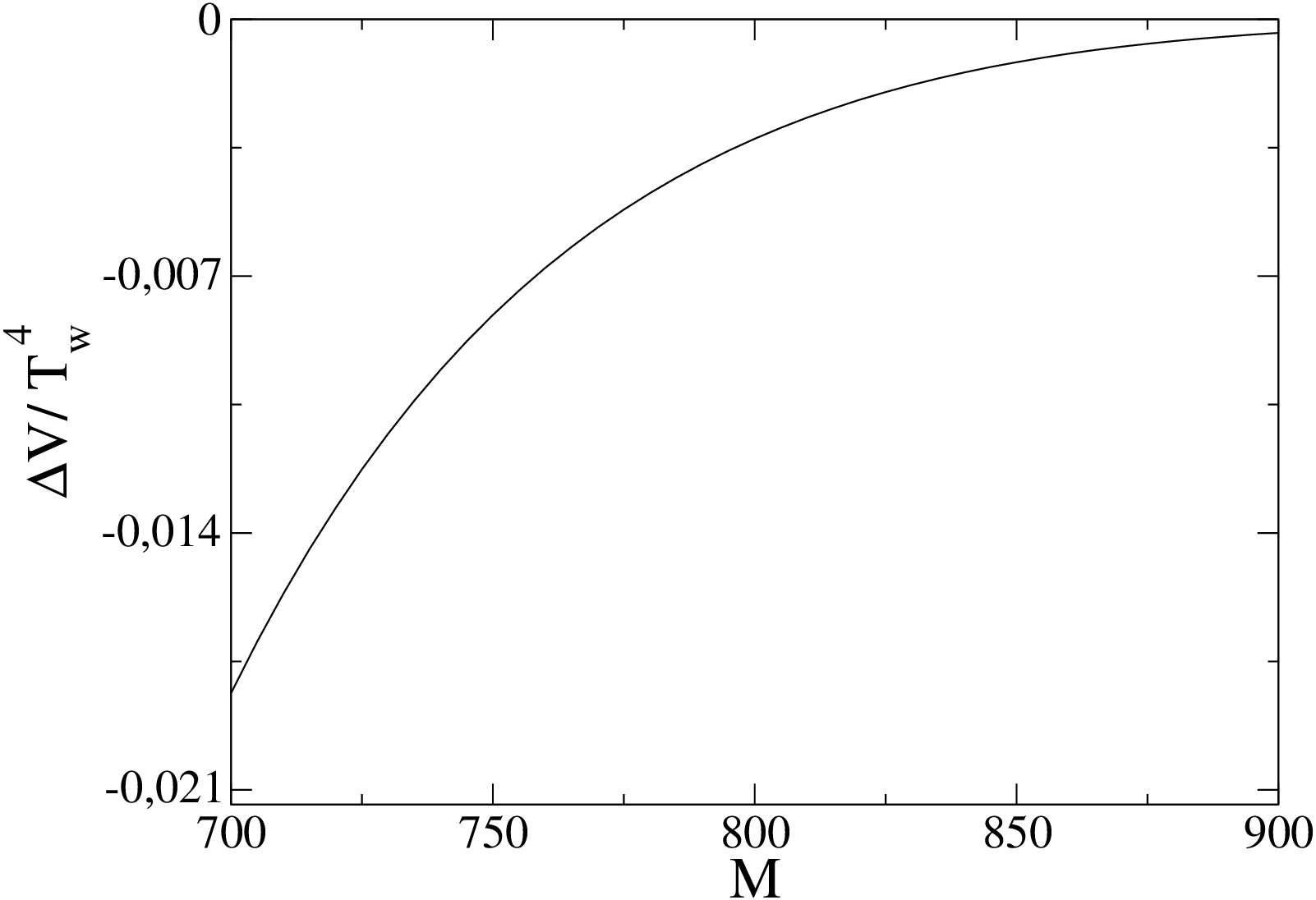}
\\[20pt]
{color{white}.}\hspace{-22mm}  
\includegraphics[width=0.46\textwidth]{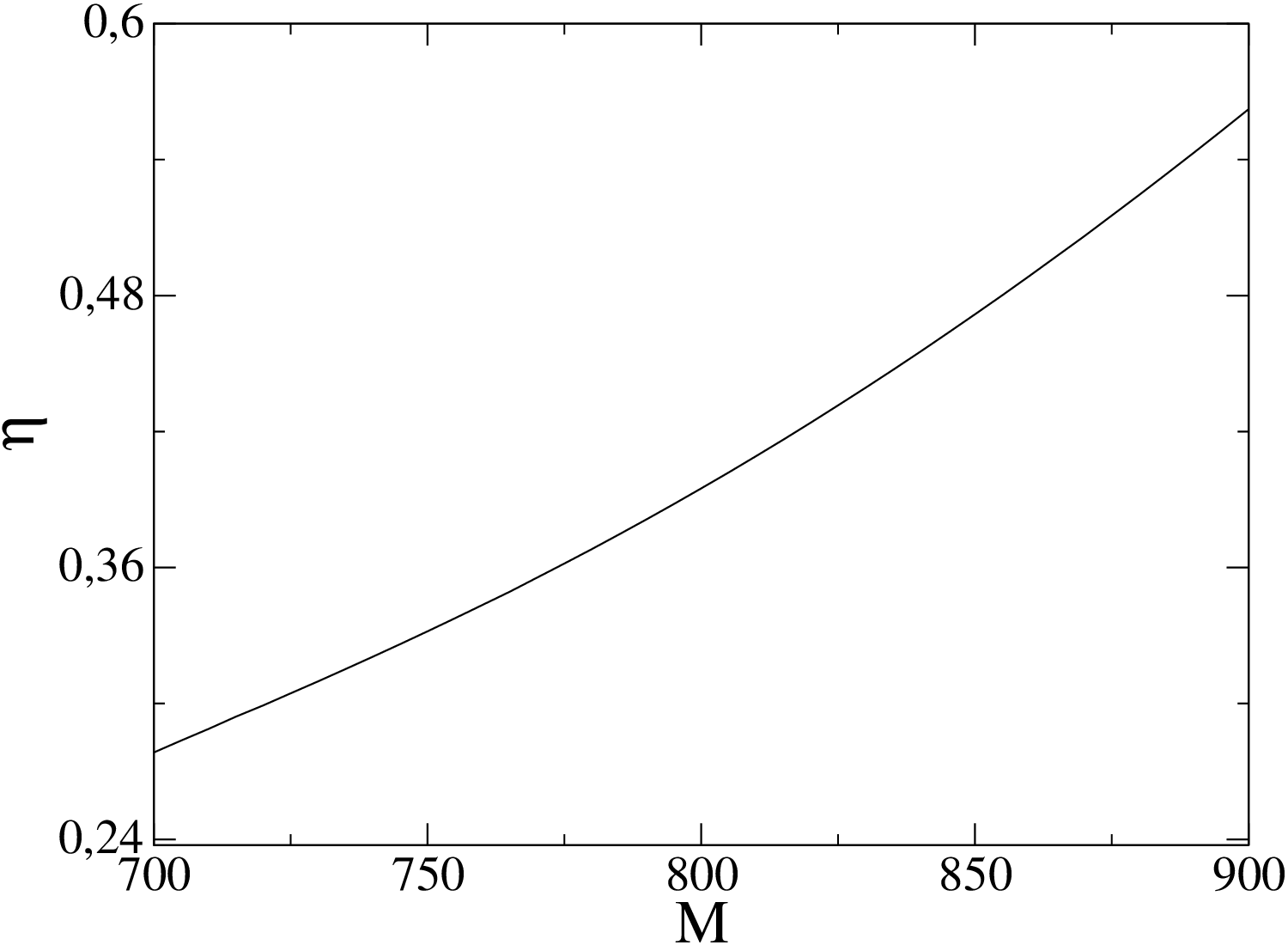}\qquad\quad
\hspace{2.mm} 
\includegraphics[width=0.463\textwidth]{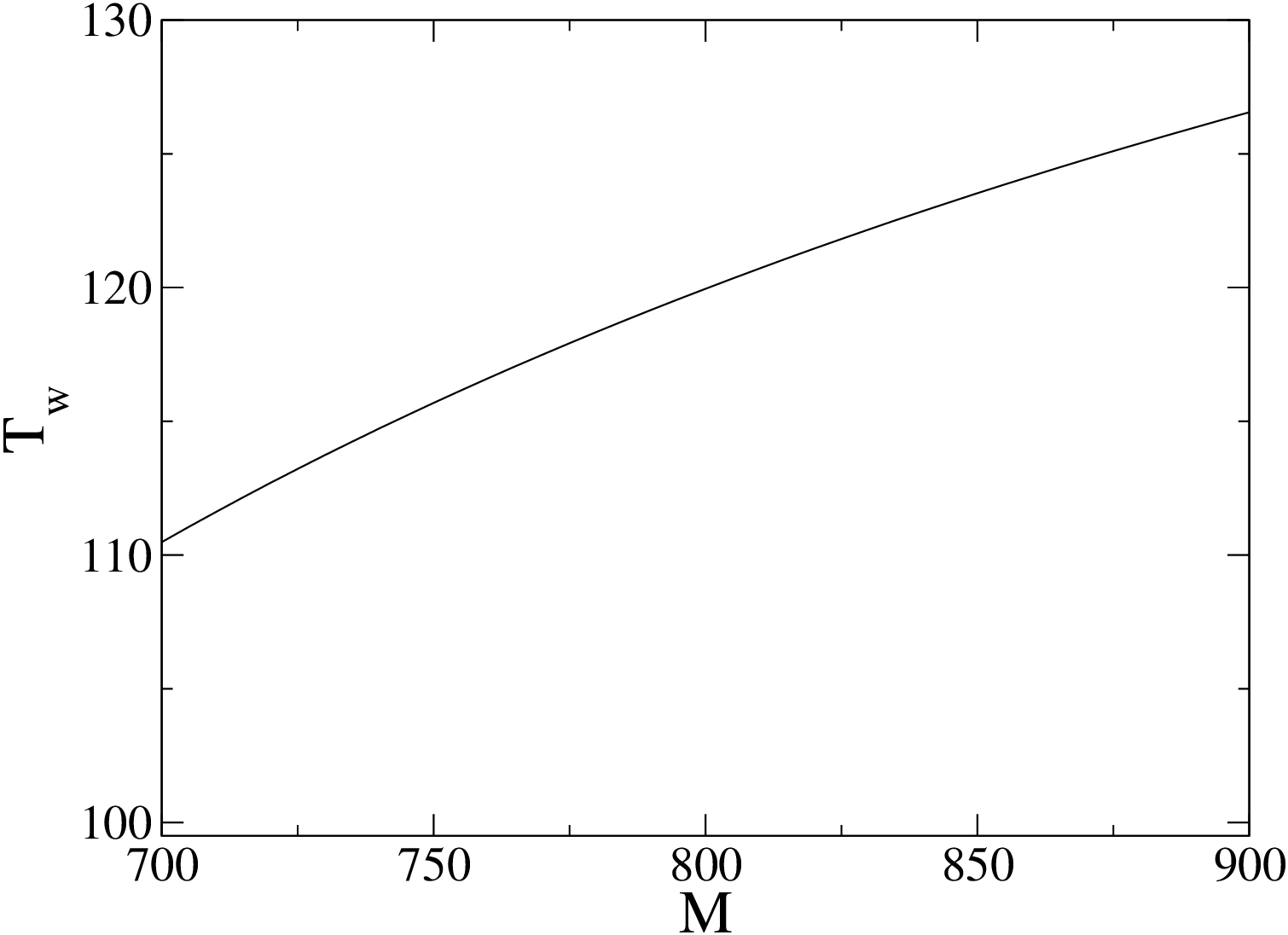}
        \caption{
\small Characteristics of the phase transition in the SM with a low cutoff}
    \label{fig:phi6}
\end{figure}

Our numerical results are displayed in figure~\ref{fig:phi6}. They are
very similar to those obtained in ref.~\cite{Huber:2013kj} by means of
the phenomenological approach based on eq.~\eqref{eq:phenoOld}. The
discrepancy is indeed smaller than a few percent, as table~\ref{tab2}
shows. In comparison, the wall velocity we obtain is slightly smaller
(larger) for light (heavier) new physics, namely, $M\sim 900\,$GeV
($M\sim 700$\,GeV). In particular, the wall thickness is small
and the wall velocity is rather close to the speed of sound, which
reduces the friction.  At the same time, we find slightly weaker phase
transitions in our potential, what reduces the wall velocity.

\begin{table}[H]
\centering
\begin{tabular}{|c|c|c|c|c|}
\hline
$M/$GeV & $\left(\frac{\phi_n}{ T_n }\right)^{hydro}$\cite{Huber:2013kj}& $ v_w^{hydro}$\,\cite{Huber:2013kj} & $\frac{\phi(T_w)}{ T_w }$& $ {v_w}$   \\
\hline
\hline
900 &  0.87  & 0.28--0.31 &  0.87 & 0.27 \\
800 &  1.30  & 0.34--0.37 &  1.29 & 0.35 \\
700 &  1.86  & 0.43--0.45 &  1.74 & 0.46 \\
\hline
\end{tabular}
\caption{
\small Comparison between hydrodynamic and microscopic approach in the SM
with a low cutoff. The hydrodynamic values are taken from
ref.~\cite{Huber:2013kj}. The two quoted values for the hydrodynamic wall
velocity correspond to two different friction coefficients.}
\label{tab2}
\end{table}

\subsection{Singlet model \label{sec:singlet}}

The third model we consider is the SM with an additional scalar.  The
extra scalar is an electroweak singlet and thus is only coupled to the
Higgs. In order to reduce the free parameters we take a model with a
manifest $Z_2$-symmetry. A peculiar feature of this model is that it
allows for very strong phase transition already in the mean field
approximations~\cite{Espinosa:2011ax}.

The Higgs potential of the model can be parametrized
as~\cite{Espinosa:2011ax}
\begin{align}
 V_{sing} (h,s,T=0)= -\frac 1 2 \mu_h \phi^2 + \frac 1 4 \lambda_h \phi^4 - \frac 1 2 \mu_s s^2
+  \frac 1 4 \lambda_s s^4 + \frac 1 4 \lambda_m \phi^2 s^2 \ .
\end{align}
Its temperature-dependent contribution is taken in the mean field
limit,
\begin{align}
V_{sing}(\phi,s,T)-V_{sing}(\phi,s,T=0)= \frac{T^2} 2 (c_h \phi^2 + c_s s^2) \ , 
\end{align}
and it is absorbed into the quadratic couplings:
\begin{align}
 V_{sing} (\phi,s,T)= - \frac 1 2 \mu_h(T) \phi^2 + \frac 1 4 \lambda_h \phi^4 - \frac 1 2 \mu_s
(T) s^2 +  \frac 1 4 \lambda_s s^4 + \frac 1 4 \lambda_m \phi^2 s^2 \ . \nn
\end{align}
In this parametrization, the strength of the phase transition at the
critical temperature $T_c$ turns out to be
\begin{align}
\frac{\phi_0(T_c)}{T_c} = \sqrt{\frac{v_{0}^2}{T_c^2}-\frac{c_h}{\lambda_h}}~.
\end{align}

Two of the five free parameters in the potential are fixed by imposing
$\phi_0(T=0)=v_{0}$ and $m_h=125\,$GeV. The remaining three
parameters can be expressed as a function of the singlet mass at zero
temperature $m_s$, the ratio $\phi_0(T_c)/T_c$ and the coupling
$\lambda_m$.

We restrict ourselves to the parameter space where the singlet
acquires a VEV before the electroweak phase transition. Because of
baryogenesis, we also require the transition to be rather
strong and the wall velocity to be subsonic. In spite of these
constraints, the allowed parameter space is still too large to allow
for an extensive analysis. We then focus on some benchmark points
where the free energies in the broken and unbroken phases at $T=T_n$
are not much different [$T_n$ is determined by the two-dimensional
bounce in the $(\phi,s)$ plane]. This automatically avoids those
configurations leading to runaway bubble
expansions~\cite{BodekerMoore}. At the same time, it
makes the bubble walls relatively thick.

Our benchmark points are listed in table~\ref{tab3}. The corresponding
numerical findings are also reported. As expected, since we are
implicitly working in the thick bubble regime, the bubble wall
thickness $L_{dyna}$ deviates from its initial value $L_{nucl}$ (see
section \ref{sec:eomHiggs}).  We have also quoted the effective
friction $\eta$ calculated in the phenomenological approach
\eqref{eq:pheno2}. In this model, the latent heat and the wall thickness 
are more or less independent parameters. Hence, we can use the model to test
a regime where the wall velocity is not too large but the wall thickness is 
relatively small. As expected from the discussion in section \ref{sec:beyond}, 
we find that in this regime the non-trivial dependence 
on the wall thickness can lead to an effective friction coefficient that differs by
up to $50\%$ from the one obtained in the SM (see figure \ref{fig:SM}).
\begin{table}[H]
\centering
\begin{tabular}{|c|c|c|c|c|c|c|c|c|c|}
\hline
$\frac{\phi_c}{T_c}$& $ m_s/$GeV & $ \lambda_m $ & $\Delta V/T_w^4$ &
$v_w$ & $L_{dyna} T_w$ & $L_{nucl} T_n$ & $\eta$ \\
\hline
\hline
1.00  & 75  & 0.303 & $1.0 \cdot 10^{-4}$ 
& 0.08 & 44.3 & 182.1 & 0.58  \\
1.00 & 75  & 0.299 & $1.1 \cdot 10^{-3}$
& 0.30 & 13.1 & 31.4  & 0.48  \\
1.25 & 100 & 0.690 & $8.7 \cdot 10^{-4}$
& 0.16 & 21.7 & 52.0  & 0.49  \\
1.25 & 75  & 0.345 & $1.5 \cdot 10^{-3}$ 
& 0.19 & 14.8 & 39.0  & 0.47 \\
1.50  & 100 & 0.793 & $6.1 \cdot 10^{-3}$
& 0.32 & 10.1 & 18.8  & 0.37 \\
1.50  & 100 & 0.826 & $9.6 \cdot 10^{-4}$
& 0.10 & 25.6 & 85.9  & 0.46  \\
\hline
\end{tabular}
\caption{\small 
Parameters of the phase transition in the singlet extension of the SM}
\label{tab3}
\end{table}

\section{Conclusion\label{sec:dis}}

Considerable progress has been made recently to determine the asymptotic wall
velocity of bubbles generated during a first-order phase
transition.  Two different procedures have been developed to address
this issue: the full Boltzmann treatment~\cite{MoAndPro,John:2000zq}
and the so-called phenomenological approach~\cite{Ignatius:1993qn,
Megevand:2009ut, Megevand:2009gh, Sopena:2010zz, Huber:2011aa,
Megevand:2012rt, Megevand:2013hwa, Huber:2013kj}.
In the former case, the dynamics of the fluid components is determined close to the interface between the two plasma phases
and used in the equation of motion of the Higgs. In the latter case the equation of motion of the Higgs is 
supplemented by a phenomenological friction term without considering its microscopic origin. 
In the present paper we have described how to extend the regime
of applicability of the Boltzmann approach and how to make contact to the phenomenological
approach from first principles.

Concerning the Boltzmann treatment, we have adopted the
Schwinger--Keldysh formalism to rederive the fluid and background
equations of motion including the correct relativistic behavior. In contrast to former work in the
literature~\cite{MoAndPro}, we have not assumed small wall velocities
but only small deviations from thermal equilibrium. Small deviations do not necessitate small wall velocities, 
as we have explained in the main text and showed in some examples. The Boltzmann
treatment with these new equations can hence be applied also to models
leading to supersonic detonation fronts. 

Subsequently, the new fluid and background equations have been solved
numerically assuming a thermal bath populated by a Standard Model-like
particle content at electroweak scales. In the Boltzmann approach, the Higgs equation contains 
two qualitatively different contributions from the plasma.
 
The first contribution comes from the particle species that are 
driven out-of-equilibrium due to interactions with the wall. 
It depends parametrially only on the Higgs wall thickness, the 
strength of the phase transition (in terms of $\phi/T$) and the wall velocity.
This term is parametrized in the phenomenological approach. 
It turns out that usually the phenomenological approach is well justified as long as the
(Lorentz-contracted) wall thickness is much larger than the mean free path of 
the particles in the plasma. In this case the relaxation-time approximation 
can be used to solve the Boltzmann equations, what justifies the phenomenological 
approach. Above criterion amounts for a SM-like particle content to a constraint on the wall thickness in terms 
of the temperature, $L \, T \gg 20$. If the phase transition produces thinner walls, 
the friction can be reduced considerably. For example we find cases with $30\%$ less friction
at $L \, T \simeq 10$. However, overall this first term in the Boltzmann equation is reproduced quite well.

The second term comes from the majority of particle species that is not driven out of equilibrium 
directly but nevertheless their temperature and velocity changes due to the latent heat that is 
released into the plasma and finally distributed under all degrees of freedom. In the phenomenological approach, this contribution is determined using local energy-momentum conservation but ignoring out-of-equilibrium effects. For small wall velocities, 
these background fields are not important
but their impact increases if the wall velocity approaches the speed of sound.
In this regime, the system changes from deflagrations 
to detonations. We have found that the out-of-equilibrium effects can have a large impact on the 
background fields. In extreme cases and depending on the latent heat, 
the Boltzmann approach can lead to no static bubble wall solutions at all.
One main difference to the phenomenologial approach is that 
this gap is quite substantial.

In order to benchmark the described improvements, we have analyzed
some specific models. The case of the Standard Model with a Higgs mass
below $70$ GeV shows that, in the non-relativistic regime with weak
phase transitions, our Boltzmann treatment agrees with former results using an expansion in small 
wall velocities~\cite{MoAndPro}. 
Some minor differences arise due to our non-linear treatment of the shock front of the wall.
Next, the Standard Model with a low cutoff allows for a comparison between
our outcomes and previous results obtained by means of the phenomenological
approach~\cite{Huber:2013kj}. Also here, no substantial discrepancies emerge. 
This is due to the fact that the wall velocity in these two models is relatively low.
Besides, the latent heat and the strength of the phase transition
are connected to a common scale (namely the Higgs mass and the cutoff, respectively).  
When these two quantities are decoupled from each other, the effective friction 
coefficient can vary more strongly even for relatively small wall velocities. 
This happens, if the model parameters allow to make the Higgs bubble wall thinner (without approaching the 
speed of sound). This can occur for instance in the singlet extension of
the Standard Model, that we discussed as a last model.

\vskip 0.3 cm

In summary, the Boltzmann approach to bubble wall velocities was so far limited to 
small wall velocities. In this regime phenomenological models to bubble wall friction 
lead to quite good results (once the friction coefficient is known and the bubble wall is not too thin, $L \, T \gg 20$). 
In the present work, we generalized the Boltzmann approach to larger wall velocity. In this regime, 
the phenomenological approach does not perform so well. The main reason is that the latent heat
of the phase transition is first released into the particle species that couple strongly to the Higgs and
then distributed under all remaining degrees of freedom by scatterings. 
The latter process is not represented well in the phenomenological approaches.
Therefore the full Boltzmann treatment has to be used when the wall velocity
approaches the speed of sound.

\section*{Acknowledgements}
This work was supported by the German Science Foundation (DFG) under the Collaborative
Research Center (SFB) 676 Particles, Strings and the Early Universe.
GN was supported in part by the National Science Foundation under Grant No. NSF PHY11-25915.

\appendix

\section{Lorentz properties and dimensional analysis of the kinetic equations \label{app:relations}}

The relation (\ref{eq:J_expansion}) displays certain relations between
the equilibrium densities that are not explicit in their
definitions. Assume that $u^\mu$ is a four vector that is not
constraint to the normalization $u^2 = 1$. Then one has
\be
\partial_{u_{\nu}}  J_0^{\mu} = \beta \bar T_0^{\mu\nu} \, .   
\ee
At the same time, Lorentz invariance implies for $J^{\mu}$ the form $J^{\mu} = u^\mu \, n$. 
The equilibrium distributions are a function of the combination $u^\mu \, \beta$ only, such that 
derivatives with respect to $u^\mu$ can be written in terms of derivatives with 
respect to $T$. For example 
\bea
\beta \bar T_0^{\mu\nu} &=& \partial_{u_\nu}  J_0^{\mu} = g^{\mu\nu} n - u^\mu u^\nu \,  \partial_T (T n) \, \nn \\
&=&  g^{\mu\nu} n - u^\mu u^\nu ( n + T \partial_T n) \, . 
\eea
Furthermore
\be
\beta u_\nu \bar T_0^{\mu\nu} = - u^\mu \, T \partial_T n = - u^\mu \, T \partial_T J_0^\mu \, .
\ee
This is consistent with (\ref{eq:J_expansion}), since $\delta \tau$ encodes just the fluctuations in the 
temperature.

Likewise one finds
\be
\beta \bar J_0^{\mu} = \partial_{u_\mu}  N_0 = - u^\mu \, T \partial_T N \, ,
\ee
and 
\be
\beta u_\mu \bar J_0^{\mu} =  - T \partial_T N \, .
\ee
Finally, using the parametrization for the energy momentum tensor
\be
T^{\mu\nu} \equiv u^\mu u^\nu\, \omega - g^{\mu\nu} \, p \, ,
\ee
with the enthalpy $\omega$ and the pressure $p$ one obtains
\be
\beta \bar M_0^{\mu\nu\lambda} =
(g^{\mu\lambda} u^\nu + g^{\nu\lambda} u^\mu) \omega - u^\mu u^\nu u^\lambda (2 \omega + T\partial_T \omega)
+ g^{\mu\nu} u^\lambda T \partial_T p \, .
\ee
The density $M$ is by construction symmetric in the three indices what implies the obvious relation 
$T \partial_T p = \omega$. Also in this case, contraction with the velocity reproduces the derivative with 
respect to temperature
\be
\beta u_\lambda \bar M_0^{\mu\nu\lambda} = - u^\mu u^\nu T \partial_T \omega + g^{\mu\nu} T \partial_T p
= - T \partial_T T_0^{\mu\nu}.
\ee
In conclusion, all the barred densities can be expressed in the usual densities and their temperature derivatives.
In terms of the out-of-equilibrium densities this implies
\be
J^\mu = J_0^\mu - T \partial_T J_0^\mu \delta \tau - u^\mu T \partial_T N \delta \mu
+ \delta u_\nu \, n  \, ,
\ee
and for the energy momentum tensor
\bea
T^{\mu\nu} &=& T_0^{\mu\nu} + \bar T^{\mu\nu}_0 \delta \mu
+ \beta \, \bar M_0^{\mu\nu\lambda} (\delta \tau u_\lambda + \delta u_\lambda) \nn \\
&=&  T_0^{\mu\nu} - T \partial_T T_0^{\mu\nu} \delta \tau \nn \\
&& + ( g^{\mu\nu} n - u^\mu u^\nu ( n + T \partial_T n) ) \delta \mu \nn \\
&& - (\delta u^\mu u^\nu +  u^\mu \delta u^\nu) \omega \, .
\eea

In fact, there are more consistency relations hidden related to the mass dependence that can be made 
explicit by analyzing the dimensionality of the various functions. For example, the four current 
is of dimension three, so it fulfills the relation
\be
2 m^2 \partial_{m^2} J^\mu = T \partial_T J^\mu - 3 J^\mu \, .
\ee
Analogously, one finds for the energy-momentum tensor the relation
\be
\label{eq:T_m_derivative}
2 m^2 \partial_{m^2} T^{\mu\nu} = T \partial_T T^{\mu\nu} - 4 T^{\mu\nu} \, .
\ee
Actually, these relations hold also for the different components of the densities, namely the 
enthalpy $\omega$ and the pressure $p$ (dimension $4$) as well as the densities $n$ (dimension $3$) 
and $N$ (dimension $2$). These  relations are important to establish a local equilibrium in case of a static 
wall. The equation for the current then reads
\be
\partial_\mu J_0^\mu = \partial_\mu m^2 u^\mu \partial_{m^2} n = {\rm coll} = 0\, ,
\ee
which is automatically fulfilled for static walls due to $\partial_\mu m^2 u^\mu = 0$.
On the other hand, the corresponding relation for the energy-momentum tensor reads
\be
\partial_\mu T_0^{\mu\nu} + \frac12 \partial_\nu m^2 N_0 = {\rm coll} = 0 \, ,
\ee
with
\be
\partial_\mu T_0^{\mu\nu} = \partial_\mu m^2 ( u^\mu u^\nu \partial_{m^2} \omega - g^{\mu\nu} \partial_{m^2} p) \, .
\ee
The first term vanishes again automatically for a static wall. The second cancels against the term involving $N$
only thanks to (\ref{eq:T_m_derivative}) and the relation $g_{\mu\nu} T^{\mu\nu} = m^2 N$. These relations are not very illuminating and also to remove derivatives with respect to $m$ using these equations are not really simplifying the system. 
Nevertheless, these relations are important for two reasons. First, if one wants to achieve explicit energy momentum conservation, these relations help to guide which terms are important. Second, these relations also lead to cancellations
in the non-equilibrium case. The easiest way to see this is by looking at the original equations for the Wightman functions
(\ref{eq:KB}). If the fluid ansatz is used, 
\be
G^< =  \frac{2\pi\delta(p^2 - m^2)}{1 \pm \exp(X)} \, , 
\ee
all the derivatives acting on the mass (which only show up in the on-shell delta function) are canceled by corresponding momentum-derivatives acting on the on-shell delta-function. The only terms that remain involve derivatives acting on $X$.

\section{Linearized collision terms\label{sec:coll}}

The matrices $\Gamma_W, \Gamma_t$ containing the collision terms appearing in the fluid equations in the wall frame are given by
\begin{align}
\Gamma_q &= \begin{pmatrix}\Gamma_{\mu_{q1}}  && \Gamma_{\delta T_{q1}}  && 0 \\
\Gamma_{\mu_{q2}}  && \Gamma_{\delta T_{q2}}  && 0 \\
0 && 0 && \Gamma_{v_{q}}  \end{pmatrix} \ ,
\end{align}
where the index $q$ stands for the particle species.
The following linearized collision terms were take from \cite{MoAndPro}. The numerical values for bosons and fermions are
\begin{align}
&\Gamma_{\mu_{f1}} = 0.00899 \, T\nonumber ,&& \Gamma_{\mu_{b1}}= 0.00521 \, T \ ,  \\ 
&\Gamma_{\mu_{f2}}= 0.01752 \, T \nonumber ,&& \Gamma_{\mu_{b2}}= 0.01012 \, T \ ,\\
&\Gamma_{\delta T_{f1}}= 0.01752 \, T ,&& \Gamma_{\delta T_{b1}}= 0.01012 \, T \ , \\
&\Gamma_{\delta T_{f2}}= 0.06906 \, T ,\nn && \Gamma_{\delta T_{b2}}= 0.03686 \, T \ , \\
&\Gamma_{v_{f}}= 0.03499 \, T ,\nonumber && \Gamma_{v_{b}}= 0.01614 \, T \ .
\end{align}
The corresponding diagrams have been calculated in the leading-log approximation. In the case of the W-bosons this approximation works very well since the gauge coupling is small. The uncertainties of the top quark collision terms
are much larger due to their color charge. The errors were estimated to be up to 50\% \cite{MoAndPro}. In order to check the impact on the friction we multiplied the matrix $\Gamma_t$ by a factor of $\chi = 0.5-1.5$.
The friction from the fluctuations in the thick wall regime scales with approximately $\frac{1+\chi}{2}$. The reason for this is that the contribution to the friction from top quarks and W-bosons is of the same order.  
The impact on the background contribution is not as easy to assess, but much smaller than the effect on the fluctuations.
From this we conclude that even if the collision terms of the top quarks are off by 50\% the resulting correction to the wall velocity is just of order 25\%.


\begin{thebibliography}{99}



\bibitem{Witten:1984rs}
  E.~Witten,
  ``Cosmic Separation of Phases,''
  Phys.\ Rev.\ D {\bf 30} (1984) 272.

\bibitem{Kosowsky:1991ua}
  A.~Kosowsky, M.~S.~Turner and R.~Watkins,
  Phys.\ Rev.\ D {\bf 45} (1992) 4514.

\bibitem{Kosowsky:1992rz}
  A.~Kosowsky, M.~S.~Turner and R.~Watkins,
  Phys.\ Rev.\ Lett.\  {\bf 69} (1992) 2026.

\bibitem{Kosowsky:1992vn}
  A.~Kosowsky and M.~S.~Turner,
  Phys.\ Rev.\ D {\bf 47} (1993) 4372
  [astro-ph/9211004].

\bibitem{Kamionkowski:1993fg}
  M.~Kamionkowski, A.~Kosowsky and M.~S.~Turner,
  Phys.\ Rev.\ D {\bf 49} (1994) 2837
  [astro-ph/9310044].

\bibitem{Huber:2008hg}
  S.~J.~Huber and T.~Konstandin,
  JCAP {\bf 0809} (2008) 022
  [arXiv:0806.1828 [hep-ph]].

\bibitem{Kuzmin:1985mm}
  V.~A.~Kuzmin, V.~A.~Rubakov and M.~E.~Shaposhnikov,
  Phys.\ Lett.\ B {\bf 155} (1985) 36.

\bibitem{Vachaspati:1991nm}
  T.~Vachaspati,
  Phys.\ Lett.\ B {\bf 265} (1991) 258.

\bibitem{Coleman:1977py}
  S.~R.~Coleman,
  Phys.\ Rev.\ D {\bf 15} (1977) 2929
   [Erratum-ibid.\ D {\bf 16} (1977) 1248].

\bibitem{Callan:1977pt}
  C.~G.~Callan, Jr. and S.~R.~Coleman,
  Phys.\ Rev.\ D {\bf 16} (1977) 1762.

\bibitem{Linde:1980tt}
  A.~D.~Linde,
  Phys.\ Lett.\ B {\bf 100} (1981) 37.

\bibitem{Dine:1992wr}
  M.~Dine, R.~G.~Leigh, P.~Y.~Huet, A.~D.~Linde and D.~A.~Linde,
  Phys.\ Rev.\ D {\bf 46} (1992) 550
  [hep-ph/9203203].

\bibitem{Liu:1992tn}
  B.~-H.~Liu, L.~D.~McLerran and N.~Turok,
  Phys.\ Rev.\ D {\bf 46} (1992) 2668.

\bibitem{Moore:1995ua}
  G.~D.~Moore and T.~Prokopec,
  Phys.\ Rev.\ Lett.\  {\bf 75} (1995) 777
  [hep-ph/9503296].

\bibitem{MoAndPro}
  G.~D.~Moore and T.~Prokopec,
  Phys.\ Rev.\ D\ {\bf 52} (1995) 7182
  [hep-ph/9506475].

\bibitem{John:2000zq}
  P.~John and M.~G.~Schmidt,
  Nucl.\ Phys.\ B {\bf 598} (2001) 291
   [Erratum-ibid.\ B {\bf 648} (2003) 449]
  [hep-ph/0002050].



\bibitem{Ignatius:1993qn}
  J.~Ignatius, K.~Kajantie, H.~Kurki-Suonio and M.~Laine,
  Phys.\ Rev.\ D {\bf 49} (1994) 3854
  [astro-ph/9309059].

\bibitem{Megevand:2009ut}
  A.~Megevand and A.~D.~Sanchez,
  Nucl.\ Phys.\ B {\bf 820} (2009) 47
  [arXiv:0904.1753 [hep-ph]].

\bibitem{Megevand:2009gh}
  A.~Megevand and A.~D.~Sanchez,
  Nucl.\ Phys.\ B {\bf 825} (2010) 151
  [arXiv:0908.3663 [hep-ph]].

\bibitem{Sopena:2010zz}
  M.~Sopena and S.~J.~Huber,
  J.\ Phys.\ Conf.\ Ser.\  {\bf 259} (2010) 012048.

\bibitem{Huber:2011aa}
  S.~J.~Huber and M.~Sopena,
  Phys.\ Rev.\ D {\bf 85} (2012) 103507
  [arXiv:1112.1888 [hep-ph]].

\bibitem{Megevand:2012rt}
  A.~Megevand and A.~D.~Sanchez,
  Nucl.\ Phys.\ B {\bf 865} (2012) 217
  [arXiv:1206.2339 [astro-ph.CO]].

\bibitem{Huber:2013kj}
  S.~J.~Huber and M.~Sopena,
  arXiv:1302.1044 [hep-ph].

\bibitem{Megevand:2013hwa}
  A.~Megevand,
  JCAP {\bf 1307} (2013) 045
  [arXiv:1303.4233 [astro-ph.CO]].


\bibitem{Hindmarsh:2013xza}
  M.~Hindmarsh, S.~J.~Huber, K.~Rummukainen and D.~J.~Weir,
  Phys.\ Rev.\ Lett.\  {\bf 112} (2014) 041301
  [arXiv:1304.2433 [hep-ph]].

\bibitem{Giblin:2014qia}
  J.~T.~Giblin and J.~B.~Mertens,
  arXiv:1405.4005 [astro-ph.CO].


\bibitem{Carena:2008vj}
  M.~Carena, G.~Nardini, M.~Quiros and C.~E.~M.~Wagner,
  Nucl.\ Phys.\ B {\bf 812} (2009) 243
  [arXiv:0809.3760 [hep-ph]].


\bibitem{Laine:2012jy}
  M.~Laine, G.~Nardini and K.~Rummukainen,
  JCAP {\bf 1301} (2013) 011
  [arXiv:1211.7344 [hep-ph]].


\bibitem{Cohen:2012zza}
  T.~Cohen, D.~E.~Morrissey and A.~Pierce,
  Phys.\ Rev.\ D {\bf 86} (2012) 013009
  [arXiv:1203.2924 [hep-ph]].

\bibitem{Curtin:2012aa}
  D.~Curtin, P.~Jaiswal and P.~Meade,
  JHEP {\bf 1208} (2012) 005
  [arXiv:1203.2932 [hep-ph]].

\bibitem{Carena:2012np}
  M.~Carena, G.~Nardini, M.~Quiros and C.~E.~M.~Wagner,
  JHEP {\bf 1302} (2013) 001
  [arXiv:1207.6330 [hep-ph]].

\bibitem{DeSimone:2011ek}
  A.~De Simone, G.~Nardini, M.~Quiros and A.~Riotto,
  JCAP {\bf 1110} (2011) 030
  [arXiv:1107.4317 [hep-ph]].


\bibitem{Barger:2007im}
  V.~Barger, P.~Langacker, M.~McCaskey, M.~J.~Ramsey-Musolf and G.~Shaughnessy,
  Phys.\ Rev.\ D {\bf 77} (2008) 035005
  [arXiv:0706.4311 [hep-ph]].

\bibitem{Ashoorioon:2009nf}
  A.~Ashoorioon and T.~Konstandin,
  JHEP {\bf 0907} (2009) 086
  [arXiv:0904.0353 [hep-ph]].

\bibitem{No:2013wsa}
  J.~M.~No and M.~Ramsey-Musolf,
  Phys.\ Rev.\ D {\bf 89} (2014) 095031
  [arXiv:1310.6035 [hep-ph]].

\bibitem{Ram:LHCsingl}
 S.~Profumo, M.~J.~Ramsey-Musolf, C.~L.~Wainwright and P.~Winslow,
  arXiv:1407.5342 [hep-ph].


\bibitem{Espinosa:2011ax}
  J.~R.~Espinosa, T.~Konstandin and F.~Riva,
  Nucl.\ Phys.\ B {\bf 854} (2012) 592
  [arXiv:1107.5441 [hep-ph]].

\bibitem{Kainulainen:2001cn}
  K.~Kainulainen, T.~Prokopec, M.~G.~Schmidt and S.~Weinstock,
  JHEP {\bf 0106} (2001) 031
  [hep-ph/0105295].

\bibitem{Prokopec:2003pj}
  T.~Prokopec, M.~G.~Schmidt and S.~Weinstock,
  Annals Phys.\  {\bf 314} (2004) 208
  [hep-ph/0312110].

\bibitem{Prokopec:2004ic}
  T.~Prokopec, M.~G.~Schmidt and S.~Weinstock,
  Annals Phys.\  {\bf 314} (2004) 267
  [hep-ph/0406140].

\bibitem{Konstandin:2013caa}
  T.~Konstandin,
  Phys.\ Usp.\  {\bf 56} (2013) 747
   [Usp.\ Fiz.\ Nauk {\bf 183} (2013) 785]
  [arXiv:1302.6713 [hep-ph]].




\bibitem{Enqvist:1991xw}
  K.~Enqvist, J.~Ignatius, K.~Kajantie and K.~Rummukainen,
  Phys.\ Rev.\ D {\bf 45} (1992) 3415.


\bibitem{Gyulassy:1983rq}
  M.~Gyulassy, K.~Kajantie, H.~Kurki-Suonio and L.~D.~McLerran,
  Nucl.\ Phys.\ B {\bf 237} (1984) 477.

\bibitem{Konstandin:2010dm}
  T.~Konstandin and J.~M.~No,
  JCAP {\bf 1102} (2011) 008
  [arXiv:1011.3735 [hep-ph]].

\bibitem{energybudget}
  J.~R.~Espinosa, T.~Konstandin, J.~M.~No and G.~Servant,
  JCAP {\bf 1006} (2010) 028
  [arXiv:1004.4187 [hep-ph]].

\bibitem{BodekerMoore}
  D.~Bodeker and G.~D.~Moore,
  JCAP {\bf 0905} (2009) 009
  [arXiv:0903.4099 [hep-ph]].




\bibitem{Aad:2012tfa}
  G.~Aad {\it et al.}  [ATLAS Collaboration],
  Phys.\ Lett.\ B {\bf 716} (2012) 1
  [arXiv:1207.7214 [hep-ex]].

\bibitem{Chatrchyan:2012ufa}
  S.~Chatrchyan {\it et al.}  [CMS Collaboration],
  Phys.\ Lett.\ B {\bf 716} (2012) 30
  [arXiv:1207.7235 [hep-ex]].



\bibitem{Kajantie:1996mn}
  K.~Kajantie, M.~Laine, K.~Rummukainen and M.~E.~Shaposhnikov,
  Phys.\ Rev.\ Lett.\  {\bf 77} (1996) 2887
  [hep-ph/9605288].

\bibitem{Aoki:1999fi}
  Y.~Aoki, F.~Csikor, Z.~Fodor and A.~Ukawa,
  Phys.\ Rev.\ D {\bf 60} (1999) 013001
  [hep-lat/9901021].



\bibitem{Quiros:1999jp}
  M.~Quiros,
  hep-ph/9901312.


\bibitem{Megevand:2013yua}
  A.~Megevand and F.~A.~Membiela,
  Phys.\ Rev.\ D {\bf 89} (2014) 103507
  [arXiv:1311.2453 [astro-ph.CO]].


\bibitem{Megevand:2014yua}
  A.~Megevand and F.~A.~Membiela,
  Phys.\ Rev.\ D {\bf 89} (2014) 103503
  [arXiv:1402.5791 [astro-ph.CO]].



\bibitem{Zhang1}
  X.~-M.~Zhang,
  Phys.\ Rev.\ D {\bf 47} (1993) 3065
  [hep-ph/9301277].



\bibitem{Grojean:2004xa}
  C.~Grojean, G.~Servant and J.~D.~Wells,
  Phys.\ Rev.\ D {\bf 71} (2005) 036001
  [hep-ph/0407019].

\bibitem{Steph}
 D.~Bodeker, L.~Fromme, S.~J.~Huber and M.~Seniuch,
  JHEP {\bf 0502} (2005) 026
  [hep-ph/0412366].



\bibitem{Zhang2}
X.~Zhang, B.~L.~Young and S.~K.~Lee,
  Phys.\ Rev.\ D {\bf 51} (1995) 5327
  [hep-ph/9406322].

\end{thebibliography}
\end{document}